\documentclass[aps,prd,nofootinbib,superscriptaddress]{revtex4-2}

\usepackage{amsmath,amssymb,amsfonts,bm}
\usepackage{graphicx}
\graphicspath{{figures/}{../manuscript/figures/}}
\usepackage{booktabs}
\usepackage{multirow}
\usepackage{xcolor}
\usepackage{placeins}
\usepackage{ragged2e}
\usepackage{hyperref}
\hypersetup{colorlinks=true,linkcolor=blue,citecolor=blue,urlcolor=blue}

\newcommand{\AU}{\mathrm{AU}}
\newcommand{\pc}{\mathrm{pc}}
\newcommand{\um}{\,\mu\mathrm{m}}

\newcommand{\rsun}{R_\odot}
\newcommand{\msun}{M_\odot}
\newcommand{\rg}{r_g}
\newcommand{\snr}{\mathrm{SNR}}
\newcommand{\snrC}{\mathrm{SNR}_{C}}
\newcommand{\snrR}{\mathrm{SNR}_{R}}
\newcommand{\data}{\bm d}
\newcommand{\stokes}{\bm S}
\newcommand{\thetaobs}{\mu_{\rm obs}}
\newcommand{\thetastar}{\mu_\star}
\newcommand{\Dimg}{D_{\rm img}}
\newcommand{\Delimg}{\Delta_{\rm img}}
\newcommand{\Disk}{\mathcal{D}}

\newcommand{\Ht}{\mathcal{H}}
\newcommand{\cov}{\bm C}
\newcommand{\pars}{\bm\Theta}
\newcommand{\model}{\mathcal{M}}
\newcommand{\fwdmodel}{\mathcal{F}}

\begin{document}

\title{A Covariance-Aware Framework for Spatially Resolved Exoplanet\\ Biosignature Inference with the Solar Gravitational Lens}

\author{Slava G. Turyshev}
\affiliation{Jet Propulsion Laboratory, California Institute of Technology,\\
4800 Oak Grove Drive, Pasadena, CA 91109-0899, USA}

\date{October 23, 2026}

\begin{abstract}
Assessing possible life on an exoplanet requires spatial, spectral, temporal, and environmental context. We develop a covariance-aware Solar Gravitational Lens (SGL) framework for regional biosignature inference from reconstructed Einstein-ring measurements. A controlled $0.45$--$2.40\um$ Stokes-$I$ experiment uses ten illuminated cells, 128 channels, and $R=70$ to compare resolved and exactly projected disk spectra with common parameters and priors. Across 96 synthetic scenes, the joint gas, vegetation, and mineral block gains $1.52\pm0.03$ nats after marginalizing cloud and calibration/response parameters; adding independently trained surrogate-discrepancy covariance gives $1.49\pm0.03$ nats. Held-out mismatch nevertheless produces linearized shifts up to $6.08$ posterior standard deviations; preserved precision does not establish unbiased recovery or evidence for life. For fixed Gaussian hypotheses, the expected log-likelihood-ratio loss under disk integration equals half the squared whitened contrast discarded by the projection. A separate $0.40$--$20\um$ continuum calculation provides absolute photon rates and observing-time estimates for an Earth-radius planet at $30\,\pc$, observed from $650\,\AU$. It includes source-averaged SGL amplification, finite-aperture extraction, coronal and host-star photons, detector noise, and a declared calibration floor. A reconstruction-covariance bracket reduces an $8\times8$ coadd gain from $7.77$ to $3.00$, costing $6.7$ times the exposure in the statistical-noise limit. The framework connects regional spectroscopy and co-location tests to the physical and instrumental validation needed to assess exoplanet biospheres and possible biological activity.
\end{abstract}

\maketitle

\section{Introduction}

The first persuasive detection of life beyond the Solar System will almost certainly be remote, indirect, and statistical.  A telescope will not see organisms.  It will measure radiation from a star after that radiation has been reflected, scattered, absorbed, emitted, polarized, blurred, and mixed by a planet and by the observing system.  The scientific question is therefore not whether O$_2$, O$_3$, CH$_4$, a red spectral slope, or a seasonal modulation is individually biological.  The question is whether the full set of measurements, with uncertainties and model alternatives included, is better explained by a living planet than by abiotic photochemistry, geology, weather, viewing geometry, and instrumental systematics.

The exoplanet-biosignature literature has converged on this contextual view.  Oxygen and ozone are powerful because they can indicate long-lived atmospheric redox disequilibrium, but abiotic O$_2$ and O$_3$ can be produced by water loss, CO$_2$ photolysis, suppressed recombination, low noncondensible inventories, or weak surface sinks \cite{DomagalGoldman2014,Harman2015,LugerBarnes2015,Meadows2018}.  Reduced gases such as CH$_4$, N$_2$O, and organosulfur compounds can strengthen a biological interpretation, yet they have their own abiotic sources and detectability limits \cite{Schwieterman2018,Seager2016BiosignatureGases}.  Surface pigments, including the terrestrial vegetation red edge, are scientifically attractive because they can be spatially local and chemically tied to energy capture, but mineral spectra, clouds, photometric dilution, and alternative metabolisms complicate interpretation \cite{Arnold2002,MontanesRodriguez2006,Tinetti2006,Fujii2018Biosignature}.  Circular spectropolarimetry is potentially distinctive because biological homochirality can generate sign-changing circular-polarization features, but the expected signals are small and instrumentally demanding \cite{Sparks2009,Patty2019,Stam2008}.  A credible biosignature-inference claim must therefore be a model-comparison problem with explicit false-positive and false-negative paths \cite{Catling2018}.  Complementary nonequilibrium-gate models connect biosphere emergence and persistence to observable margins such as atmospheric disequilibrium, seasonal coherence, surface reflectance, polarimetry, and climate stability \cite{Turyshev2026Nonequilibrium}.

Conventional direct-imaging missions are the natural first stage in this program.  A coronagraph, starshade, or mid-infrared interferometer can discover and characterize nearby terrestrial planets with disk-integrated spectra.  Such spectra can identify atmospheric H$_2$O, CO$_2$, O$_2$, O$_3$, CH$_4$, CO, clouds, and broad surface color.  However, the disk operator collapses ocean, land, ice, deserts, clouds, hazes, surface pigments, minerals, and atmospheric path lengths into one number per wavelength and epoch.  A local biosignature with amplitude $A_{\rm loc}$, clear-sky visibility $c$, and areal fraction $f$ appears disk-integrated with amplitude approximately $f c A_{\rm loc}$.  At fixed continuum count rate and noise, the dilution penalty is roughly $(fc)^{-2}$; here $fc$ is the feature-bearing flux fraction.  More importantly, disk integration hides covariance: a weak band can be a gas abundance, a cloud-height effect, a surface mixture, or a path-length effect.  Rotational light curves and spin-orbit tomography recover low-order spatial information \cite{CowanAgol2008,FujiiKawahara2012,CowanFujii2018}, but they do not provide regional spectra at 100--300 km scales for Earth analogs at tens of parsecs.

This complementarity defines the scientific sequence.  A Habitable Worlds Observatory (HWO)-class coronagraph or starshade is the natural route to optical/near-IR disk spectra of nearby terrestrial planets, while a Large Interferometer For Exoplanets (LIFE)-class mid-IR nulling interferometer is the natural route to disk-integrated thermal spectra in the $5$--$20\,\mu{\rm m}$ range \cite{Astro2020,Quanz2022LIFEI,Hansen2022LIFEIV}.  The Solar Gravitational Lens (SGL) is a complementary resolved-context observatory for selected, previously characterized planets.  HWO and LIFE can identify promising worlds and measure disk-integrated biosignature context, but do not directly resolve regional co-location at the 100--300 km scales considered here.  The proposed SGL measurements add regional spectra, surface/cloud/gas covariance separation, temporal recurrence, and co-location tests.  The single-planet calculation below is a reference case; observing additional planets requires a separate target-access and mission-allocation analysis.

The scale gap is severe.  An Earth-radius planet has angular diameter
\[
  \theta_\oplus = \frac{2R_\oplus}{z_0}\simeq 2.84\,\mu{\rm as}\left(\frac{30\,{\rm pc}}{z_0}\right),
\]
and a 128-element diameter map corresponds to an angular sample of only
$\theta_\oplus/128\simeq0.022\,\mu{\rm as}\,(30\,{\rm pc}/z_0)$.  A conventional filled aperture or interferometric baseline would need
\[
  B\sim\frac{\lambda}{\theta_\oplus/128}\simeq4.6\times10^3\,{\rm km}\left(\frac{\lambda}{0.5\,\mu{\rm m}}\right)\left(\frac{z_0}{30\,{\rm pc}}\right)
\]
for a 100 km-class reflected-light element, and $\simeq9.2\times10^4\,{\rm km}$ at $10\,\mu{\rm m}$ for the same angular sampling.  Even at $10\,{\rm pc}$ the corresponding baselines are roughly $1.5\times10^3\,{\rm km}$ at $0.5\,\mu{\rm m}$ and $3.1\times10^4\,{\rm km}$ at $10\,\mu{\rm m}$, before high-contrast suppression, spectral multiplexing, calibration, and multi-epoch surface mapping are included.  Interferometry can synthesize angular resolution in principle, but a facility that simultaneously provides these baselines, exoplanet dynamic range, broad spectral coverage, and stable regional spectroscopy is not part of any currently credible near- or mid-term architecture.  The SGL replaces that physical baseline with the Sun's gravitational lens, at the cost of focal-region access beyond $\simeq550\,\AU$, scanning, coronal calibration, and a demanding inverse problem.

The SGL changes the observational regime.  The Sun's gravitational field produces a strong-interference focal region beyond $\rsun^2/(2\rg)\simeq547.8\,\AU$, where $\rg=2G\msun/c^2$ \cite{Einstein1936,Eshleman1979,Turyshev2017Wave,TuryshevToth2017Diffraction}.  A spacecraft in the focal region near the optical axis of a target observes the target light as an Einstein ring around the Sun.  For an extended source the SGL does not produce a normal focal-plane image; it projects the planet into a compressed image cylinder, and a spacecraft must scan the image plane and solve an inverse problem \cite{TuryshevToth2020Photometric,TothTuryshev2021Recovery,TuryshevToth2022MNRAS}.  Spectrally resolved SGL calculations further show that the useful bandpass is not set by an arbitrary wavelength cutoff.  It is determined by the planet spectral radiance, source- and aperture-averaged SGL amplification, solar-corona and host-star shot noise, thermal and zodiacal backgrounds, detector noise, ring-extraction throughput, and whether solar suppression is provided by an internal coronagraph or an external occulter.  A conventional internal coronagraph tied to a meter-class aperture is naturally optical/short-near-infrared (short-near-IR) limited, whereas an external occulter can decouple solar suppression from the telescope diffraction limit and opens a plausible optical-to-mid-infrared (mid-IR) SGL observing mode \cite{TuryshevToth2022Spectral}.

The thesis of this paper is that the highest astrobiological value of the SGL is spatially resolved and spatially correlated spectroscopy.  The SGL science program has three linked modes.  First, optical imaging with a sufficiently large telescope and internal coronagraph, or an equivalent solar-suppression architecture, maps albedo, clouds, oceans, ice, continents, glint candidates, and rotation.  Second, full-band spectroscopy from reflected light through the short-IR transition and thermal-IR regimes uses the appropriate internal-coronagraph or external-occulter architecture to measure atmospheric and climate diagnostics.  Third, spatially correlated spectroscopy interprets those spectra together with location, surface class, cloud state, illumination geometry, temporal variability, and co-location with other diagnostics.  The most compelling long-term SGL biosignature-context product is therefore not reflected light alone, and not thermal emission alone, but a registered reflected-plus-emitted spectral cube: reflected-light maps identify surfaces, clouds, water/glint candidates, Rayleigh slope, O$_2$/O$_4$, and pigment/mineral context, while thermal-emission spectra constrain temperature, cloud-top height, heat redistribution, H$_2$O, CO$_2$, O$_3$, CH$_4$, N$_2$O, and climate consistency.  The SGL therefore does not merely take a picture; it tests whether atmospheric, surface, temporal, and spatial evidence form a coherent case for habitability and possible biological activity, and whether a complex-biosphere interpretation is more plausible than conservative abiotic alternatives.  A broad-band map identifies heterogeneity; a resolved spectral cube identifies what each region is made of; polarimetry constrains scattering, clouds, aerosols, glint, and possibly chirality; temporal sampling separates rotation, weather, seasons, and recurrent surface or atmospheric variability.  The object of inference is therefore
\begin{equation}
  \stokes(x,y,\lambda,t)=\left[I,Q,U,V\right](x,y,\lambda,t),
\end{equation}
not an isolated two-dimensional image or a single disk spectrum.  A resolved SGL data set can ask whether O$_2$ and CH$_4$ coexist in a wet atmosphere, whether H$_2$O absorption and ocean glint co-locate, whether a pigment-like edge is restricted to low-cloud land or shallow-water environments, whether the candidate pigment recurs seasonally, whether a Stokes-$V$ feature is co-located with that surface spectrum, and whether an abiotic oxygen or mineral-mimic model explains the full cube.  The simulations below, however, are Stokes-$I$ reflected-light simulations.  Linear and circular polarization are retained in the notation because they define important future observables and calibration requirements; they are not treated as demonstrated products of the numerical experiment.

This paper develops that program as a quantitative framework and controlled simulation study using a simplified surrogate forward model.  The scope is requirement tracing: defining the data product, inference problem, dominant covariance terms, and observing conditions that a flight-level model must close.  Within that scope, the controlled population diagnostics, adversarial mineral template, and scalar SGL operator are stress tests of information flow rather than external biosphere-population or optical-mission forecasts.  The framework combines three methodological elements: scalar SGL imaging benchmarks that connect the monopole lens to scanning, calibration, metrology, and inversion requirements; a full-band $0.40$--$20\um$ SGL spectral-performance calculation that makes the wavelength-dependent planet spectrum, corona, throughput, detector, thermal-background, and occulter problem explicit; and high-contrast-spectroscopy inference methods in which forward modeling, cross-correlation or molecular mapping, and empirically measured covariance determine the significance of extracted spectra rather than independent per-channel SNR values \cite{Soummer2012KLIP,Pueyo2016KLIPFM,Snellen2015HDSHCI,GrecoBrandt2016,BrogiLine2019,Ruffio2018Bayesian,Turyshev2026SGLExoimage,TuryshevToth2022Spectral}.  

The results are separated into three levels.  First, the model-independent results are geometric and statistical: disk integration dilutes localized signals and hides spatial covariance.  Second, the conditional results are obtained from the stated surrogate forward model, scalar SGL operator, noise prescription, and covariance model; they include the Stokes-$I$ reflected-light population information audit and parameter-block Fisher gains.  Third, mission forecasts require independent vector radiative transfer, broader geological and biological spectral libraries, physical cloud fields, blind out-of-template population tests, and a flight-relevant SGL optical and reconstruction model.

\begin{figure}[tbp]
\noindent\makebox[\textwidth][c]{\includegraphics[width=0.88\textwidth]{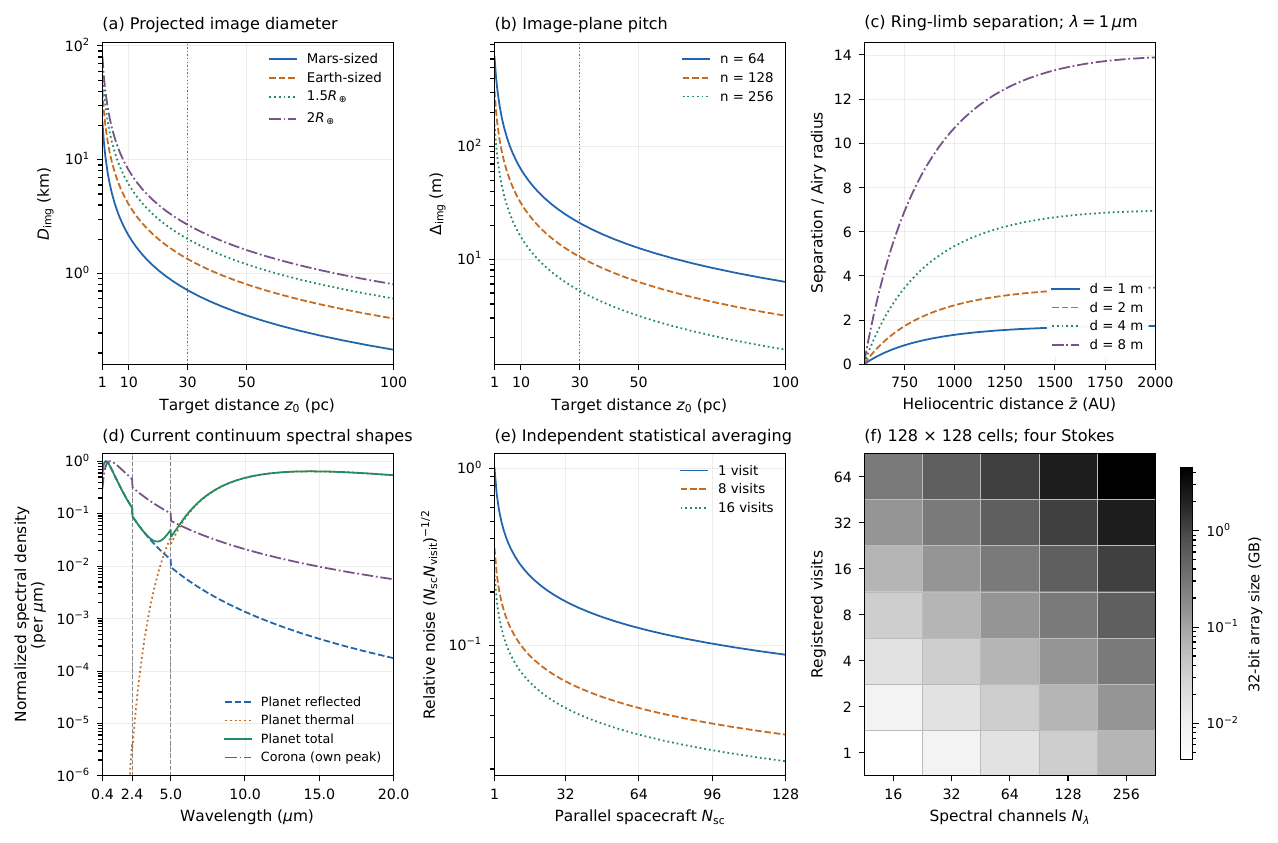}}
  \caption{\justifying Quantitative SGL spectroscopic-observatory scalings. Panels (a)--(b) give image diameter and sampling pitch at $650\,\AU$; the dotted line marks $30\,\pc$, and the Mars-sized illustration has radius $0.532R_\oplus$. Panel (c) gives ring--limb separation in Airy radii at $1\um$. Panel (d) uses the current detected continuum spectral densities: planet components share the total-planet peak normalization $2.448\times10^3\,{\rm e^-\,s^{-1}\,\mu m^{-1}}$, whereas the corona uses its own peak, $3.833\times10^8\,{\rm e^-\,s^{-1}\,\mu m^{-1}}$. Shapes use separate normalizations; dashed lines mark throughput-arm boundaries on the linear wavelength axis. Panel (e) shows $(N_{\rm sc}N_{\rm visit})^{-1/2}$ statistical-noise scaling at fixed elapsed raster duration per visit under ideal parallel sampling; pooling visits increases campaign duration. Panel (f) gives the 32-bit array size $4\times128^2\times4N_\lambda N_{\rm visit}/10^9$ GB for $128\times128$ cells and four Stokes components. One 128-channel visit occupies $33.55$ MB, excluding covariance, raw frames, masks, metadata, and calibration products.}
  \label{fig:framework}
\end{figure}

We connect SGL measurements to regional spectra while tracking errors shared across pixels, wavelengths, and epochs. The information audit measures conditional improvement in parameter precision, not evidence for life. Figures \ref{fig:framework} and \ref{fig:spectral_cube}
summarize the geometry and reconstruction; Table \ref{tab:notation}
defines the notation.

Table \ref{tab:status_hierarchy} distinguishes analytical results, controlled numerical calculations, architecture-level estimates, and requirements.  Numerical calculations are conditional on the stated forward model and covariance.  Architecture-level estimates use explicit photon budgets and extraction geometry, with assumed solar suppression and reconstruction penalties.  Validation requirements identify the additional evidence needed for mission-performance forecasts.

\begin{table*}[tbp]
\caption{\justifying Scientific status of the analytical results, controlled experiments, architecture estimates, and future observing requirements. Numerical consistency within a stated model is distinct from physical validation of an SGL mission or biosphere retrieval.}
\label{tab:status_hierarchy}
\begin{tabular}{@{}p{0.16\textwidth}p{0.29\textwidth}p{0.24\textwidth}p{0.26\textwidth}@{}}
\toprule
Status & Quantities & Interpretation & Further validation\\
\midrule
Analytical result & Disk dilution, Fisher ordering under exact projection, Gaussian contrast-loss identity & Statistical/geometric consequences under stated assumptions & Astrophysical relevance depends on realistic contrasts and covariance.\\
Controlled numerical reference & Ten-cell reflected-light population; explicit Jacobians, priors, covariance and null controls & Executed conditional precision experiment; class templates and normalized noise are assumed & Independent RT, broader surfaces/clouds, nonlinear bias and interval coverage.\\
Scalar-map / contrast references & Finite scalar radiance/index maps; graph-response spectral contrasts; prescribed regional covariance brackets & Executed response, index, and fixed-hypothesis contrast calculations & Independent scenes, physical reconstruction, and flight-relevant injection recovery.\\
Physical continuum reference & Absolute $0.40$--$20\um$ rates, SNR, exact exposure roots, background sensitivity & Executed source/optics/detector bookkeeping with ideal solar suppression and assumed scalar inverse & Occulter diffraction, measured coronal residuals, scan-dependent reconstruction, thermal molecular RT.\\
Requirements projection & Stokes $Q/U/V$, thermal retrieval, sub-ppm stability, navigation and cadence & Quantitative design drivers; not demonstrated capabilities & Laboratory/on-orbit calibration, vector RT, full wave optics and operations.\\
\bottomrule
\end{tabular}
\end{table*}

The paper is organized as follows: Section \ref{sec:framework} defines biosignature assessment as statistical model comparison and describes why imaging and spectroscopy are complementary rather than competing SGL products.  Section \ref{sec:planet} specifies the planet, atmosphere, cloud, surface, biological, geological, phase, and temporal assumptions.  Section \ref{sec:sgl} gives the SGL measurement model, with notation audited to avoid symbol collisions: $\thetastar$ and $\thetaobs$ denote illumination and viewing cosines, while $\bar\mu_{\rm SGL}$ denotes the SGL gain; $\Dimg$ is the image-cylinder diameter, $\Delimg$ is the image-plane pitch, and $\Disk$ is the disk-integration operator.  Section \ref{sec:covspectro} develops the covariance-aware spectral-cube inference machinery adapted from high-contrast spectroscopy.  Section \ref{sec:retrieval} gives the retrieval model and population information audit.  Section \ref{sec:simulations} describes the controlled simulations and their interpretation.  Section \ref{sec:results} presents spectral cubes, internal retrieval-consistency checks, region-conditioned spectra, false-positive stress tests, detectability trends, and a population-level conditional information-gain audit.  Sections \ref{sec:discussion} and \ref{sec:conclusions} identify robust conclusions, model-dependent conclusions, validation requirements, and mission implications.
Appendix~\ref{app:traceability} discusses numerical traceability.

\begin{table*}[tbp]
\caption{\justifying Notation, abbreviations, and key symbols used repeatedly in the manuscript.  The table separates illumination/viewing cosines from SGL amplification, identifies the wavelength-dependent photon-budget quantities used in the full-band spectral-performance calculation, and defines the covariance branches that control reconstruction and information-gain claims.}
\label{tab:notation}
\begin{tabular}{@{}p{0.21\textwidth}p{0.25\textwidth}p{0.44\textwidth}@{}}
\toprule
Symbol or abbreviation & Meaning & Typical use in this paper \\
\midrule
SGL & Solar Gravitational Lens & Solar gravitational focal region and Einstein-ring measurement geometry. \\
HWO, LIFE & Habitable Worlds Observatory; Large Interferometer For Exoplanets & Precursor disk-spectroscopy classes used for opportunity-cost comparison: optical/near-IR coronagraph or starshade spectra and mid-IR nulling-interferometer spectra. \\
IR, short-IR, mid-IR, thermal-IR & infrared; short-, mid-, and thermal-infrared regimes & Wavelength-regime labels in the $0.40$--$20\,\mu{\rm m}$ observability calculation. \\
RT & radiative transfer & Used for the reflected-light surrogate model and for the required future vector radiative-transfer validation. \\
LSF, IFS & line-spread function; integral-field spectrograph & LSF sets the effective resolving power; simultaneous IFS acquisition avoids sequential-channel dwell penalties. \\
SNR & signal-to-noise ratio & Used for convolved-ring statistics $\mathrm{SNR}_{C,a}$ and recovered-pixel statistics $\mathrm{SNR}_{R,a}$. \\
$\stokes(x,y,\lambda,t)=[I,Q,U,V]$ & Stokes spectral cube & Formal SGL data product; only Stokes-$I$ reflected light is simulated numerically. \\
$\lambda$, $R$, $N_\lambda$ & wavelength, resolving power, channel count & Spectral-performance and spectral-cube notation; logarithmic channelization is used; each displayed axis identifies its wavelength scale. \\
$\mu_\star$, $\mu_{\rm obs}$ & stellar-illumination and observer-viewing cosines & Reflected-light geometry and airmass; these are not SGL magnification factors. \\
$\bar\mu_{\rm SGL}(\lambda)$ & peak point-source SGL gain & Ideal monopole peak amplification $\propto\lambda^{-1}$; finite-aperture and extended-source averages are distinct. \\
$\bar z$, $z_0$ & SGL heliocentric distance and target distance & Image-cylinder compression, count-rate normalization, and mission-time scalings. \\
$D_{\rm img}$, $\Delta_{\rm img}$ & image-cylinder diameter and image-plane pitch & Scan geometry for a projected exoplanet image, Eqs. (\ref{eq:dimg})--(\ref{eq:deltaimg}). \\
$d$, $n$ & telescope diameter and linear raster size & Ring collection, scalar reconstruction penalty, metrology, and full-raster time scaling. \\
$q_{\rm refl}(\lambda)$, $q_{\rm th}(\lambda)$, $q_p(\lambda)$ & reflected, thermal, and total planet photon spectral density & Full-band signal model following the wavelength-dependent SGL spectral framework in Ref.~\cite{TuryshevToth2022Spectral}. \\
$q_{\rm cor}(\lambda)$ & solar-corona photon spectral density & Dominant wavelength-dependent background and shot-noise term in the SGL ring measurement. \\
$\eta_{s,a}$, ${\cal T}(\lambda)$ & relative correction; nominal response & $\eta_{s,a}=1$ at the reference setting; ${\cal T}$ includes wavelength-dependent throughput and spatial/extraction response. \\
$K_{\rm arch}$, $K_{\rm occ}$ & architecture extraction scale; external-occulter value & Dimensionless normalization; $K_{\rm occ}=1$ when source integration, extraction, throughput, and detector efficiency are explicit. \\
$Q^{\rm exo}_a$, $Q^{\rm bg}_a$ & channel-integrated signal and background rates & Inputs to the per-channel SNR and dwell-time calculations for spectral channel $a$. \\
$R_{\rm TOA}$ & top-of-atmosphere reflectance & Reflected-light surrogate-model output before SGL convolution; not an external validation product. \\
$\mathcal D$ & disk-integration operator & Maps a resolved cube or covariance to a disk-integrated spectrum. \\
$\mathcal H_{\rm SGL}$ & SGL measurement operator & Scalar convolution for the map test; graph response in Eq.~(\ref{eq:reference_response}) for information tests. A physical mission requires wave-optical ring extraction. \\
$C_{\rm rec}$, $C_{\rm RT}$, $C_{\rm SGL}$ & reconstruction, RT model-form, and SGL model-form covariance & Load-bearing covariance branches that must be measured or validated before a mission forecast. \\
$\rho_C$ & disk-only residual-covariance multiplier & Defined by Eq. (\ref{eq:rhoc}); $\rho_C=1$ is the exact common-data comparison. \\ 
$F_{\rm prior}$ & prior Fisher matrix & Common Gaussian coordinate priors with numerical widths in Table~\ref{tab:fisher_numerical_definition}. \\
$\Delta I_m[\mathcal F,C]$ & conditional Fisher-information audit & Requirement tracing for a specified forward model $\mathcal F$ and covariance $C$, not a universal separability forecast. \\
\bottomrule
\end{tabular}
\end{table*}

\section{Scientific framework}\label{sec:framework}

\subsection{Biosignature inference as model comparison}

Let $\model_L$ denote a class of living-planet models and $\model_A$ a set of abiotic alternatives.  The data $\data$ may include images, spectra, polarimetry, light curves, stellar context, and prior orbital information.  The relevant evidential quantity is the posterior odds,
\begin{equation}
  \frac{P(\model_L|\data)}{P(\model_A|\data)}
  =\frac{P(\model_L)}{P(\model_A)}
   \frac{P(\data|\model_L)}{P(\data|\model_A)}
  \equiv \frac{P(\model_L)}{P(\model_A)}K_{LA},
  \label{eq:odds}
\end{equation}
where $K_{LA}$ is a Bayes factor.  The likelihood must include nuisance parameters $\pars$: stellar spectrum, phase, cloud state, aerosol opacity, surface classes, gas columns, calibration terms, SGL optical response, and reconstruction covariance.  Thus
\begin{equation}
  P(\data|\model)=\int P(\data|\pars,\model)P(\pars|\model)d\pars .
  \label{eq:evidence}
\end{equation}
Eq. (\ref{eq:evidence}) is the formal reason that isolated detections are dangerous.  The contribution of a strong O$_2$ band to $K_{LA}$ depends on marginalization over abiotic O$_2$ production, dry atmospheres, O$_4$, CO, H$_2$O loss, cloud altitude, pressure, and stellar UV.  A red edge likewise requires mineral slopes, cloud leakage, surface mixture, phase, and instrument color to be included.  Marginalization can increase or decrease the evidence ratio.  A weak Stokes-$V$ feature can be decisive only if instrument polarization, scattering, and mineral circular polarization are controlled at or below the signal level.

\subsection{Why imaging and resolved spectroscopy are complementary}

Imaging-only data estimate morphology: albedo structure, cloud cover, ocean-glint candidates, surface heterogeneity, rotation, and possibly persistent continents.  Spectroscopy-only data estimate composition: gases, broad surface colors, and temperature or pressure proxies if the bandpass is broad enough.  The biosignature-inference problem requires both.  A surface pigment without atmospheric and climatic context is not enough; atmospheric disequilibrium without spatial and environmental context can be false-positive-prone.  Resolved spectroscopy supplies class-conditioned spectra,
\begin{equation}
  \bar S_c(\lambda,t)=\frac{\sum_p w_{pc}(t)S_p(\lambda,t)}{\sum_p w_{pc}(t)},
  \label{eq:class_spectrum}
\end{equation}
where $p$ indexes reconstructed surface elements and $c$ is a class such as low-cloud land, ocean, desert/mineral, ice, cloud top, glint region, or candidate pigment patch.  The weights $w_{pc}$ can be hard masks or posterior class probabilities.  Eq. (\ref{eq:class_spectrum}) is central: it changes a disk-integrated degeneracy into an observed distribution of spectra conditioned on environment.

For a local surface signal with amplitude $A_{\rm loc}$ and effective continuum-flux fraction $fc$ (surface fraction $f$ times clear-sky visibility $c$ in the uniform-brightness approximation), disk integration gives
\begin{equation}
  A_{\rm disk}\simeq f c A_{\rm loc},\qquad
  \frac{t_{\rm disk}}{t_{\rm resolved}}\simeq \frac{1}{(fc)^2},
  \label{eq:dilution}
\end{equation}
at equal continuum count rate and noise for the same feature significance.  This isolates dilution; it does not equate the photon budgets or reconstruction costs of disk and resolved observations.  For $f=0.05$ and $c=0.5$, the conditional penalty is $\sim1600$; actual exposure ratios require the branch-specific photon and covariance model.

There is an important formal point.  If the resolved data vector contains the disk spectrum as a deterministic projection, $\data_{\rm disk}=\Disk\data_{\rm res}$, and the two cases use common priors and a consistent covariance model, then the disk operator is a lossy linear map.  The Fisher matrix for the disk product cannot exceed that of the resolved product in the positive-semidefinite ordering, so a non-negative conditional Fisher gain is structurally expected.  Positivity of $\Delta I_m$ is therefore not the scientific result.  The scientific content is the block decomposition of the gain, its conditional magnitude under a stated covariance, and its failure modes under null controls and model-form covariance inflation.

The controlled comparison propagates the same reconstructed data and covariance into the disk product. A separate sensitivity sweep adds a stated disk-only residual covariance. This separates the information carried by spatial structure from gains introduced by a different noise assumption.

The claim hierarchy separates geometric dilution, conditional surrogate-model precision, and biological interpretation. Equation (\ref{eq:dilution}) gives the limiting dilution law. The audit $\Delta I_m[\fwdmodel,C]$ identifies constrained gas, surface, mineral, cloud/path, and calibration/SGL blocks under the adopted response and covariance. Extending these gains to biological/abiotic separability requires the independent validation in Sec.~\ref{sec:discussion}.

\section{Planet, atmosphere, surface, and biosignature model}\label{sec:planet}

\subsection{Geometry, wavelength grid, and fiducial target}

The fiducial planet is Earth-radius, observed at phase angle $\alpha=60^\circ$, orbiting a solar-type star.  The observer-viewing cosine is
\begin{equation}
  \thetaobs(x,y)=\cos\theta_{\rm obs},
\end{equation}
and the stellar-illumination cosine is
\begin{equation}
  \thetastar(x,y)=\cos\theta_\star .
\end{equation}
A projected pixel contributes when $\thetaobs>0$ and $\thetastar>0$.  The slant airmass used in the forward model is
\begin{equation}
  m(x,y)=\frac{1}{\max(\thetaobs,\mu_{\rm min})}
        +\frac{1}{\max(\thetastar,\mu_{\rm min})},
  \label{eq:airmass}
\end{equation}
with $\mu_{\rm min}=0.08$.  This cutoff is a surrogate-model parameter, not a physical limb treatment.

The SGL spectral-observability calculation in this paper is defined on the unified band
\begin{equation}
  0.40\leq\lambda\leq20\um .
  \label{eq:full_band}
\end{equation}
Across this range the photon budget changes character.  From $0.40$ to about $2.4\um$ the planet is treated primarily as a reflected-light source.  From $2.4$ to about $5\um$ reflected sunlight, thermal emission, detector background, and thermal-instrument terms all become comparable architecture variables.  From about $5$ to $20\um$ an Earth-temperature planet is self-luminous and the key observables are thermal climate, cloud-top temperature, and mid-IR molecular absorption or emission.  The peak point-source gain decreases as $\lambda^{-1}$; the extended-source average additionally depends on source geometry. The planet and background spectra, extraction, throughput, detector noise, and reconstruction covariance determine the useful bandpass.

The controlled retrieval and population simulations use the reflected-light subset of Eq. (\ref{eq:full_band}),
\begin{equation}
  0.45\leq\lambda\leq2.40\um,
  \qquad N_\lambda=128,
  \qquad \Delta\ln\lambda=\frac{\ln(2.40/0.45)}{128}=0.013078 .
  \label{eq:reflected_grid}
\end{equation}
The 129 channel edges are $\lambda_a^-=0.45\exp(a\Delta\ln\lambda)\um$, $a=0,\ldots,128$, with geometric channel centers. The sampling scale is $R_{\rm samp}=76.46$; a Gaussian $R_{\rm LSF}=70$ response spans $1.09$ channel spacings per FWHM. We evaluate the nonlinear reflectance on 16 logarithmic subbins per channel, convolve the normalized intensity with a Gaussian in $\ln\lambda$ padded by eight standard deviations, and integrate with exact subbin wavelength widths. Refinement to 64 subbins changes the tested gains by at most $1.5\times10^{-5}$ nat. The band covers O$_2$, O$_3$, H$_2$O, CH$_4$, CO$_2$, CO, and O$_4$ templates; Appendix~\ref{app:traceability} gives their numerical definitions. Figure~\ref{fig:spectral_model} separates these idealized reflected-light templates from the absolute full-band continuum calculation.

The same full-band SGL radiometric framework evaluates the short-IR and thermal-IR extensions.  These extensions include CH$_4$ near $3.3$ and $7.7\um$, CO$_2$ near $4.3$ and $15\um$, H$_2$O near $2.7$ and $6.3\um$, O$_3$ near $9.6\um$, and N$_2$O near $7.8$ and $17\um$.  They are not inserted into the reflected-light population information audit because that audit does not contain a validated thermal-emission radiative-transfer model, temperature-pressure structure, mid-IR cloud physics, or detector/background covariance.  They are instead treated as architecture-dependent observing modes governed by the wavelength-resolved SGL count model.  Table \ref{tab:fiducial} lists the fiducial numerical assumptions, and Table \ref{tab:wavelength_modes} separates the reflected-light, short-IR transition, and thermal-IR wavelength regimes used in the observability analysis.

\begin{table*}[tbp]
\caption{\justifying Distinct numerical and architectural references. The normalized Fisher experiment is independent of the absolute photon-count and scalar raster benchmarks.}
\label{tab:fiducial}
\begin{tabular}{@{}p{0.20\textwidth}p{0.27\textwidth}p{0.24\textwidth}p{0.23\textwidth}@{}}
\toprule
Calculation & Measurement & Reference & Purpose \\
\midrule
Reflected-light information experiment & $0.45$--$2.40\um$ normalized Stokes $I$, 128 channels, $R=70$, ten illuminated projected cells & Four spectral compressions; raw Gaussian noise $\sigma_0=0.004$ at normalized intensity $0.10$ & Conditional Fisher gains with explicit spatial response, common priors, and exact disk covariance. \\
Scalar imaging/raster benchmark & Reflected-light broadband counts and $128\times128$ raster & $d=1\,{\rm m}$, $30\,\pc$, $650\,\AU$; Eqs.~(\ref{eq:qexo})--(\ref{eq:cw}) & Reconstruction scalings, regional covariance brackets, and illustrative mission-time laws. \\
Full-band continuum reference & $0.40$--$20\um$ reflected plus emitted photons; ideal external solar suppression & $d=0.4\,{\rm m}$, $30\,\pc$, $650\,\AU$; $R=70$ (Fig.~\ref{fig:spectral_performance_envelope}) and $R=50$ (Fig.~\ref{fig:thermal_branch_trade}); $K_{\rm occ}=1$ & Absolute channel rates, continuum SNR, exact read-scheduled dwell, and background sensitivity. \\
Thermal/polarimetric extensions & Regional molecular bands, temperature, and Stokes $Q/U/V$ & Require thermal/vector RT, calibrated optical response, and time-dependent scanning & Future biosphere-context measurements; no thermal or polarized retrieval is simulated. \\
\bottomrule
\end{tabular}
\end{table*}

The $0.4\,{\rm m}$ continuum reference uses a common physical normalization for both spectral configurations, with only the channelization changed. Table~\ref{tab:radiometric_references} supplies absolute rates, throughput, detector inputs, and exposures. The information experiment instead fixes a dimensionless noise scale to isolate the effect of spatial projection; no physical duration is assigned to that normalized experiment.

\subsection{Reflected-light surrogate forward model}

The numerical reflected-light branch uses an intermediate-fidelity surrogate forward model.  Here, surrogate means a simplified deterministic radiative-transfer calculation with tunable coefficients that is used for controlled information-flow tests; it is more physical than a linear color mixture, but it is not a line-by-line vector multiple-scattering model.  Its purpose is to test information routing, retrieval conditioning, and false-positive behavior under stated assumptions, not to validate absolute gas abundances or cloud-radiative transfer.  All population and detectability metrics below are conditional on this surrogate model and should be read as internal diagnostics unless and until the same retrieval is reproduced with vector multiple-scattering radiative transfer.  The total gas optical depth in a pixel is
\begin{equation}
  \tau_g(x,y,\lambda)=\sum_j X_j(x,y)\widehat\tau_j(\lambda),
  \label{eq:gas_tau}
\end{equation}
where $j\in\{\mathrm{O_2,O_3,H_2O,CH_4,CO_2,CO,O_4}\}$ and $X_j$ are dimensionless amplitudes of explicit Gaussian optical-depth templates. $X_{\mathrm{O_4}}$ is an oxygen-pair template amplitude, not an independent molecular column. The templates are evaluated before the nonlinear transfer in Eqs.~(\ref{eq:tclear})--(\ref{eq:rtoa}); the intensity is then convolved and integrated into channels. These idealized bands do not represent physical opacity tables or establish gas abundances. That interpretation requires HITRAN or correlated-$k$ inputs and converged radiative transfer \cite{Gordon2022HITRAN,LacisOinas1991}.

Rayleigh and aerosol optical depths are modeled as
\begin{align}
  \tau_R(\lambda)&=0.055\Big(\frac{0.55\um}{\lambda}\Big)^4,\label{eq:rayleigh}\\
  \tau_a(\lambda)&=0.025\Big(\frac{0.55\um}{\lambda}\Big)^{1.3}.\label{eq:aerosol}
\end{align}
The clear-column top-of-atmosphere reflectance is
\begin{equation}
  R_{\rm clr}=R_{\rm path}+T_{\rm clr}\frac{A_s}{1-S_{\rm atm}A_s},
  \label{eq:clear_rt}
\end{equation}
where
\begin{align}
  T_{\rm clr}&=\exp\Big[-m\Big(\tau_g+0.15\tau_R+0.25\tau_a\Big)\Big],\label{eq:tclear}\\
  R_{\rm path}&=P_R(\alpha)\Big[0.035\Big(\frac{0.55\um}{\lambda}\Big)^4+0.012\Big(\frac{0.55\um}{\lambda}\Big)^{1.3}\Big],\label{eq:rpath}\\
  S_{\rm atm}&={\rm clip}\Big[0.05+0.12\Big(\frac{0.55\um}{\lambda}\Big)^4,0.02,0.55\Big].\label{eq:satm}
\end{align}
Here $A_s(x,y,\lambda,t)$ is the surface reflectance, $P_R(\alpha)=(3/4)(1+\cos^2\alpha)$ is the Rayleigh phase factor, and $S_{\rm atm}$ is an approximate atmospheric spherical albedo.  The numerical factors in Eqs. (\ref{eq:tclear})--(\ref{eq:satm}) are declared surrogate coefficients, fixed before evaluating the information gains; they are not derived from first principles.

Clouds are subpixel cloudy columns with fraction $f_c(x,y,t)$ and optical depth $\tau_c(x,y,t)$.  Cloud reflectance is
\begin{equation}
  A_c(x,y,\lambda,t)=0.78\big[1-\exp(-0.19\tau_c)\big]C_\lambda,
\end{equation}
where $C_\lambda=1-0.02\ln[\lambda/(0.55\um)]$ is the declared cloud color term.  The cloudy-column transmission uses a reduced gas path above the cloud top,
\begin{equation}
  T_{\rm cld}=\exp\big[-0.42e^{\theta_p}m(\tau_g+0.10\tau_R)\big].
\end{equation}
The total reflected-light surrogate model is
\begin{equation}
  R_{\rm TOA}=(1-f_c)R_{\rm clr}+f_c\Big[R_{\rm path}+T_{\rm cld}\frac{A_c}{1-0.25S_{\rm atm}A_c}\Big].
  \label{eq:rtoa}
\end{equation}
For projected pixels, the specific intensity before SGL propagation is
\begin{equation}
  I(x,y,\lambda,t)=\frac{F_\star(\lambda,t)}{\pi}\,\thetastar R_{\rm TOA}(x,y,\lambda,t).
  \label{eq:intensity}
\end{equation}
Here $F_\star$ is incident stellar spectral irradiance and $R_{\rm TOA}$ a reflectance factor.  Observer foreshortening enters once through $dA_{\rm proj}=\thetaobs dA$ when integrating surface-area elements.
The omitted physics is scientifically important: vertical pressure-temperature structure, correlated-$k$ or line-by-line gas absorption, full vector multiple scattering, cloud altitude distributions, ocean bidirectional reflectance distribution function (BRDF) and glint, three-dimensional cloud shadowing, and thermal emission.  These omissions define the validation path discussed in Sec. \ref{sec:discussion}.  They also limit the interpretation of the numerical performance metrics.  Formally, the physical top-of-atmosphere reflectance should be written
\begin{equation}
  R_{\rm phys}=R_{\rm TOA}^{B}(\pars)+\delta R_{\rm RT},
\end{equation}
where $R_{\rm TOA}^{B}$ is the baseline analysis model. Its mean discrepancy and centered covariance require calibration with independent radiative transfer or data. Here a second, explicitly perturbed surrogate supplies a controlled discrepancy test; it is not an independent RT validation. The mean discrepancy is kept separate from its covariance.

\subsection{Structural forward-model mismatch and RT model-form covariance}\label{sec:model_mismatch}

Model B supplies the baseline spectra and Jacobians. Model A changes the clear-column gas optical depth and the cloudy-column path factor to
\begin{align}
 \tau_{g,\mathrm{clr}}^A&=\tau_g[1+0.04\tanh\tau_g],\nonumber\\
 p_{\rm cld}^A&=0.42e^{\theta_p}[1+0.15\tanh h_c],
 \label{eq:model_a_cloud_shielding}
\end{align}
and replaces linear surface mixing by the bounded prescription
\begin{equation}
 A_s^A=1-\exp\left[\sum_k w_k\ln(1-A_k)\right],\qquad
 w_k\geq0,\quad\sum_kw_k=1.
 \label{eq:model_a_intimate_mixing}
\end{equation}
Here $h_c$ is the dimensionless scene field defined in Appendix~\ref{app:traceability}. These perturbations test structural sensitivity, not a physical particulate-mixture law. From 24 independent training scenes, define $e_q=f_A(0,q)-f_B(0,q)$, $\bar e=24^{-1}\sum_qe_q$, and
\begin{equation}
 C_{m,AB}^{\rm RT}=\frac{1}{23}\sum_{q=1}^{24}(e_{m,q}-\bar e_m)(e_{m,q}-\bar e_m)^T.
 \label{eq:ab_rt_covariance}
\end{equation}
The same training covariance is projected into each spectral and disk mode. The comparison adds it to the baseline covariance while retaining Model B's mean and Jacobian. For a local prior centered at the expansion point, a held-out residual $e$ produces the first-order posterior shift $\delta\pars\simeq F^{-1}J^T(C^{\rm eff})^{-1}e$. Across the 96 evaluation scenes, resolved shifts have RMS $0.618$ local posterior standard deviations and reach $6.08$ for one vegetation-edge coordinate; subtracting the training mean gives $0.597$ and $5.82$. For the same sixth land-photosynthetic scene and vegetation coordinate, the disk shifts are $1.78$ and $1.43$ posterior standard deviations before and after mean correction. These are first-order local displacements, not nonlinear retrieval errors or interval coverage; narrower resolved posteriors remain sensitive to model bias.

\subsection{Surface spectra, biology, and geological false positives}

The framework considers ocean, vegetation, desert, basalt, ice, mineral-edge, and oxidized surfaces, motivated by terrestrial biosignature and spectral-library studies \cite{Kiang2007,Hegde2015,Clark2007,Baldridge2009}. The executed population reference uses three explicit components: an ocean--land background, a vegetation edge, and one mineral mimic. Their coefficients and mixture fields are given in Appendix~\ref{app:traceability}. The vegetation component is
\begin{equation}
\begin{split}
  A_{\rm veg}(x,y,\lambda,t)&=A_{\rm vis}(\lambda)+b_{\rm veg}(x,y,t)A_{\rm edge}(\lambda),\\
  A_{\rm edge}(\lambda)&=\Big[1+\exp\Big(-\frac{\lambda-0.705\um}{0.018\um}\Big)\Big]^{-1}.
\end{split}
  \label{eq:rededge}
\end{equation}
The reference has $A_{\rm vis}=0.04$ and fiducial $b_{\rm veg}=0.45$, varied in a bounded logit coordinate. The mineral template has a broader rise centered at $0.72\um$ and a Gaussian absorption near $0.9\um$. Both are idealized end members. Spatial fractions are fixed within each local Fisher evaluation, so the result is conditional on known scene morphology. Seasonal recurrence and broader geological discrimination remain additional tests, not simulated features in the information audit.

The biological interpretation is contextual.  A red edge is suggestive only if it is localized in plausible habitats, persists under cloud-aware regional spectra, is spectrally sharper or otherwise different from mineral alternatives, recurs in time, and is not better explained by geology.  The worked biological case is intentionally Earth-like: an atmosphere- and surface-modifying biosphere with land photosynthesis.  This is a scientifically valuable target class, not a universal model of life.  A marine, subsurface, microbial-mat, low-productivity, pre-oxygenic, or nonphotosynthetic biosphere may produce no detectable surface pigment and may remain cryptic to this measurement strategy.  The simulations therefore include both a land-photosynthetic case and a cryptic inhabited case with weak surface expression.

Circular polarization is modeled as a possible but aspirational channel.  We write
\begin{equation}
\begin{split}
  \frac{V}{I}(x,y,\lambda,t)=&\;\epsilon_\chi B(x,y,t)\psi_\chi(x,y,\lambda,t)
 +\epsilon_{\rm min}M(x,y)\psi_{\rm min}(\lambda),
\end{split}
  \label{eq:stokes_v}
\end{equation}
Here $B$ and $M$ are biological and mineral masks, and $\psi_\chi=(\partial A_{\rm veg}/\partial\lambda)/\max_\lambda|\partial A_{\rm veg}/\partial\lambda|$, set to zero when the denominator vanishes.  Both $\psi_\chi$ and the mineral mimic $\psi_{\rm min}$ have unit peak absolute amplitude.  The adopted scale is $\epsilon_\chi=3\times10^{-4}$ for biological patches and $\epsilon_{\rm min}=10^{-4}$ for the adversarial mineral term.  For ideal analyzer-count statistics, $\sigma_{V/I}\simeq[\eta_{\rm pol}\sqrt{f_V}\snr_I]^{-1}$, where $\snr_I$ refers to the total reference photon budget, $f_V$ is its fraction allocated to circular polarimetry, and $\eta_{\rm pol}$ is the modulation efficiency.  The examples adopt $f_V=1/2$; the resulting $\sqrt{2}$ is an allocation choice, not a universal dual-beam penalty.  An $m\sigma$ detection of a fractional circular-polarization amplitude $P_c$ therefore requires
\begin{equation}
  \snr_I\gtrsim \frac{m\sqrt{2}}{\eta_{\rm pol}P_c}.
  \label{eq:v_snr}
\end{equation}
For $m=5$, $P_c=10^{-3}$, and $\eta_{\rm pol}=0.7$, Eq. (\ref{eq:v_snr}) gives $\snr_I\simeq1.0\times10^4$ before instrumental-polarization, coronal-subtraction, and deconvolution systematics.  The fiducial biological amplitude in Eq. (\ref{eq:stokes_v}) is smaller, $P_c\simeq\epsilon_\chi=3\times10^{-4}$, for which the same calculation gives $\snr_I\simeq3.4\times10^4$.  We therefore do not treat Stokes $V$ as a guaranteed pillar of the evidence budget; we treat it as a high-value, systematics-limited discriminator if it can be calibrated.

\begin{figure}[tbp]
\noindent\makebox[\textwidth][c]{\includegraphics[width=\textwidth]{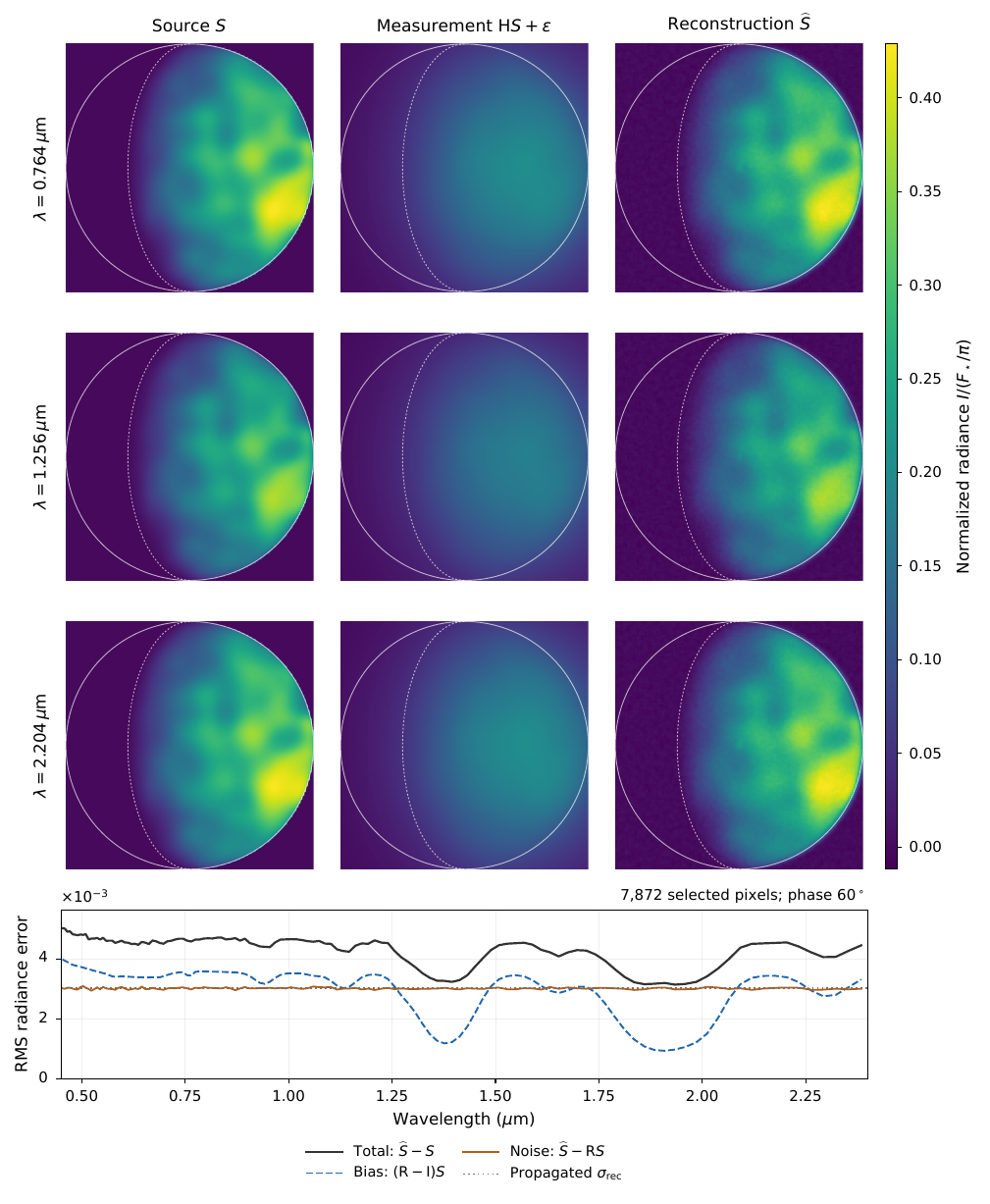}}
  \caption{\justifying Finite scalar spectral-cube reference at the channel centers nearest $0.76$, $1.25$, and $2.20\,\mu\mathrm m$. Columns: source $S$, measurement $\mathsf HS+\epsilon$, and Wiener reconstruction $\widehat S$, on a common radiance scale. The $128\times128$ fields retain blur/noise outside the illuminated support; outlines mark the limb and $60^\circ$ terminator. Lower panel: RMS total error, regularization bias and response-centered noise over $128$ channels and $7,872$ selected pixels, with covariance-predicted $\sigma_{\rm rec}=0.003015$. Appendix~A defines the scene, scalar blur, noise, inverse, and boundaries. This scalar test is distinct from the ten-cell Fisher reference and does not establish wave-optical SGL reconstruction performance.}
  \label{fig:spectral_cube}
\end{figure}

\begin{figure}[tbp]
\noindent\makebox[\textwidth][c]{\includegraphics[width=0.88\textwidth]{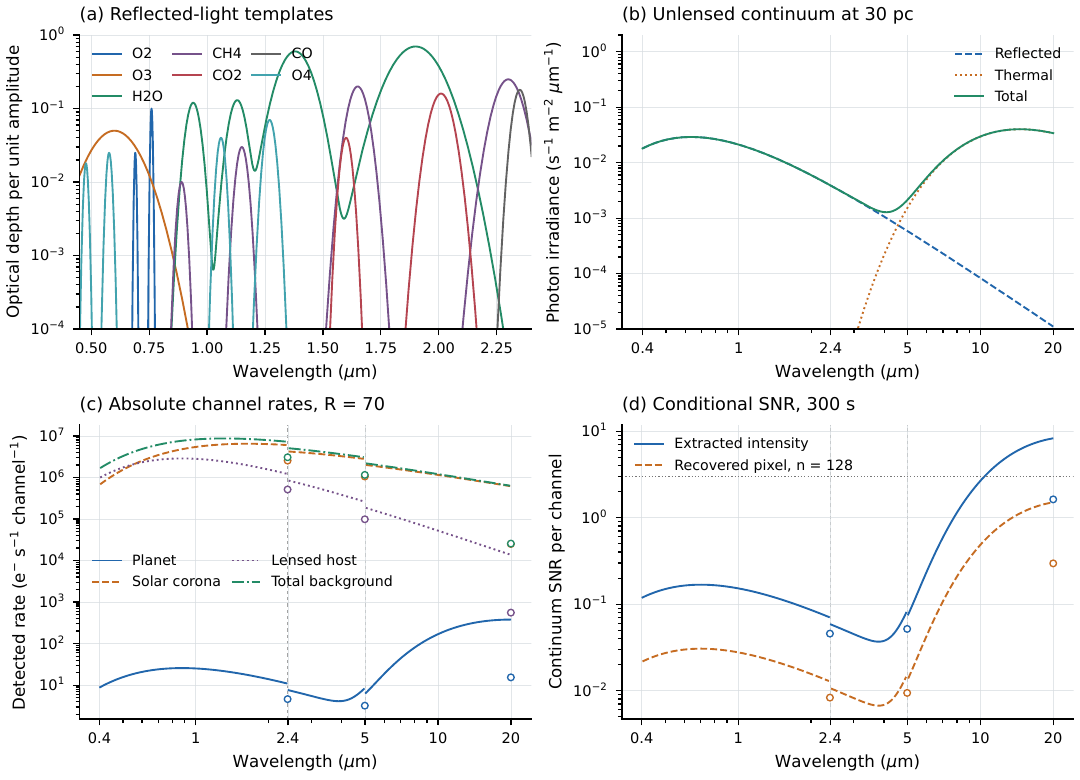}}
  \caption{\justifying Two distinct spectral references. (a) Gaussian optical-depth templates for the $0.45$--$2.40\um$ Fisher experiment, before transfer and instrumental convolution. (b) Unlensed continuum photon irradiance of the Sun/Earth analogue at $30\,\pc$, separating reflected and thermal photons. (c) Absolute $R=70$ channel rates for planet, corona, unmasked host, and total background. (d) Extracted-intensity and $n=128$ recovered-pixel SNR for $300\,{\rm s}$ per sample. Panels (b)--(d) use the $0.4\,{\rm m}$ ideal solar-suppression reference with finite-source gain, Airy extraction, detector terms, and the coronal floor. Full-band axes are logarithmic. Dashed lines mark detector-arm boundaries; hollow circles retain partial-width bins. Continuum-channel SNR is not molecular-detection significance.}
  \label{fig:spectral_model}
\end{figure}

\section{SGL and instrumental measurement model}\label{sec:sgl}

\subsection{Image-plane compression and SGL gain}

For a source-plane coordinate $\bm x'$ and image-plane coordinate $\bm x_0$, the SGL image-cylinder compression is
\begin{equation}
  \bm x_0\simeq -\frac{\bar z}{z_0}\bm x',
  \label{eq:compression}
\end{equation}
where $\bar z$ is heliocentric observer distance and $z_0$ is target distance.  For an Earth-radius planet,
\begin{equation}
  \Dimg=2R_\oplus\frac{\bar z}{z_0}
  =1.338\left(\frac{\bar z}{650\,\AU}\right)
  \Big(\frac{30\,\pc}{z_0}\Big)\, {\rm km}.
  \label{eq:dimg}
\end{equation}
For a linear raster size $n$, the image-plane pitch is
\begin{equation}
  \Delimg=\frac{\Dimg}{n}.
  \label{eq:deltaimg}
\end{equation}
At $n=128$ in the fiducial $30\,\pc$, $650\,\AU$ case, $\Delimg=10.46\,{\rm m}$ and the source-plane pixel scale is $2R_\oplus/n=99.6\,{\rm km}$.

The ideal monopole SGL gain is denoted here by
\begin{equation}
\mu_{\rm SGL}(\rho)=  \bar\mu_{\rm SGL}(\lambda) J_0^2\!\Big(k\rho\sqrt{\frac{2r_g}{\bar z}}\Big),\qquad
  \bar\mu_{\rm SGL}(\lambda)=\frac{4\pi^2\rg}{\lambda}.
  \label{eq:sgl_gain}
\end{equation}
 Here $k=2\pi/\lambda$ and $r_g=2GM_\odot/c^2$.  For $\lambda=1\um$, Eq. (\ref{eq:sgl_gain}) gives $\bar\mu_{\rm SGL}\simeq1.17\times10^{11}$.  The finite-aperture, aperture-averaged scalar kernel has a long tail approximately proportional to $d/(4\rho)$ in the image plane \cite{TuryshevToth2022MNRAS}.  For an on-axis point source and a uniformly collecting circular aperture of radius $a=d/2$, area averaging gives
\begin{equation}
 \langle\mu_{\rm SGL}\rangle_a=\bar\mu_{\rm SGL}[J_0^2(u)+J_1^2(u)],\qquad
 u=ka\sqrt{2r_g/\bar z}.
 \label{eq:aperture_gain}
\end{equation}
This follows by integrating $2\rho J_0^2(u\rho/a)/a^2$ over $0\leq\rho\leq a$. The bracket is chromatic; an extended-planet measurement additionally requires source integration and ring extraction.

\subsection{Wavelength coverage and observing architecture}\label{sec:bandpass}

The $0.40$--$20\um$ SGL band is treated here as one spectral observability problem with three physical regimes.  In the visible and near-IR reflected-light regime ($0.40$--$2.4\um$ in the performance model, with the numerical reflected-light cube beginning at $0.45\um$), the dominant biosignature-assessment observables are O$_2$/O$_4$, H$_2$O, CH$_4$, CO$_2$, CO, clouds, Rayleigh slope, surface color, pigment-like edges, and mineral false positives.  In the short-IR transition regime ($2.4$--$5\um$), the planet moves from reflected sunlight toward thermal emission; CH$_4$ at $3.3\um$, CO$_2$ at $4.3\um$, and H$_2$O bands become strong, while detector and thermal terms become architecture variables.  In the thermal-IR regime ($5$--$20\um$), an Earth-temperature planet is self-luminous and the strongest diagnostics are H$_2$O near $6.3\um$, O$_3$ near $9.6\um$, CO$_2$ near $15\um$, CH$_4$ near $7.7\um$, N$_2$O near $7.8$ and $17\um$, cloud-top temperature, and climate context.

The emitted-light regime is not scientifically secondary.  For an Earth--Sun analog, a reflected-light planet/star contrast is approximately
\begin{equation}
  C_{\rm refl}\simeq A_g\Phi(\alpha)\left(\frac{R_p}{a_p}\right)^2\simeq 2\times10^{-10}
  \left(\frac{A_g\Phi}{0.1}\right),
  \label{eq:reflected_contrast}
\end{equation}
whereas the disk thermal contrast is
\begin{equation}
  C_{\rm th}(\lambda)\simeq \left(\frac{R_p}{R_\star}\right)^2
  \frac{B_\lambda(T_p)}{B_\lambda(T_\star)} .
  \label{eq:thermal_contrast}
\end{equation}
For $T_p=252\,{\rm K}$ and $T_\star=5772\,{\rm K}$, Eq. (\ref{eq:thermal_contrast}) gives $C_{\rm th}\simeq8\times10^{-8}$ at $10\um$ and $3.4\times10^{-7}$ at $15\um$, hundreds to thousands of times larger than the reflected-light contrast in Eq. (\ref{eq:reflected_contrast}).  The SGL measurement is not a conventional star-contrast measurement--the dominant local backgrounds are the solar corona, detector and telescope backgrounds, zodiacal/exozodiacal light, and reconstruction covariance--but the planet's own thermal photon spectrum can still make the $5$--$20\um$ regime an unusually strong source of climate and molecular information if the long-wavelength observing architecture closes.

The useful bandpass is architecture-dependent.  A conventional internal coronagraph must suppress solar light while the Einstein ring is separated from the solar limb by a diffraction scale that worsens as $\lambda/d$.  For meter-class apertures this naturally favors optical and short-near-infrared (short-near-IR) observations and becomes unfavorable in the mid-IR.  An external occulter or equivalent annular ring-extraction system decouples solar suppression from the telescope diffraction limit and can preserve throughput into the mid-IR, but it replaces an internal-coronagraph problem with formation flying, occulter leakage, cryogenic detector, thermal-background, and calibration covariance problems.  The analysis below therefore uses the full $0.40$--$20\um$ count model for spectral performance, while restricting the synthetic population retrievals to the $0.45$--$2.40\um$ reflected-light cube that the current reflected-light surrogate model can support.  Operationally, this separates the first-generation imaging problem from the full spectroscopic and polarimetric problem: high-fidelity spatial maps are expected first in the optical and near-IR where ring/limb separation and detector backgrounds are more favorable, whereas annular spectroscopy and targeted Stokes measurements can use the full $0.40$--$20\um$ band only if an external occulter or equivalent ring-extraction architecture supplies the required solar suppression, calibration stability, detector performance, and thermal-background covariance.  The full-band material should therefore be read as an architecture-level observability layer.  It demonstrates common photon-budget and covariance bookkeeping for reflected, transition, and thermal regimes; it does not demonstrate an optical-through-thermal retrieval, a thermal-IR biosignature retrieval, or full-Stokes recovery.  

\begin{figure}[tbp]
\noindent\makebox[\textwidth][c]{\includegraphics[width=0.88\textwidth]{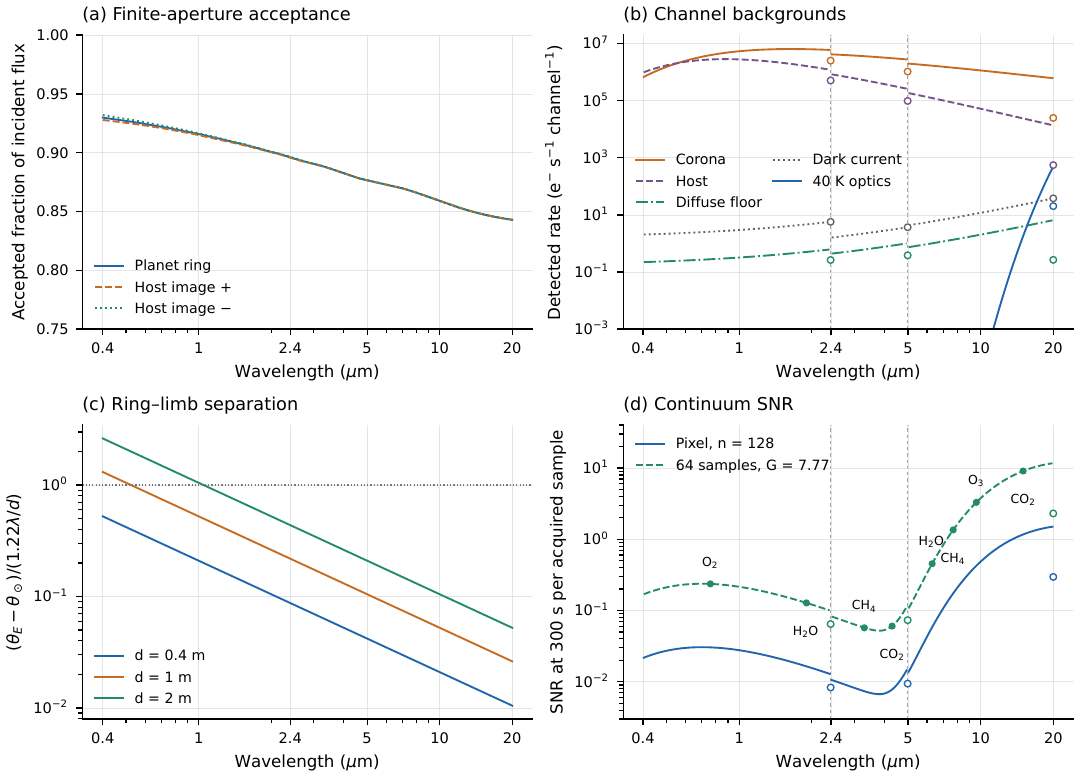}}
  \caption{\justifying Optical acceptance and background budget of the continuum reference. (a) The normalized Airy acceptance for the planet ring and the two lensed host-star images in the circular extraction aperture. (b) Absolute $R=70$ coronal, host, diffuse-sky, dark-current, and $40\,{\rm K}$ optics rates. (c) Geometric ring--solar-limb separation in units of $1.22\lambda/d$ for three apertures; this diagnoses separation, not coronagraph throughput or SNR. (d) Continuum recovered-pixel and $64$-sample regional SNR at $300\,{\rm s}$ per sample, with $G_{\rm reg}=7.77$ and an unaveraged common coronal floor. Filled markers locate diagnostic channels; hollow circles retain partial bins. These are not molecular detections. Wavelength axes are logarithmic; the solar photosphere is ideally suppressed.}
  \label{fig:extended_wave}
\end{figure}

For the full-band blackbody reference, the unlensed photon irradiance at the observer is the sum of reflected stellar radiation and thermal emission,
\begin{align}
  q_{p}(\lambda) &= q_{\rm refl}(\lambda)+q_{\rm th}(\lambda),\nonumber\\
  q_{\rm refl}(\lambda) &=
  \frac{\lambda}{hc}\pi\Big(\frac{R_p}{z_0}\Big)^2
  A_g(\lambda)\Phi(\alpha)\Big(\frac{R_\star}{a_p}\Big)^2
  B_\lambda(T_\star)\,T_{\rm atm}^{\rm refl}(\lambda),\nonumber\\
  q_{\rm th}(\lambda) &=
  \frac{\lambda}{hc}\pi\Big(\frac{R_p}{z_0}\Big)^2
  \epsilon_p(\lambda)B_\lambda(T_p)\,T_{\rm atm}^{\rm th}(\lambda),
  \label{eq:planet_fullband_spectrum}
\end{align}
Here $B_\lambda$ is spectral radiance per micrometre; $\lambda/(hc)$ uses wavelength in metres, giving $q$ in photons\,s$^{-1}$\,m$^{-2}$\,$\mu$m$^{-1}$. The new continuum reference has $T_\star=5772\,{\rm K}$, $T_p=252\,{\rm K}$, $A_g=0.20$, $a_p=1\,\AU$, and $R_\star=R_\odot$. Its Lambert phase function is $\Phi(\alpha)=[\sin\alpha+(\pi-\alpha)\cos\alpha]/\pi$, so $\Phi(60^\circ)=0.608998$; $\epsilon_p=T_{\rm atm}^{\rm refl}=T_{\rm atm}^{\rm th}=1$. Molecular absorption is not imposed on this continuum. The equivalent uniform source disk reproduces the phase-integrated flux, rather than the resolved Lambert brightness pattern. Figure~\ref{fig:extended_wave} shows the extraction and background terms.

\begin{table*}[tbp]
\caption{\justifying SGL wavelength regimes for biosignature-inference observations in the unified $0.40$--$20\,\mu{\rm m}$ spectral-performance model.  The reflected-light branch is simulated numerically in the population audit; the short-IR and thermal-IR branches are evaluated as architecture-dependent extensions governed by the same wavelength-resolved count model.}
\label{tab:wavelength_modes}
\begin{tabular}{@{}p{0.15\textwidth}p{0.12\textwidth}p{0.25\textwidth}p{0.38\textwidth}@{}}
\toprule
Regime & Nominal range & Primary observables & Principal technical requirement \\
\midrule
Visible reflected & $0.40$--$1.2\,\mu{\rm m}$ & O$_2$, O$_4$, O$_3$, H$_2$O, Rayleigh slope, clouds, pigments & Coronal calibration, ring-limb separation, and leakage control; internal coronagraphy may be possible if throughput closes. \\
Near-IR reflected & $1.2$--$2.4\,\mu{\rm m}$ & H$_2$O, CH$_4$, CO$_2$, CO, O$_4$, surface/mineral context & Simultaneous IFS or annular spectrograph; external occulter or larger aperture becomes increasingly attractive. \\
Short-IR transition & $2.4$--$5\,\mu{\rm m}$ & CH$_4$ 3.3, CO$_2$ 4.3, H$_2$O 2.7, reflected/thermal separation & External solar suppression, cooled detectors, thermal-background modeling, and joint reflected/thermal retrieval. \\
Thermal IR & $5$--$20\,\mu{\rm m}$ & H$_2$O 6.3, CH$_4$ 7.7, O$_3$ 9.6, CO$_2$ 15, N$_2$O 7.8/17, temperature and cloud-top context & External occulter or equivalent ring extraction, cryogenic optics, detector thermal covariance, and validated thermal-emission retrieval. \\
\bottomrule
\end{tabular}
\end{table*}

\subsection{Photon-count model and notation audit}

For one image-plane sample $p$, wavelength channel $a$, and Stokes intensity ($s=I$), the expected detector counts are
\begin{equation}
\begin{split}
  N_{pas}=t_{pa}\Big\{&\eta_{s,a}^{\rm exo}Q_{a}^{\rm exo}\,[\Ht_{a,s}\stokes_s]_p
  +\eta_{s,a}^{\rm cor}Q_{a}^{\rm cor}
  +Q_{pa}^{\star}+Q_{pa}^{\rm zodi} 
  +Q_{a}^{\rm dark}+Q_{a}^{\rm th}\Big\}+N_{pa}^{\rm bias} .
\end{split}
  \label{eq:counts}
\end{equation}
The response $[\Ht_{a,I}\stokes_I]_p$ is dimensionless relative to the fiducial radiance defining $Q_a^{\rm exo}$; that normalization is included in $\Ht_{a,I}$.  All terms inside braces are count rates. The $\eta$ factors are relative state/calibration corrections, unity at the reference setting; nominal throughputs already included in Eq. (\ref{eq:spectral_count_integrals}) are not applied twice.  The corresponding Gaussian variance is
\begin{equation}
\begin{split}
  \sigma^2_{pas}=t_{pa}\Big\{&\eta_{s,a}^{\rm exo}Q_{a}^{\rm exo}[\Ht_{a,s}\stokes_s]_p
  +\eta_{s,a}^{\rm cor}Q_{a}^{\rm cor}
  +Q_{pa}^{\star}+Q_{pa}^{\rm zodi} 
  +Q_{a}^{\rm dark}+Q_{a}^{\rm th}\Big\}
  +N_{\rm pix}N_{\rm fr}\sigma_{r,a}^2+\sigma^2_{\rm sys,pas} .
\end{split}
  \label{eq:variance}
\end{equation}
Polarimetric models must propagate nonnegative analyzer-count covariance through the demodulation matrix to obtain signed Stokes covariance.  Eqs. (\ref{eq:counts})--(\ref{eq:variance}) also include non-photon errors.  Solar-corona photon noise is irreducible after coronal photons enter the extraction statistic, while $\sigma_{\rm sys}$ represents residual coronal subtraction, detector calibration, pointing, stellar leakage, exozodiacal light, thermal-background subtraction, wavelength-solution errors, and template-marginalization covariance.

The detected rates are finite-channel integrals. Write $\tau(\lambda)$ for optical throughput, $\eta(\lambda)$ for quantum efficiency, and $\mathcal E(\theta,\lambda)$ for the fraction of a normalized circular-pupil Airy pattern from direction $\theta$ falling inside $\theta_{\rm ap}=\theta_E+1.22\lambda/d$. With $\theta_E=\sqrt{2r_g/\bar z}$ and $\theta_\odot=R_\odot/\bar z$,
\begin{align}
 Q_a^{\rm exo}&=K_{\rm occ}A_{\rm tel}\int_a\tau\eta\,
       \overline\mu_p\mathcal E(\theta_E,\lambda)q_p(\lambda)d\lambda,\nonumber\\
 Q_a^{\rm cor}&=A_{\rm tel}\int_a\tau\eta\frac{\lambda}{hc}
       \int_{\theta_\odot}^{\infty}2\pi\theta B_{\rm cor}(\theta,\lambda)
       \mathcal E(\theta,\lambda)d\theta\,d\lambda,\nonumber\\
 Q_a^{\rm bg}&=Q_a^{\rm cor}+Q_a^\star+Q_a^{\rm zodi}+Q_a^{\rm dark}+Q_a^{\rm th}.
 \label{eq:spectral_count_integrals}
\end{align}
Here $A_{\rm tel}=\pi d^2/4=0.125664\,{\rm m^2}$ for $d=0.4\,{\rm m}$. Rates are detected electrons per second; optical throughput, quantum efficiency, and extraction enter once. The source- and pupil-averaged gain at the central image-plane sample is the large-uniform-source limit \cite{TuryshevToth2022Spectral},
\begin{equation}
 \overline\mu_p=\frac{2\theta_Ez_0}{R_p}
  \left\langle\frac{2}{\pi}E\!\left[(r/R_{\rm img})^2\right]\right\rangle_{\rm pupil}
  =2.26485\times10^6,
 \label{eq:extended_source_reference}
\end{equation}
where $E$ is the complete elliptic integral of the second kind with its parameter shown, and $R_{\rm img}=669.23\,{\rm m}$. Finite source-disk and pupil integration of the $J_0^2$ kernel agrees within $2.4\times10^{-4}$ at the five checked wavelengths from $0.4$ to $20\um$. This average replaces the peak point-source gain; it is not an additional magnification factor.

With the optical factors explicit, the reference normalization is
\begin{equation}
 K_{\rm occ}^{(j)}\equiv
 \frac{Q_{{\rm exo},\mathcal B}^{\rm ref,(j)}}
 {A_{\rm tel}\int_{\mathcal B}\tau\eta\overline\mu_p
 \mathcal E(\theta_E,\lambda)q_p(\lambda)d\lambda}=1,
 \quad j\in\{6,7\}.
 \label{eq:occulter_calibration}
\end{equation}
The numerator is the predicted rate from the same specified source and optics, not an empirical calibration datum. Table~\ref{tab:radiometric_references} supplies both configurations and absolute check values. The coronal radiance uses Eq.~(33) of Ref.~\cite{TuryshevToth2022Spectral},
\begin{equation}
 B_{\rm cor}=10^{-6}\frac{\lambda}{0.925\um}B_\lambda(5772\,{\rm K})
 \left[3.670\left(\frac{\theta_\odot}{\theta}\right)^{18}
 +1.939\left(\frac{\theta_\odot}{\theta}\right)^{7.8}
 +0.0551\left(\frac{\theta_\odot}{\theta}\right)^{2.5}\right].
 \label{eq:corona_reference}
\end{equation}
Its infrared continuation is an empirical reference prescription. The extraction includes light diffracted from outside the aperture. Ideal rejection of the solar photosphere is assumed; it does not reject the host star. At projected separation $a_p\sin60^\circ/z_0$, the host's two images have magnifications $28.3474$ and $27.3474$ and lie outside the solar disk. Both enter with their own Airy acceptance \cite{TuryshevToth2022MNRAS}. Subtracting their mean leaves their shot noise. The nonzero solar-transmission sensitivity adds a photospheric leakage term $Q_a^{\odot}$ to $Q_a^{\rm bg}$; it is zero in the baseline. A uniform $0.10\,{\rm MJy\,sr^{-1}}$ diffuse floor and effective $40\,{\rm K}$, emissivity-$0.10$ optics complete the reference. Structured host exozodiacal emission and a physical occulter diffraction response are not modeled.

The extraction uses unit weights. For weights $w_k$ and independent detector rates $r_k$, the Poisson and read terms instead become $t\sum_kw_k^2r_k$ and $N_{\rm fr}\sigma_r^2\sum_kw_k^2$. The adopted systematic term is a fully correlated $0.1$-ppm coronal rate residual, $s_a=10^{-7}Q_a^{\rm cor}$. It is a design assumption whose effect is tested explicitly, not demonstrated instrument stability.

The scalar aperture-integrated rates used in earlier SGL imaging studies are retained only as a broadband spatial-reconstruction benchmark,
\begin{align}
  Q_{\rm exo}^{\rm sc}&=8.01\times10^4
  \Big(\frac{d}{1\,{\rm m}}\Big)^2
  \Big(\frac{650\,\AU}{\bar z}\Big)^{1/2}
  \Big(\frac{30\,\pc}{z_0}\Big)
  \Big(\frac{\lambda}{1\um}\Big)\, {\rm s}^{-1},\label{eq:qexo}\\
  Q_{\rm cor}^{\rm sc}&=6.20\times10^9
  \Big(\frac{d}{1\,{\rm m}}\Big)^2
  \Big(\frac{650\,\AU}{\bar z}\Big)^2
  \Big(\frac{\lambda}{1\um}\Big)\, {\rm s}^{-1}.\label{eq:qcor}
\end{align}
These expressions are useful for checking the image-plane dwell, deconvolution penalty, and mission-time scaling; they are not used as the wavelength-resolved spectral-performance model.  The convolved-image scalar SNR for dwell time $t$ is
\begin{equation}
  \snrC=\frac{Q_{\rm exo}^{\rm sc}t}{[(Q_{\rm exo}^{\rm sc}+Q_{\rm cor}^{\rm sc})t]^{1/2}}.
  \label{eq:snrc}
\end{equation}
For $t=1800\,{\rm s}$, Eq. (\ref{eq:snrc}) gives $\snrC=43.16$ in the scalar broadband benchmark.  The spectral-performance reference assumes simultaneous dispersive Stokes-$I$ acquisition; its per-channel dilution is included in Eq. (\ref{eq:spectral_count_integrals}). The scalar broadband normalization does not set those channel rates.  If the instrument observes channels sequentially, the dwell time must be multiplied by the channel count or by the appropriate spectral duty-cycle factor.

The scalar reconstruction penalty used as a consistency check is
\begin{equation}
  \frac{\snrR}{\snrC}\simeq c_W\frac{\Delimg}{dn},\qquad c_W=0.891 .
  \label{eq:cw}
\end{equation}
The coefficient $c_W=\pi/[4\ln(\sqrt{2}+1)]\simeq0.891$ is an
approximate analytical result for the scalar aperture-averaged model,
tested numerically in Ref.~\cite{TothTuryshev2021Recovery}.
It is not a universal SGL constant and is used here only as a
consistency check, not a substitute for wave-optical reconstruction.

For the fiducial $n=128$, $d=1\,{\rm m}$, $\bar z=650\,\AU$, and $z_0=30\,\pc$ reflected-light scalar case, Eq. (\ref{eq:cw}) gives $g_n\simeq0.073$.  The broadband $1800\,{\rm s}$ convolved statistic with $\snr_C\simeq43$ therefore becomes only $\snr_R\simeq3$ per recovered broadband pixel.  The full-band external-occulter diagnostic in Fig. \ref{fig:spectral_model}d shows the analogous channelized behavior for the architecture-level branch: individual reconstructed spectral pixels are low-SNR in much of the band, even when the convolved Einstein-ring statistic is useful.  The science case therefore relies on covariance-aware regional coaddition, class conditioning, and template marginalization, and the efficiency of those operations is bounded by $C_{\rm rec}$, the least-validated covariance branch in this framework.  A full ring-sector, chromatic, wave-optical SGL inverse could increase pixel-pixel correlations and reduce the coherent coaddition gain; this is a primary validation requirement, not a minor implementation detail.

\begin{figure}[tbp]
\noindent\makebox[\textwidth][c]{\includegraphics[width=0.86\textwidth]{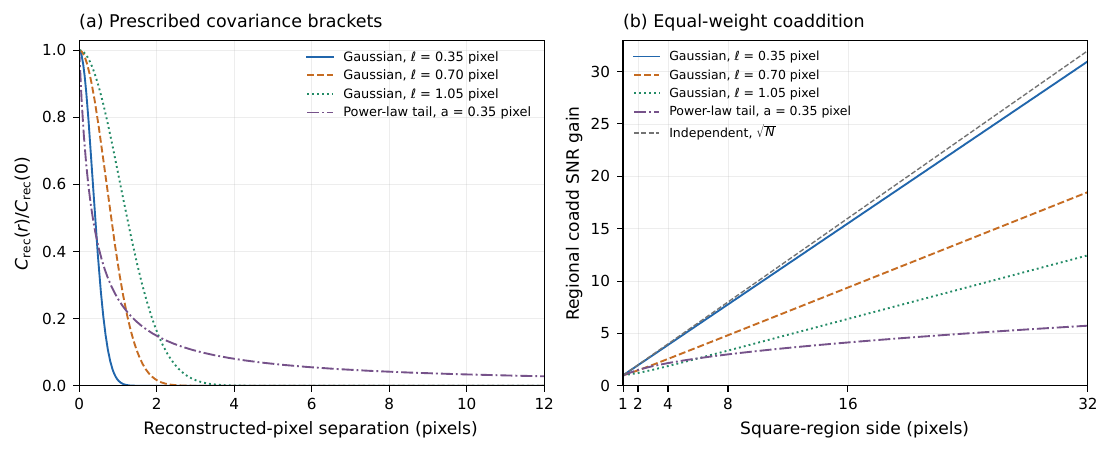}}
  \caption{\justifying Reconstruction-covariance bracket for regional spectral coaddition.  Panel (a) shows normalized residual correlation functions for a scalar Gaussian covariance, two- and three-times-longer Gaussian correlation lengths, and an illustrative heavy-tailed power-law covariance motivated by the long-tail behavior that may appear in a chromatic ring-sector SGL inverse.  The reproduced curves use $K(r)=\exp[-r^2/(2\ell^2)]$ with $\ell=0.35,0.70,1.05$ pixels and $K(r)=(1+r/0.35)^{-1}$.  Panel (b) uses equal weights and equal signal/variance in each square region; these are covariance brackets, not measured SGL residuals.  For an $8\times8$ region the scalar Gaussian gain is $7.77$, while the two- and three-times Gaussian brackets give $4.83$ and $3.36$ and the power-law tail gives $3.00$.  The Gaussian brackets are controlled sensitivity tests, not conservative upper bounds on degradation; a flight-relevant wave-optical reconstruction could produce still different $C_{\rm rec}$.}
  \label{fig:reconstruction_covariance}
\end{figure}

To quantify this dependence without claiming a flight-level wave-optical inverse, Fig. \ref{fig:reconstruction_covariance} brackets $C_{\rm rec}$ with three Gaussian cases and one power-law-tail bracket: residual-correlation lengths equal to the scalar benchmark and to two and three times that value, plus an illustrative power-law tail designed to mimic the possibility that a chromatic ring-sector inverse retains a long $d/(4\rho)$-like residual.  For an $8\times8$ region, the scalar Gaussian covariance gives a coadd gain of $7.77$, close to the uncorrelated $\sqrt{64}=8$ limit.  Doubling and tripling the Gaussian correlation length reduce the gain to $4.83$ and $3.36$, or $0.62$ and $0.43$ times the scalar SNR gain, with dwell-time penalties of $2.6$ and $5.3$.  The power-law-tail bracket gives a gain of $3.00$, or $0.39$ times the scalar SNR gain, corresponding to a $6.7$-fold dwell penalty.  The Gaussian cases are therefore not claimed to be conservative upper bounds on degradation; they are controlled sensitivity tests that may understate the effect of heavy-tailed SGL reconstruction covariance.  A full wave-optical SGL+telescope+occulter injection-recovery test is required before regional spectroscopic performance can be claimed as a mission forecast.

\subsection{Spectral-performance envelope and resolution trades}\label{sec:spectral_performance}

\begin{figure}[tbp]
\noindent\makebox[\textwidth][c]{\includegraphics[width=0.88\textwidth]{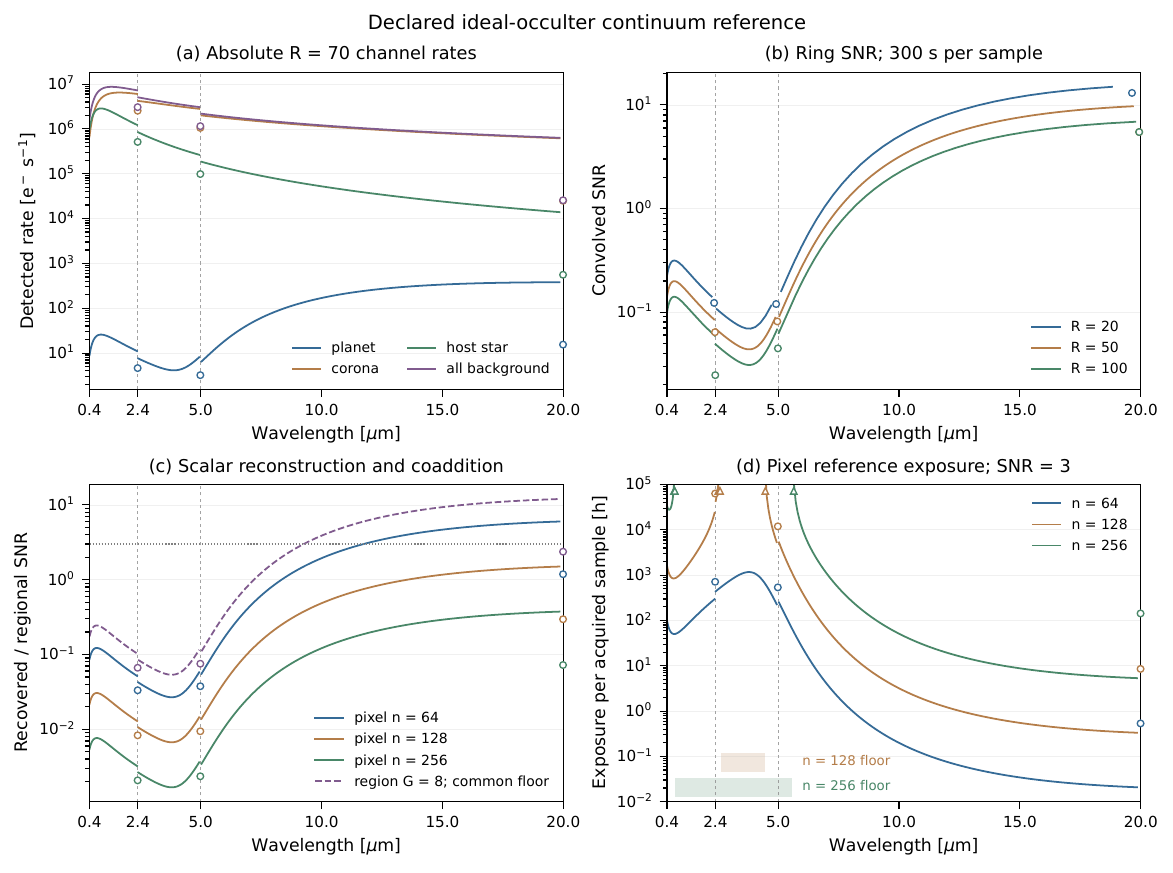}}
  \caption{\justifying Absolute continuum sensitivity for the $0.4\,{\rm m}$ ideal solar-suppression reference at $30\,\pc$ and $650\,\AU$. (a) Detected $R=70$ planet, corona, unmasked-host, and total-background channel rates in electrons per second. (b) Extracted-ring SNR for $R=20,50,100$ at $300\,{\rm s}$ per sample. (c) Recovered-pixel SNR at $R=70$ for $n=64,128,256$ and the $n=128$ regional reference with statistical gain $G=8$; the common floor is not averaged down. (d) Per-sample exposures for channel SNR $3$, including $60\,{\rm s}$ subframes ($30\,{\rm s}$ for $n=256$). Shaded strips identify floor-limited intervals; upward triangles mark finite exposures above $10^5$ h. Hollow circles retain partial-width bins; dashed lines mark detector-arm boundaries. Wavelength axes are linear. Table~\ref{tab:radiometric_references} gives the normalization. These continuum-channel estimates use an assumed scalar reconstruction response; they are not molecular retrieval times.}
  \label{fig:spectral_performance_envelope}
\end{figure}

\begin{figure}[tbp]
\noindent\makebox[\textwidth][c]{\includegraphics[width=0.88\textwidth]{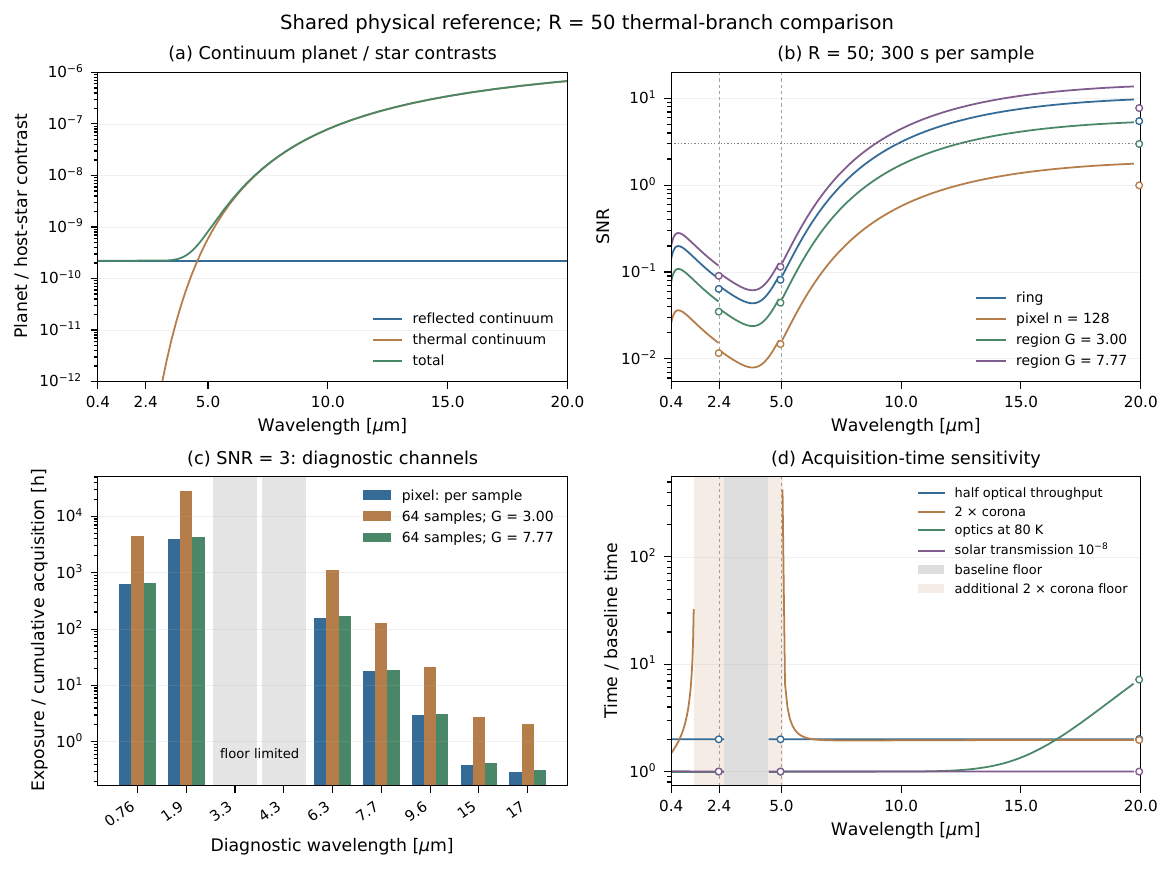}}
  \caption{\justifying Thermal-branch comparison using the continuum reference of Fig.~\ref{fig:spectral_performance_envelope}, with $R=50$. (a) Reflected, emitted, and total planet/host contrasts. (b) Ring, $n=128$ recovered-pixel, and regional SNR at $300\,{\rm s}$ per sample, with $g_n=0.181993$ and statistical gains $3.00$ and $7.77$. (c) Pixel exposures and cumulative $64$-sample regional acquisitions for channel SNR $3$; shaded $3.3$ and $4.3\um$ channels cannot reach the target under the $0.1$-ppm common floor. (d) Regional acquisition-time ratios, $G=7.77$, for half optical throughput, twice coronal radiance, $80\,{\rm K}$ optics, or $10^{-8}$ photosphere transmission. Gray shading marks the baseline floor; orange adds the doubled-corona limit. Ratios require both times finite; read schedules and the common floor are retained. Full-band axes are linear. Regional totals exclude overheads and wider sampling required by a nonlocal inverse; they are not thermal molecular-detection times.}
  \label{fig:thermal_branch_trade}
\end{figure}

The spectral resolving power of an SGL biosignature-inference observation is selected by the instrument, not by the gravitational lens alone.  The lens supplies a wavelength-dependent gain, diffraction kernel, image-plane compression, and coronal background; the spectrograph or filter bank chooses the channelization.  The useful resolving power is therefore the value for which the recovered, covariance-propagated spectrum still has adequate SNR after photons have been divided among wavelength channels, image-plane samples, Stokes states, visits, and nuisance modes.  To make this trade explicit over the full $0.40$--$20\um$ band, we define logarithmic channels
\begin{equation}
  \lambda_a^{\pm}=\lambda_a\exp\left(\pm\frac{1}{2R}\right),\qquad
  \Delta\lambda_a\simeq\frac{\lambda_a}{R} .
  \label{eq:channel_bounds}
\end{equation}
The logarithmic grid is an internal numerical representation of constant resolving power.  Figures~\ref{fig:spectral_performance_envelope}--\ref{fig:thermal_branch_trade} use linear wavelength axes; the supporting full-band component plots use logarithmic axes.  The channel-integrated rates are those in Eq. (\ref{eq:spectral_count_integrals}).  The convolved Einstein-ring spectral statistic is
\begin{equation}
  \snr_{C,a}=
  \frac{Q_{a}^{\rm exo}t_a}
  {\left[(Q_{a}^{\rm exo}+Q_{a}^{\rm bg})t_a
  +N_{\rm pix}N_{\rm fr}\sigma_{r,a}^2+\sigma_{\rm sys,a}^2\right]^{1/2}} .
  \label{eq:channel_snr}
\end{equation}
Eq.~(\ref{eq:channel_snr}) is a spectral-performance diagnostic, not a biosignature statistic.  It describes the SNR before spatial deconvolution and before marginalization over continuum, cloud, mineral, calibration, wavelength-solution, detector-background, and line-spread-function modes.  For a scalar reconstruction check we propagate Eq. (\ref{eq:cw}) channel by channel,
\begin{equation}
  \snr_{R,a}\simeq g_n\snr_{C,a},\qquad
  g_n=c_W\frac{\Delimg}{dn}=c_W\frac{\Dimg}{d n^2} .
  \label{eq:channel_reconstruction_penalty}
\end{equation}
The distinction between $\snr_{C,a}$ and $\snr_{R,a}$ is essential.  The former is an annular ring-extraction statistic; the latter is a recovered spatial element after inversion of the long-tailed SGL kernel.  The science-facing regional spectrum is still another quantity.  For fixed class or regional weights $w_{p{\cal R}}$, the recovered spectrum and covariance are
\begin{align}
  \bar S_{{\cal R},a} &=
  \frac{\sum_p w_{p{\cal R}}\widehat S_{pa}}{\sum_p w_{p{\cal R}}}, \nonumber\\
  \sigma^2_{{\cal R},a} &=
  \frac{\bm w_{\cal R}^{T}C_a\bm w_{\cal R}}
       {(\bm w_{\cal R}^{T}\bm 1)^2},\qquad
  \snr_{{\cal R},a}=\frac{\bar S_{{\cal R},a}}{\sigma_{{\cal R},a}} .
  \label{eq:regional_snr}
\end{align}
Weights inferred from the same noisy cube require propagation of their uncertainty and selection dependence.
Thus class-conditioned spectra improve SNR only to the extent that the reconstruction covariance permits coherent coaddition; the gain is not generally $N_{\rm pix}^{1/2}$ because neighboring reconstructed pixels are correlated.

The performance curves in Fig.~\ref{fig:spectral_performance_envelope} integrate the explicit reflected-plus-thermal continuum and all accepted backgrounds. Mid-IR sensitivity improves because of the photon budget, not a sharper inverse: $g_n$ remains the conditional scalar factor of Eq.~(\ref{eq:channel_reconstruction_penalty}). The source-averaged gain and extraction are calculated before channel integration. These central-sample rates and the prescribed $g_n$ do not constitute a chromatic wave-optical raster reconstruction.

Under the simplifying assumptions of simultaneous spectral acquisition, smooth spectra across a channel, and the corona-dominated scalar benchmark of Eqs.~(\ref{eq:qexo})--(\ref{eq:qcor}), the aperture dependence follows directly from the stated count rates and from the scalar reconstruction penalty.  Equations~(\ref{eq:qexo})--(\ref{eq:qcor}) give $Q_{\rm exo}^{\rm sc}\propto d^2$ and $Q_{\rm cor}^{\rm sc}\propto d^2$, so the convolved annular statistic scales as $\snr_{C,a}^{\rm sc}\propto d$ in the coronal-background limit.  Equation~(\ref{eq:channel_reconstruction_penalty}) gives $g_n\propto d^{-1}$.  The leading aperture factors therefore cancel in the scalar recovered-pixel statistic,
\begin{equation}
  \snr_{R,a}^{\rm sc}\propto
  \frac{\bar z^{3/2}}{z_0^2}
  \frac{1}{n^2}
  R^{-1/2}t_a^{1/2}\,{\cal F}_a(\lambda),
  \label{eq:performance_scaling}
\end{equation}
where ${\cal F}_a(\lambda)$ contains the wavelength-dependent planet spectrum, corona spectrum, throughput, detector, thermal-background, and branch assumptions other than the scalar aperture factors that cancel above.  For fixed recovered-pixel SNR in this scalar aperture-averaged benchmark,
\begin{equation}
  t_a^{\rm sc}\propto R n^4 z_0^4 \bar z^{-3}{\cal F}_a^{-2}(\lambda) .
  \label{eq:dwell_scaling}
\end{equation}
A full raster contains $n^2$ image-plane samples, so the corresponding scalar reference scaling is
\begin{equation}
  T_{\rm cube}^{\rm sc}\propto
  R n^6 z_0^4\bar z^{-3}
  \Big(\frac{M_pN_\phi}{N_{\rm sc}}\Big) f_{\rm oh}{\cal F}_a^{-2}(\lambda) .
  \label{eq:cube_time_scaling}
\end{equation}
Eqs.~(\ref{eq:dwell_scaling})--(\ref{eq:cube_time_scaling}) are scalar consistency scalings, not universal aperture design laws.  In this benchmark, increasing $d$ simultaneously increases the convolved photon statistic and strengthens the approximate analytical scalar deconvolution penalty, so recovered-pixel dwell is approximately aperture neutral at fixed wavelength factor.  A branch whose optical design fixes the reconstruction covariance while increasing the effective collecting area would instead place the collecting-area dependence inside ${\cal F}_a$ and, in the background-limited limit, can approach $T\propto A_{\rm eff}^{-1}$.  That favorable aperture scaling must be demonstrated with the appropriate internal-coronagraph or external-occulter throughput, leakage, detector, thermal-background, and $C_{\rm rec}$ model; it should not be inferred from the scalar $c_W$ benchmark.  The robust common trends are that doubling $R$ costs approximately a factor of two in dwell time, doubling the linear spatial grid costs approximately 16 per sample and 64 for a full raster, moving from 30 pc to 10 pc improves the scalar recovered-pixel time scale by roughly $3^4\simeq81$, and operating farther down the focal line reduces the scalar dwell approximately as $\bar z^{-3}$ before cruise, navigation, and thermal constraints are applied.  Table \ref{tab:spectral_modes} summarizes the observing modes implied by these full-band spectral-performance and reconstruction scalings.

For the $R=70$, $n=128$ continuum reference, recovered-channel SNR $3$ requires $3.19$ h at $10\um$ and $0.550$ h at $15\um$ per acquired sample; the corresponding $64$-sample regional totals at $G=7.77$ are $3.38$ and $0.584$ h. At $6.3$ and $7.7\um$, the pixel exposures are $223$ and $24.8$ h. At $3.3$ and $4.3\um$, the adopted common coronal floor prevents SNR $3$ at any finite exposure. These are continuum estimates at diagnostic wavelengths. Molecular-band inference also requires continuum, temperature, cloud, line-spread-function, and model-discrepancy marginalization. Figure~\ref{fig:thermal_branch_trade} uses the same physical normalization with $R=50$ channels; its sensitivity sweep tests throughput, corona, optics temperature, and photospheric leakage.

For a regional spectrum, $G_{\rm reg}$ is the covariance-limited SNR gain at a common exposure per acquired sample.  When the count variance is proportional to exposure and $G_{\rm reg}$ is constant,
\begin{equation}
 t_{\rm samp}=\frac{t_{\rm pixel}}{G_{\rm reg}^{2}},\qquad
 T_{\rm region}=N_{\rm samp}t_{\rm samp}
              =\frac{N_{\rm samp}t_{\rm pixel}}{G_{\rm reg}^{2}} .
 \label{eq:regional_acquisition_time}
\end{equation}
Here $t_{\rm pixel}$ is exposure per acquired sample and $N_{\rm samp}$ counts acquired samples, not reconstructed pixels. In the statistical limit, a hypothetical $20$ h pixel reference and $64$ samples give $21.2$--$142$ h total for $G_{\rm reg}=7.77$--$3.00$; independent coaddition with $G=8$ gives $T_{\rm region}=t_{\rm pixel}$. The executed reference retains integer read counts and a nonaveraging floor:
\begin{equation}
 \snr_{{\cal R},a}(t)=\frac{g_nQ_a^{\rm exo}t}
 {\{[(Q_a^{\rm exo}+Q_a^{\rm bg})t+N_{\rm fr}V_{r,a}]/G_{\rm reg}^2+s_a^2t^2\}^{1/2}},
 \quad N_{\rm fr}=\lceil t/t_{\rm sub}\rceil .
 \label{eq:regional_noise_reference}
\end{equation}
Here $V_{r,a}=N_{\rm pix}\sigma_{r,a}^2$ and $s_a=10^{-7}Q_a^{\rm cor}$. We use $t_{\rm sub}=60\,{\rm s}$ for ring, $n\le128$, and regional products, and $30\,{\rm s}$ for $n=256$; the $300\,{\rm s}$ reference is accumulated exposure. Pixel results set $G=1$; regional time is $64t$ after solving this equation, not an unrestricted $G^2$ improvement. A full inverse may require more than 64 samples. The scalar $z_0^4$ law is a separate limiting benchmark, not a distance law for every background or systematic regime.

\begin{table*}[tbp]
\caption{\justifying Computed continuum sensitivity at selected diagnostic wavelengths, $R=70$, $n=128$, target recovered-channel SNR $3$. Regional totals include 64 acquired samples, integer read counts, and the common coronal floor. They exclude overheads and are not molecular-band detection times. The $R=50$ comparison in Fig.~\ref{fig:thermal_branch_trade} uses wider channels.}
\label{tab:thermal_case}
\begin{tabular}{@{}cccccc@{}}
\toprule
$\lambda$ [$\mu$m] & Context & Pixel SNR, $300$ s & $t_{\rm pixel}$ [h] & $T_{\rm region},G=7.77$ [h] & $T_{\rm region},G=3.00$ [h]\\
\midrule
$3.3,4.3$ & CH$_4$, CO$_2$ & $<0.01$ & Floor limited & Floor limited & Floor limited\\
$6.3$ & H$_2$O & $0.0584$ & $223.$ & $236.$ & $1586$ \\
$7.7$ & CH$_4$/N$_2$O & $0.174$ & $24.8$ & $26.3$ & $176.$ \\
$9.6$ & O$_3$ & $0.424$ & $4.16$ & $4.41$ & $29.6$ \\
$10$ & Thermal continuum & $0.485$ & $3.19$ & $3.38$ & $22.7$ \\
$15$ & CO$_2$ & $1.17$ & $0.550$ & $0.584$ & $3.91$ \\
$17$ & N$_2$O & $1.34$ & $0.415$ & $0.441$ & $2.95$ \\
\bottomrule
\end{tabular}
\end{table*}

\begin{table*}[tbp]
\caption{\justifying Illustrative observing products motivated by the $0.40$--$20\um$ continuum budget and scalar reconstruction scalings. The population audit implements ten reflected-light cells at $R=70$ with four compressions of 128 channels. The listed map sizes and additional observing modes are architecture examples; thermal and polarimetric retrievals require separate validation.}
\label{tab:spectral_modes}
\begin{tabular}{@{}p{0.13\textwidth}p{0.09\textwidth}p{0.13\textwidth}p{0.24\textwidth}p{0.34\textwidth}@{}}
\toprule
Mode & Illustrative $R$ & Spatial product & Architecture driver & Primary scientific use \\
\midrule
Broadband acquisition map & $1$--$10$ & $64^2$--$128^2$ & Internal coronagraph if leakage and ring-limb separation close; occulter if available & Disk registration, cloud/ocean/land heterogeneity, phase, and rotation solution. \\
Moderate-$R$ reflected-light cube & $50$--$100$ & $64^2$--$128^2$ & Simultaneous IFS or dispersive annular ring spectrograph & O$_2$, O$_4$, O$_3$, H$_2$O, CH$_4$, CO$_2$, CO, pigments, minerals, and cloud-aware regional spectra. \\
Region-coadded spectroscopy & $100$--$300$ & masks/classes & Covariance-aware coaddition of selected low-cloud or surface-class regions & Targeted molecular or surface tests after a map and class posterior exist. \\
Short-IR transition spectra & $20$--$100$ & regions or coarse maps & External occulter, cooled detectors, and reflected/thermal separation & CH$_4$ 3.3, CO$_2$ 4.3, H$_2$O 2.7, and transition to thermal emission. \\
Thermal-IR targeted branch & $20$--$50$ & $16^2$--$64^2$ or regions & External occulter, cryogenic optics, detector and telescope thermal control & O$_3$ 9.6, CO$_2$ 15, H$_2$O 6.3, CH$_4$ 7.7, N$_2$O, climate, and cloud-top temperature. \\
Full-band polarimetry & low--moderate $R$ & selected regions & Modulation stability, cross-Stokes calibration, and regional coaddition over the accessible band & Linear polarization constrains scattering, clouds, aerosols, and glint; circular spectropolarimetry is a high-value homochirality test after intensity and linear-polarization systematics close. \\
\bottomrule
\end{tabular}
\end{table*}

The disk-integrated data operator is denoted by the calligraphic symbol $\Disk$; the ordinary symbol $D$ is reserved for diameters such as $\Dimg$:
\begin{equation}
  \data_{a}^{\rm disk}=\Disk[\stokes_a]+\epsilon_a
  =\sum_p W_p\stokes_{pa}+\epsilon_a .
  \label{eq:diskop}
\end{equation}
The resolved data are
\begin{equation}
  \data_{pa}^{\rm res}=[\Ht_a\stokes_a]_p+\epsilon_{pa} .
  \label{eq:resop}
\end{equation}
The comparison between Eqs. (\ref{eq:diskop}) and (\ref{eq:resop}) is the central statistical comparison in the paper.

\section{Covariance-aware resolved spectroscopy}\label{sec:covspectro}

The most important lesson from high-contrast spectroscopy, molecular mapping, and forward-model matched-filter work is that the science-facing quantity is not a scalar per-channel SNR \cite{Soummer2012KLIP,Pueyo2016KLIPFM,Snellen2015HDSHCI,GrecoBrandt2016,BrogiLine2019,Ruffio2018Bayesian}.  It is the covariance of the extracted spectrum after propagation through the instrument, extraction operator, calibration model, nuisance projection, and astrophysical forward model.  The SGL generalization replaces the one-dimensional spectral vector by a spatial-spectral-temporal Stokes cube.  We therefore index a calibrated SGL data product by spatial element $p$, wavelength channel $a$, epoch $e$, and Stokes component $s$:
\begin{equation}
  \widehat{\stokes}_{paes}=[\mathsf R\stokes]_{paes}+n_{paes},
  \qquad
  {\rm Cov}(n_{paes},n_{p'a'e's'})=C_{(paes)(p'a'e's')} .
  \label{eq:cube_covariance}
\end{equation}
For background-subtracted measurements $y=\mathsf H\stokes+\epsilon$ with zero-mean $\epsilon$ and a fixed linear reconstruction $\mathsf W$, the resolution matrix is $\mathsf R=\mathsf W\mathsf H$ and the propagated noise covariance is $C=\mathsf W C_\epsilon\mathsf W^T$.  Regularization generally gives $\mathsf R\ne I$: its bias $(\mathsf R-I)\stokes$ belongs in the forward mean and Jacobian, not in centered noise covariance.  Retrievals therefore require both $\mathsf R$ and $C$.  Photon and detector noise acquire off-diagonal correlations under reconstruction; coronal residuals, calibration, wavelength/LSF errors, and astrophysical uncertainty add further shared modes.

A minimal branch-resolved covariance roll-up for SGL spectral cubes is
\begin{equation}
\begin{split}
 C ={}& C_{\rm ph}+C_{\rm rec}+C_{\rm cor}+C_{\rm cal}+C_\lambda+C_{\rm LSF}
      +C_{\rm ext}+C_{\rm pol} 
    +C_{\rm astro}+C_{\rm RT}+C_{\rm prior} .
\end{split}
\label{eq:cov_rollup}
\end{equation}
Here $C_{\rm ph}$ contains exoplanet, corona, zodiacal, detector, and leakage shot noise; $C_{\rm rec}$ is the reconstruction covariance induced by the SGL inverse and regularization; $C_{\rm cor}$ describes residual structured coronal subtraction; $C_{\rm cal}$ contains flat-field, gain, throughput, and shared visit-level modes; $C_\lambda$ and $C_{\rm LSF}$ are derivative-weighted wavelength and line-spread-function terms; $C_{\rm ext}$ describes ring-sector extraction and detector-to-spectrum mixing; $C_{\rm pol}$ describes polarimetric leakage and modulation uncertainty; $C_{\rm astro}$ contains stellar, exozodiacal, cloud, and surface-model nuisance covariance; and $C_{\rm RT}$ is the radiative-transfer model-form error introduced in Sec. \ref{sec:planet}.  Equation (\ref{eq:variance}) describes count-level variance, whereas Eq. (\ref{eq:cov_rollup}) specifies covariance bookkeeping, not a measured error budget. Only the simplified subset identified in Table \ref{tab:figure_covariance_map} is implemented. Contributions must share one data space and count propagated photon noise once.  The sum assumes uncorrelated branches; shared latent modes require their cross-covariances or a joint covariance model.  $C_{\rm prior}$ denotes only uncertainty from separately marginalized latent modes, not the parameter priors added through $F_{\rm prior}$.  Unimplemented branches remain unquantified.

A compact separable representation is useful for traceability:
\begin{equation}
\begin{aligned}
 C^{(b)}_{(pae)(p'a'e')}={}&
 \sigma^{(b)}_{pae}\sigma^{(b)}_{p'a'e'}
 K_x^{(b)}(p,p')K_\lambda^{(b)}(a,a')K_t^{(b)}(e,e') 
 +\left[U_b\Lambda_b U_b^T\right]_{(pae)(p'a'e')} .
\end{aligned}
 \label{eq:branch_cov}
\end{equation}
where $b$ labels a covariance branch.  Local photon noise has nearly diagonal kernels.  Residual coronal structure and detector flat-field modes are spatially correlated.  Wavelength-solution and LSF terms are derivative-weighted in $\lambda$.  Visit-level calibration drifts are low-rank in time.  Cloud fields have finite temporal correlation and spatial coherence.  This representation makes the comparison between disk-integrated and resolved observations explicit: disk integration applies the operator $\Disk$ to the same covariance,
\begin{equation}
  C_{\rm disk}=\Disk C_{\rm res}\Disk^T + C_{\rm lost},
  \label{eq:disk_covariance}
\end{equation}
For the matched reference, $f_{d0}=\Disk f_{\rm res}$, $J_{d0}=\Disk J_{\rm res}$, and $C_{d0}=\Disk C_{\rm res}\Disk^T$, with a common parameter vector, reference point, and prior.  The reconstructed mean and $J_{\rm res}$ retain $\mathsf R=\mathsf W\mathsf H$ and its parameter dependence; regularization is not replaced by an identity response.  Here $C_{\rm lost}=0$.  A nonzero $C_{\rm lost}$ specifies additional model uncertainty; it is not generated by disk integration itself.  Changing the nuisance covariance is therefore a separate modeling assumption.

For a molecular or surface template $h_m$, fixed positive-definite $C$, full-column-rank $N$, and unrestricted linear nuisance amplitudes, the generalized-least-squares estimator is
\begin{align}
  M_N &= C^{-1}-C^{-1}N\left(N^T C^{-1}N\right)^{-1}N^T C^{-1},\label{eq:nuisance_precision}\\
  \widehat a_m &= \frac{h_m^T M_N (d-d_0)}{h_m^T M_N h_m},\label{eq:template_amp}\\
  \sigma^2(a_m) &= \left(h_m^T M_N h_m\right)^{-1},\label{eq:template_sigma}\\
  Z_m &= \frac{\widehat a_m}{\sigma(a_m)} .\label{eq:template_z}
\end{align}
The nuisance matrix $N$ must include continuum, cloud/mineral color, calibration, wavelength-shift, LSF, and relevant astrophysical modes; $h_m^T M_Nh_m>0$ is required.  After marginalizing independent zero-mean Gaussian nuisance amplitudes of covariance $\Lambda_N$, replace $N^TC^{-1}N$ by $N^TC^{-1}N+\Lambda_N^{-1}$ in Eq. (\ref{eq:nuisance_precision}); the resulting precision is $(C+N\Lambda_NN^T)^{-1}$.  Eq. (\ref{eq:template_z}) is the statistic used conceptually for the matched-filter detectability contours: Figure~\ref{fig:detectability} reports fixed-hypothesis template-contrast SNR in the surrogate model, not posterior odds for life.  The full machinery in Eqs. (\ref{eq:template_amp})--(\ref{eq:template_z}) and the posterior in Eq. (\ref{eq:posterior}) define the proposed flight-analysis framework.  The present controlled simulations exercise only a subset: local band-depth and continuum indices, covariance-propagated regional coadds, the explicit nuisance-marginalized spectral-contrast calculation, and the Fisher information audit of Eqs. (\ref{eq:fisher_cube})--(\ref{eq:block_information_gain}).  They do not constitute a full multi-template Bayesian retrieval over all Stokes components, epochs, thermal-IR channels, and nuisance models.

For physical retrievals and requirement tracing, the Fisher matrix must be understood as conditional on an adopted forward model $\fwdmodel$ and covariance model $C$.  For observing mode $m$ with mean $f_m(\pars)=\mathbb E[\data_m|\pars]$ and covariance fixed at the fiducial parameters, the local prior-augmented Fisher matrix is
\begin{equation}
  F^{(m)}_{ij}[\fwdmodel,C]
  =\left(\frac{\partial f_m}{\partial\Theta_i}\right)^T C_m^{-1}
        \left(\frac{\partial f_m}{\partial\Theta_j}\right)+F^{\rm prior}_{ij},
  \label{eq:fisher_cube}
\end{equation}
where $\Theta_i$ may be a gas-template amplitude, cloud parameter, surface-class coefficient, phase parameter, SGL optical parameter, or calibration coefficient.  The precision gain of mode $m$ relative to the declared disk reference is a conditional quantity,
\begin{equation}
  \Delta I_m[\fwdmodel,C]
  = \frac{1}{2}\ln\frac{|F_m[\fwdmodel,C]|}{|F_{\rm disk}[\fwdmodel,C]|},
  \label{eq:information_gain}
\end{equation}
for a common parameterization and common priors, with positive-definite Fisher matrices.  When disk data are a deterministic projection of the same resolved measurements with consistent covariance, $F_{\rm disk}\preceq F_{\rm res}$ and $\Delta I_{\rm res}\geq0$.  Strict positivity is not guaranteed, and non-nested imaging and spectroscopic feature sets have no automatic Fisher ordering.  Eq. (\ref{eq:information_gain}) is still preferable to raw classifier accuracy when the goal is retrieval precision rather than class labels, but it is not circularity-free: both the Jacobian $\partial f_m/\partial\pars$ and the covariance $C_m$ are generated within the same surrogate-model family in the present study.  Thus $\Delta I_m[\fwdmodel,C]$ is a conditional information audit, not an externally validated astrophysical information gain.  Its useful outputs are the parameter blocks carrying the gain, the null-test behavior, and the degradation under additional model-form covariance. With $C_m$ held fixed, define $F_{{\rm like},m}=F_m-F_{\rm prior}$. For a correctly specified likelihood and unbiased estimators satisfying the regularity conditions, the Cram\'er--Rao bound is $\mathrm{Cov}(\widehat{\pars})\succeq F_{{\rm like},m}^{-1}$ when $F_{{\rm like},m}$ is nonsingular. The prior-augmented $F_m^{-1}$ instead approximates posterior covariance in the local linear-Gaussian regime; it does not guarantee attained precision
\cite{Vallisneri2008Fisher}.

A matched-reference attribution separates spatial information from an assumed disk nuisance covariance.  With $F_r$ for the resolved data, $F_{d0}$ for their exact disk projection, and $F_{d\nu}$ for the same disk mean and Jacobian with a changed covariance, define
\begin{equation}
\begin{aligned}
 \Delta I_{\rm total}
 &=\underbrace{\frac12\ln\frac{|F_r|}{|F_{d0}|}}_{\Delta I_{\rm spatial}}
 +\underbrace{\frac12\ln\frac{|F_{d0}|}{|F_{d\nu}|}}_{\Delta I_{\rm nuisance}} .
\end{aligned}
\label{eq:information_attribution}
\end{equation}
All three use the same parameters, reference point, and prior, with fixed covariance in each local Fisher calculation.  The spatial term is nonnegative for exact projection; the nuisance term measures a reference-dependent covariance assumption and can have either sign.  The reference calculation evaluates both terms explicitly in Fig.~\ref{fig:information_gain}(c).

The same notation makes the missing model-form covariance explicit.  A flight-quality model should replace the covariance used in Eq. (\ref{eq:fisher_cube}) by
\begin{equation}
  C_m^{\rm eff}=C_m^{\rm ph}+C_m^{\rm rec}+C_m^{\rm sys}
  +\alpha_{\rm RT}C_m^{\rm RT}
  +\alpha_{\rm SGL}C_m^{\rm SGL}
  +\alpha_{\rm geo}C_m^{\rm geo},
  \label{eq:effective_covariance}
\end{equation}
where $C_m^{\rm RT}={\rm Cov}(\delta\data_m^{\rm RT})$ is the centered model-discrepancy covariance propagated into mode-$m$ data units, $C_m^{\rm SGL}$ is the wave-optical reconstruction and ring-extraction covariance, and $C_m^{\rm geo}$ represents unmodeled surface and mineral-library diversity.  The baseline matched audit sets $\alpha_{\rm RT}=\alpha_{\rm SGL}=\alpha_{\rm geo}=0$, but the model-mismatch audit uses $C_{m,AB}^{\rm RT}$ from Eq. (\ref{eq:ab_rt_covariance}) and the reconstruction bracket in Fig. \ref{fig:reconstruction_covariance} tests longer-range forms of $C_m^{\rm rec}$.  These additions test specified model errors. They reduce absolute precision, but their effect on a resolved-minus-disk gain also depends on the disk projection; a relative gain can increase if the disk product degrades more.

For a science block $A$, with $N$ containing all remaining fitted parameters and $F_{NN}$ invertible, the marginalized Fisher matrix is
\begin{equation}
  F^{(m)}_{A|N}[\fwdmodel,C^{\rm eff}]
  =F^{(m)}_{AA}-F^{(m)}_{AN}\left(F^{(m)}_{NN}\right)^{-1}F^{(m)}_{NA},
  \label{eq:schur_fisher}
\end{equation}
and define the conditional block information gain as
\begin{equation}
  \Delta I_A^{(m)}[\fwdmodel,C^{\rm eff}]
  =\frac12\ln\frac{|F^{(m)}_{A|N}[\fwdmodel,C^{\rm eff}]|}{|F^{({\rm disk})}_{A|N}[\fwdmodel,C^{\rm eff}]|} .
  \label{eq:block_information_gain}
\end{equation}
These marginal block gains need not sum to the total gain when blocks are correlated.  They quantify covariance reduction for a common parameterization and provide a requirement-tracing metric: which parameter blocks become constrained, and which nuisance modes must be controlled, when disk spectra are replaced by resolved SGL data products.  Table \ref{tab:covariance_branches} lists the covariance branches that a flight analysis must propagate.

In the local Gaussian approximation, $\Delta I_m$ is the logarithm
of the disk-to-mode-$m$ uncertainty-volume ratio. Maximizing
$\ln|F_{A|N}^{(m)}|$ at fixed observing resources provides a
D-optimal design criterion for spectral coverage, $R$, $n$, and cadence \cite{Alexanderian2023DOptimal}. No such optimization is performed here.

A complementary diagnostic compares two fixed mean signals $f_L$ and $f_A$ under a common positive-definite Gaussian covariance $C$.  Let $\delta=f_L-f_A$, $\delta_w=C^{-1/2}\delta$, $B=\Disk C^{1/2}$, and $\Pi=B^T(BB^T)^{-1}B$, with $\Disk$ full row rank and $C_{\rm disk}=\Disk C\Disk^T$.  The expected log-likelihood-ratio loss under disk integration is
\begin{equation}
 \mathbb E_L[\ln\Lambda_{\rm res}-\ln\Lambda_{\rm disk}]
 =\tfrac12\|(I-\Pi)\delta_w\|^2\geq0 .
 \label{eq:contrast_loss}
\end{equation}
Each expected log ratio is half the corresponding squared Mahalanobis distance; subtracting them gives the orthogonal-projector form above.  Equality holds when the whitened contrast lies entirely in the retained subspace.  Thus spatial structure helps distinguish these specified hypotheses only when it carries contrast discarded by disk integration.  This fixed-hypothesis identity is distinct from a nuisance-integrated Bayes factor and from the Fisher uncertainty-volume gain.  Direct numerical evaluations verify Eq.~(\ref{eq:contrast_loss}) in 30 Gaussian cases and two null controls to an absolute tolerance of $10^{-10}$ nat; these checks test the identity, not biological classification.

A production analysis should estimate $C$ from both analytical branch models and empirical ensembles.  If $S$ is the sample covariance from $N_{\rm sim}$ injection-recovery or off-source residual realizations and $T$ is the analytical roll-up in Eq. (\ref{eq:cov_rollup}), a stable inverse can be obtained with a structured shrinkage estimator,
\begin{equation}
  \widehat C_\gamma=(1-\gamma)S+\gamma T+\epsilon_{\rm floor}\,\mathrm{diag}(T),
  \label{eq:shrinkage_cov}
\end{equation}
with $0\leq\gamma\leq1$, $\epsilon_{\rm floor}\geq0$, a positive-definite result, and the numerical rank and condition number reported.  This requirement is directly inherited from covariance-aware spectroscopy: a biosignature template significance is not reproducible unless the covariance source, nuisance basis, and regularization are specified.

\section{Retrieval and inference model}\label{sec:retrieval}

\subsection{Forward model and likelihood}

Let the state vector be
\begin{equation}
  \pars=\{X_j,A_s,f_c,\tau_c,\alpha,\Omega,\eta,\beta\},
\end{equation}
where $X_j$ are dimensionless gas-template amplitudes, $A_s$ is surface reflectance, $f_c$ and $\tau_c$ describe clouds, $\alpha$ is phase angle, $\Omega$ is the spin state, $\eta$ contains SGL and instrumental response parameters, and $\beta$ contains calibration and background nuisance parameters.  Define the residual vector $\bm r(\pars,\model)=\data-F(\pars,\model)$.  The resolved likelihood is
\begin{equation}
  \ln P(\data|\pars,\model)=-\frac12\bm r^T\cov^{-1}\bm r
  -\frac12\ln|\cov|+C_0 .
  \label{eq:like}
\end{equation}
where $F$ includes Eqs. (\ref{eq:rtoa})--(\ref{eq:variance}), $C_0$ is a normalization constant, and $\cov$ is the branch-resolved cube covariance defined in Sec. \ref{sec:covspectro}.  The controlled simulations include simplified representatives of these terms but do not propagate a physical vector-RT or wave-optical SGL covariance.  Disk-integrated retrieval uses the same atmospheric and surface models but replaces the resolved operator by $\Disk$ and uses the compressed covariance in Eq. (\ref{eq:disk_covariance}).  A common-data comparison requires the same generated measurements, parameterization, priors, and consistently propagated covariance.  Disk gas bands, slopes, red-edge and mineral indices must be retained where measurable; temporal and polarimetric quantities require the corresponding observations.  Any additional $C_{\rm lost}$ must be stated separately.

The hierarchical retrieval should not treat each pixel independently.  Gas abundances are atmospheric fields with spatial priors, surface classes are correlated maps, cloud fields evolve in time, and calibration terms produce correlated residuals.  A useful posterior has the form
\begin{equation}
\begin{split}
  P(\pars,\model|\data)&\propto P(\data|\pars,\model)
  P[X_j]P[A_s|C_s]P[f_c,\tau_c|C_c] 
  \times P[\alpha,\Omega]P[\eta]P[\beta]P(\model),
\end{split}
  \label{eq:posterior}
\end{equation}
where $C_s$ and $C_c$ are spatial/temporal covariance operators for surfaces and clouds.  This is the formal expression of the paper's scientific claim: spatial resolution is valuable because it changes the covariance structure of the inference.  It makes part of the nuisance covariance observable.

\subsection{Feature extraction and population information audit}

The executed information experiment uses the normalized intensity $\mu_\star R_{\rm TOA}$ on ten equal-area illuminated cells. Coordinates are selected from $x,y\in\{-0.75,-0.25,0.25,0.75\}$ by $x^2+y^2<1$ and $\mu_\star=\tfrac12\sqrt{1-x^2-y^2}+\tfrac{\sqrt3}{2}x>0$. All selected cells are retained; there is no continuum or SNR selection. Four modes average disjoint groups of $128/k$ consecutive channels, $k=1,8,32,128$. These are exact linear spectral compressions, not independent duplicate features. No temporal, polarimetric, thermal, or band-index proxy enters this Fisher calculation.

Let $L$ be the unweighted nearest-neighbor graph Laplacian for cell separations below $0.51$, divided by its largest eigenvalue. The idealized scalar response and fixed reconstruction are
\begin{equation}
 H(\theta_b)=e^{-0.8e^{\theta_b}L},\qquad
 W=(H_0^TH_0+0.15L^TL)^{-1}H_0^T,\qquad R=WH.
 \label{eq:reference_response}
\end{equation}
This graph response is not a wave-optical SGL PSF. Its parameter dependence remains in the Jacobian. Since $W$ is square and invertible, retaining its full covariance preserves the raw-data Fisher information; regularization alone does not demonstrate information loss in this experiment.

With raw normalized noise $\sigma_0=0.004$, the covariance before spectral compression is
\begin{equation}
 C_r=\sigma_0^2(WW^T)\otimes K_\lambda,\qquad
 (K_\lambda)_{ab}=0.85\delta_{ab}+0.15e^{-|\ln(\lambda_a/\lambda_b)|/0.04}.
 \label{eq:reference_covariance}
\end{equation}
A normalized signal $0.10$ has raw SNR $25$ at unit dimensionless reference exposure. It is independent of the absolute radiometry in Figures~\ref{fig:spectral_performance_envelope}--\ref{fig:thermal_branch_trade}. If $B_k$ averages spectral groups, set $A_k=I_{10}\otimes B_k$, $J_{r,k}=A_kJ_r$, and $C_{r,k}=A_kC_rA_k^T$. The matching disk operator is $\Disk_k=\bm1_{10}^T/10\otimes I_k$, giving $J_{d,k}=\Disk_kJ_{r,k}$ and $C_{d,k}=\Disk_kC_{r,k}\Disk_k^T$. The trained discrepancy covariance is transformed by the same operators.

Each scene has the common 14-coordinate vector in Table~\ref{tab:fisher_numerical_definition}, with zero local fiducial and prior means. Gas-template amplitudes, vegetation/mineral edge amplitudes, cloud fraction/path, gray/color calibration, and blur scale are fitted; scene morphology, geometry, and cloud optical-depth fields remain fixed. Thus the resulting precision is conditional on these spatial templates. For each mode,
\begin{equation}
 F_m=J_m^TC_m^{-1}J_m+P,\qquad P=\operatorname{diag}(s_i^{-2}).
 \label{eq:mode_fisher}
\end{equation}
Central differences use $h_i=0.005$, except $0.001$ for gray gain and color. Covariance is fixed at the fiducial state, and all modes use identical priors. Half-step and complex-step checks test derivatives; changing prior widths instead changes $P$ at fixed $J$, $C$, and population. Appendix~\ref{app:traceability} gives every coefficient, seed, and population rule. Other fitted coordinates are marginalized by Eq.~(\ref{eq:schur_fisher}); no fitted nuisance is also added as an independent covariance term.

\section{Simulation suite}\label{sec:simulations}

The simulation suite separates a scalar map/index test, the ten-cell population reference, a fixed-contrast spectral-resolution sweep, and an absolute continuum budget (Table~\ref{tab:figure_covariance_map}). Their inputs and noise conventions are specified in Appendix~\ref{app:traceability}. The population uses 16 evaluation and four independent discrepancy-training scenes per class to test conditional precision. The continuum reference integrates reflected and emitted photons, finite-source gain, coronal and host photons, extraction, throughput, and detector terms. Its $R=70$ and $R=50$ configurations share $K_{\rm occ}=1$ and are independent of normalized inference noise.

Six planet classes are used: land-photosynthetic inhabited planets, cryptic inhabited planets, wet abiotic planets, abiotic oxygen-rich planets, desert/mineral worlds, and cloudy ocean worlds.  The false-positive cases are intentionally nontrivial but finite.  The abiotic oxygen case includes O$_2$/O$_3$ plus O$_4$ and CO context.  The desert/mineral case includes one broad red-edge mimic; it is an illustrative first adversarial test, not an exhaustive geological library.  The cryptic inhabited case is a false-negative stress test because it has weak surface expression and only modest disequilibrium.  The cloudy ocean case tests cloud and ocean dilution.

Evaluation scenes use NumPy PCG64 seed $824601$; the 24 training scenes use $824602$. All six classes receive weight $1/6$. Gains are computed per scene before averaging; error bars are class-stratified Monte Carlo standard errors, not astrophysical uncertainties. The baseline resolved--disk comparison uses exact covariance propagation. Figure~\ref{fig:information_gain}(c) holds the full 128-channel mode fixed and changes only the disk residual covariance:
\begin{equation}
 C_d(\rho_C)=\Disk C_r\Disk^T+(\rho_C^2-1)(0.001)^2K_\nu,\qquad
 (K_\nu)_{ab}=e^{-|\ln(\lambda_a/\lambda_b)|/0.12},\quad\rho_C\geq1.
 \label{eq:rhoc}
\end{equation}
Here $C_r$ includes the discrepancy term when that comparison is selected. The control $\rho_C$ multiplies an added disk-only residual mode; it is not a ratio of total disk and resolved errors. At $\rho_C=1$, panel (c) exactly matches panel (a)'s 128-channel reference. Values $1,1.5,2,3,4,6,8$ test the stated additional covariance assumption with identical parameters, priors, exposure convention, and scene population.

\section{Results}\label{sec:results}

\subsection{Resolved spectral cubes and retrieved property maps}

Figure~\ref{fig:spectral_cube} shows source radiance, the blurred/noisy measurement, and reconstruction for the specified scalar scene. Cloud fraction, gas path length, surface reflectance, and illumination determine its spatial structure. The wavelength-dependent error budget separates regularization bias and noise against the propagated covariance. Spectral contrasts and their covariance are required to interpret molecular or surface features.

\begin{figure}[tbp]
\noindent\makebox[\textwidth][c]{\includegraphics[width=0.82\textwidth]{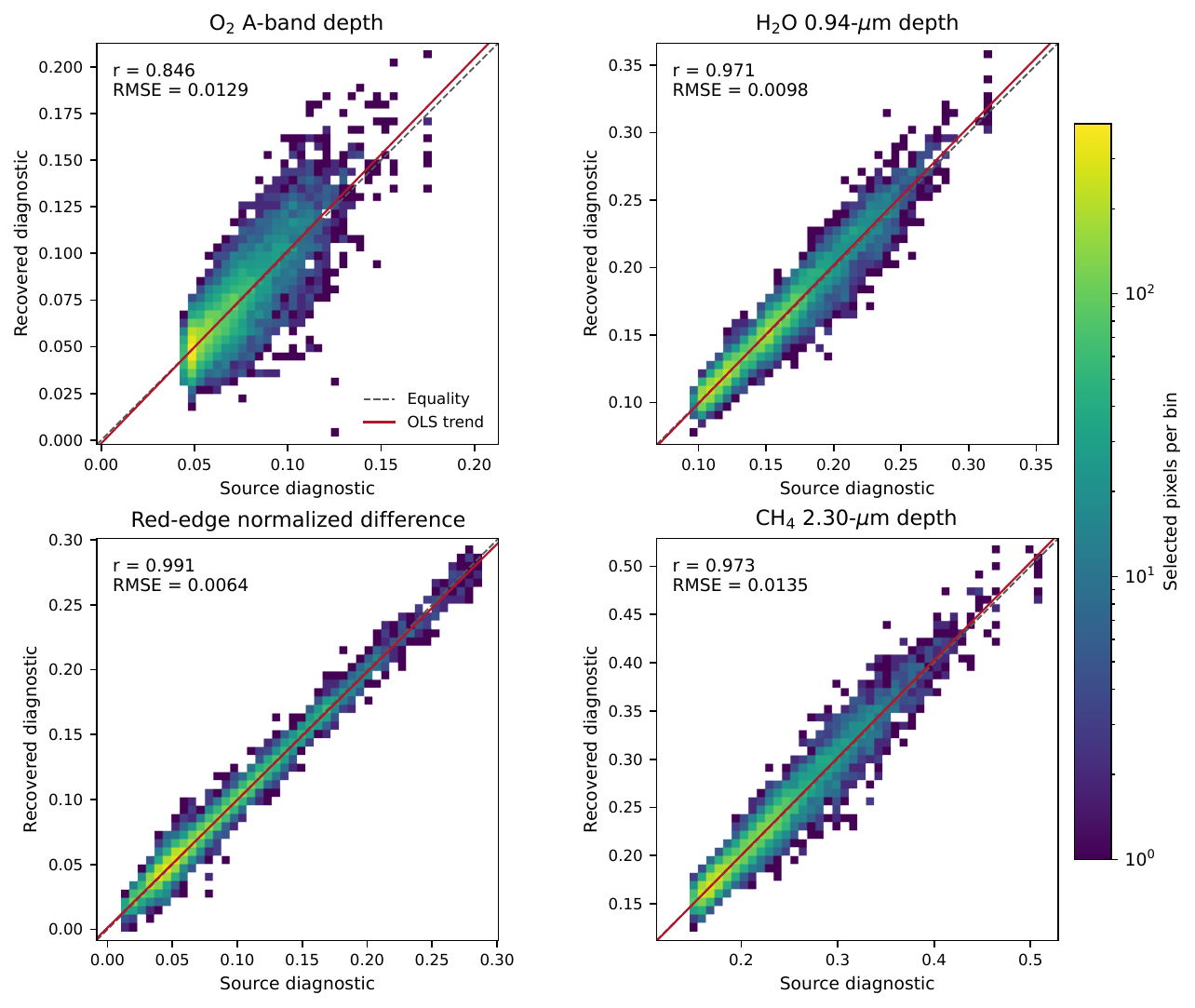}}
  \caption{\justifying Diagnostic-index recovery in the explicit scalar-cube reference of Fig.~\ref{fig:spectral_cube}. All $7,872$ geometrically selected pixels are included; Appendix~A specifies the common band/continuum weights and mask. Color shows counts per bin; the gray dashed line is equality and the red solid line is the unweighted least-squares trend. Correlations and RMSE are calculated from the unbinned source/recovered pairs. These statistics include the specified response bias and ratio noise; they are not gas-column errors or independent-radiative-transfer validation. No noisy quality cut or clipping is applied.}
  \label{fig:retrieved_maps}
\end{figure}

Figure~\ref{fig:retrieved_maps} compares source and reconstructed O$_2$, H$_2$O, red-edge, and CH$_4$ indices in the explicit scalar scene. Over the same $7,872$ geometrically selected pixels, their correlations are $0.846$, $0.971$, $0.991$, and $0.973$, with RMSE $0.0129$, $0.0098$, $0.0064$, and $0.0135$, respectively. These spectral diagnostics include response bias and ratio noise; they are not physical gas columns. Appendix~A defines every window, weight, mask, and numerical convention. The population conclusions below use the separate full-channel likelihood without index selection; physical retrieval requires Eq.~(\ref{eq:posterior}) and independent radiative transfer.

\subsection{Regional spectra and class-conditioned biomarkers}

\begin{figure}[tbp]
\noindent\makebox[\textwidth][c]{\includegraphics[width=0.84\textwidth]{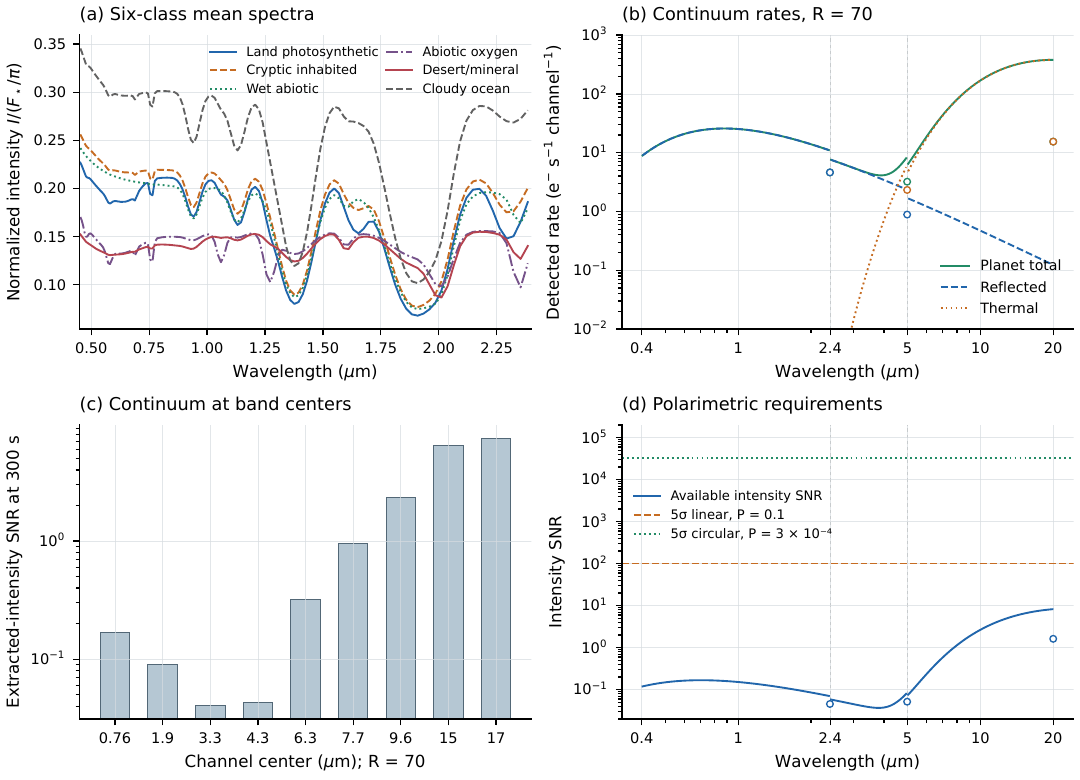}}
  \caption{\justifying Spectral context and photon requirements. (a) Mean normalized disk spectra of the 16 evaluation scenes per class in the $R=70$ reflected-light model. Individual scene Jacobians enter the Fisher calculation. (b) Absolute $R=70$ planet continuum channel rates, separating reflected and emitted photons; dashed lines mark detector arms and hollow circles partial-width bins. (c) Extracted-intensity SNR for $300\,{\rm s}$ in diagnostic-centered channels, not molecular band-depth significance. (d) Available intensity SNR and the intensity requirements for $5\sigma$ linear ($P=0.1$) and circular ($P=3\times10^{-4}$) polarization with efficiency $0.7$ and photon allocation $1/2$. No polarized measurement or retrieval is simulated.}
  \label{fig:regional_spectra}
\end{figure}

Figure~\ref{fig:regional_spectra} connects the two executed references. Panel (a) shows the mean disk-normalized spectrum of each of the six synthetic classes in the reflected-light experiment. Panels (b)--(d) use the absolute continuum budget to show the reflected/thermal photon transition, channel SNR near diagnostic wavelengths, and polarimetric requirements. The scientific purpose remains regional context: a biosignature candidate must be interpreted together with its surface, cloud, water, and atmospheric setting.

\subsection{Population information-gain audit}

The population analysis measures conditional parameter precision, not biological classification accuracy. Scene labels define atmospheric, surface, cloud, and false-positive stress tests. Equations (\ref{eq:information_gain}) and (\ref{eq:block_information_gain}) quantify total and marginal block gains after SGL reconstruction, covariance propagation, and nuisance marginalization. Nonnegative gain follows for an exact common-data disk projection; the numerical questions are its magnitude, block structure, and response to the null controls.

\begin{figure}[tbp]
\noindent\makebox[\textwidth][c]{\includegraphics[width=0.86\textwidth]{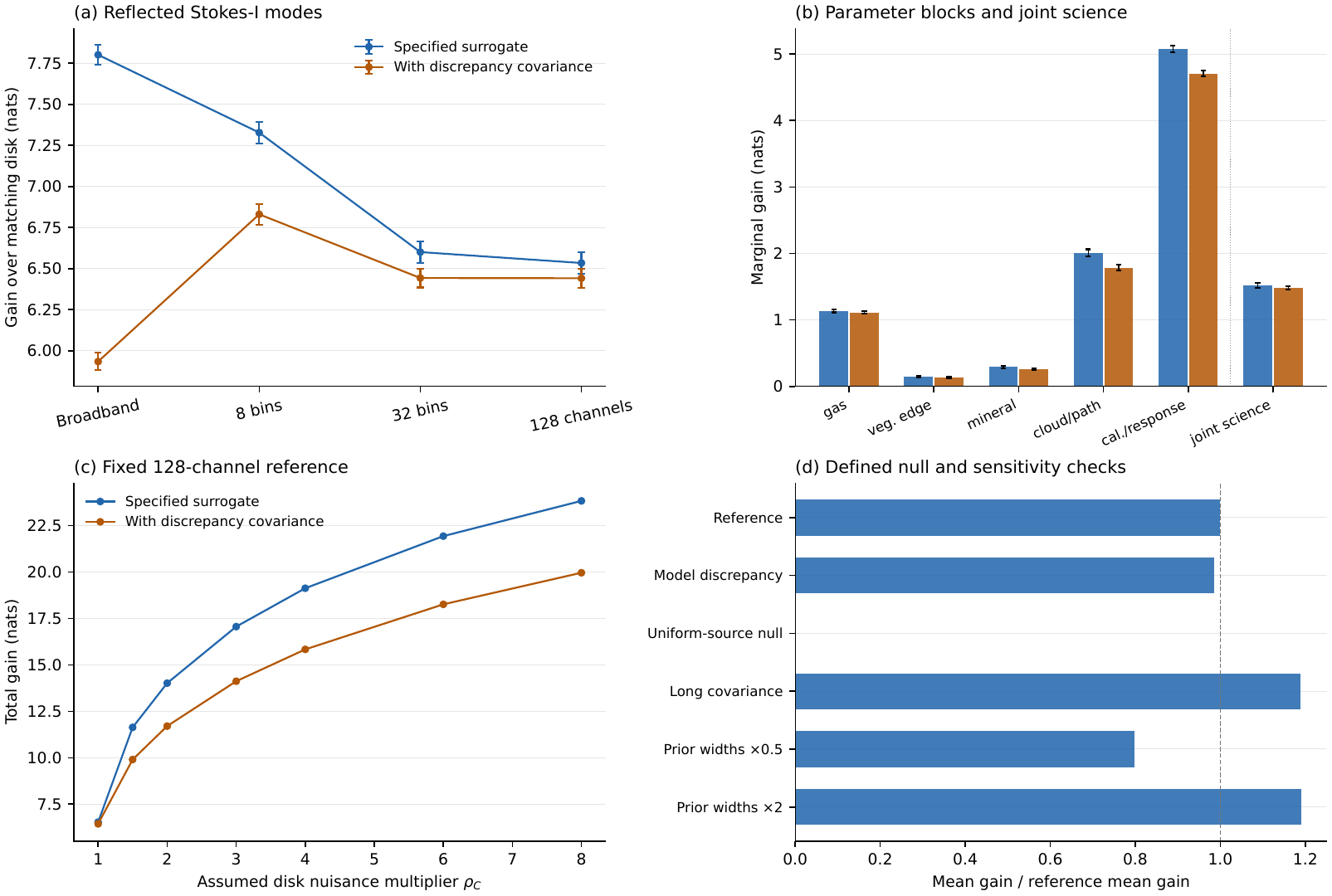}}
  \caption{\justifying Conditional Fisher-information audit for the explicit reflected-light reference. (a) Each resolved spectral compression is compared with its exact matching disk projection using common parameters and priors. Both absolute information measures increase with spectral sampling; the baseline resolved-minus-disk difference decreases. The discrepancy series first rises from broadband to eight bins. (b) Marginal block gains for 128 channels, including the joint nine-coordinate gas/vegetation/mineral block. The correlated blocks do not add to the total. (c) The identical 128-channel reference with disk-only covariance augmentation from Eq.~(\ref{eq:rhoc}); $\rho_C=1$ reproduces panel (a), while larger values include an assumed additional disk uncertainty. (d) The reference, trained discrepancy covariance, uniform-source null, long-range covariance, and prior-width controls, normalized by the baseline mean gain. The orange series adds centered discrepancy covariance trained on 24 independent scenes. Error bars in (a)--(b) and bands in (c) show one class-stratified Monte Carlo standard error over the 96 evaluation scenes. All modes contain only single-epoch Stokes $I$; the experiment demonstrates conditional precision and numerical consistency, not unbiased recovery or evidence for life.}
  \label{fig:information_gain}
\end{figure}

Figure~\ref{fig:information_gain} gives a total 128-channel gain of $6.534\pm0.064$ nats, or $6.441\pm0.058$ with independently trained discrepancy covariance. The science block comprising seven gas amplitudes, vegetation, and mineral edge parameters gains $1.518\pm0.035$ and $1.486\pm0.027$ nats after cloud and calibration/response marginalization. For gases and vegetation alone the corresponding gains are $1.272\pm0.027$ and $1.235\pm0.020$. The baseline science-block gain corresponds to a geometric-mean uncertainty-volume reduction $e^{1.518}=4.57$, not a uniform reduction in parameter standard deviations or biological model odds.

The baseline marginal blocks are calibration/response $5.075$, cloud/path $2.012$, gas $1.136$, mineral $0.292$, and vegetation edge $0.144$ nats. They are not additive. Much of the all-parameter gain therefore constrains response and cloud structure rather than biosphere-related spectral parameters. The mean resolved/disk posterior-standard-deviation ratio is $0.904$ for vegetation across all classes, but $0.451$ within the land-photosynthetic class; four classes have no vegetation, so that coordinate remains prior limited. These results condition on the adopted spatial templates and priors.

For the baseline in panel (a), relative gains for $1,8,32,128$ spectral bins are $7.801,7.328,6.600,6.534$ nats. Both absolute information quantities increase: relative to the common prior, the resolved information grows from $13.147$ to $34.420$ nats and the disk information from $5.346$ to $27.886$. The decreasing difference reflects a larger disk improvement, not an advantage of discarding spectral channels. Panel (c) uses the identical full-channel reference. At $\rho_C=4$, the baseline total $19.130$ nats is $6.534$ nats of spatial information plus $12.597$ nats introduced by the disk-only covariance assumption.

The uniform-source statistical null gives $|\Delta I|<3.3\times10^{-12}$ nat. Adding the long-range covariance of Appendix~\ref{app:traceability} increases the relative gain to $7.765$ nats because the disk product degrades more. Halving or doubling all prior widths gives $5.212$ or $7.781$ nats: prior dependence is material. These controls identify conditional information flow within the surrogate. The held-out shifts in Sec.~\ref{sec:model_mismatch} show that a preserved Fisher gain does not establish unbiased biosphere inference.

\subsection{Adversarial co-location and polarimetry}

The red-edge false-positive problem is retained as a first adversarial feature-level stress test rather than a demonstration of robust geological rejection.  The mineral population is deliberately assigned one red near-infrared slope, so a disk red-edge index alone is not sufficient within the surrogate model.  This is not an exhaustive library of minerals, soils, evaporites, hydrated surfaces, oxides, weathering states, grain sizes, mixed pixels, topographic effects, or cloud-correlated geology.  The numerical reference fits gas and surface-template amplitudes in spatially structured scenes. Habitat co-location, temporal recurrence, and Stokes-$V$ consistency are proposed extensions; none contributes to Fig.~\ref{fig:information_gain}.  ROC/AUC measures classification performance, whereas $\Delta I_m$ measures conditional parameter precision; neither validates the other. ROC/AUC within this closed surrogate family would test injected
separability, not establish discrimination on independent,
out-of-template planets.

The polarimetric channel must be interpreted conservatively.  Eq. (\ref{eq:v_snr}) shows that the required intensity SNR is very large.  The Stokes-$V$ and adversarial mineral $V/I$ amplitudes in Eq.~(\ref{eq:stokes_v}) define requirements, not propagated polarimetric measurements.  Circular polarimetry is best treated as a high-value consistency test that can strengthen a spatially and spectrally contextual interpretation only after instrumental polarization, coronal subtraction, and mineral circular-polarization mimics are controlled.

\subsection{Detectability versus integration time and spectral resolution}

\begin{figure}[tbp]
\noindent\makebox[\textwidth][c]{\includegraphics[width=0.82\textwidth]{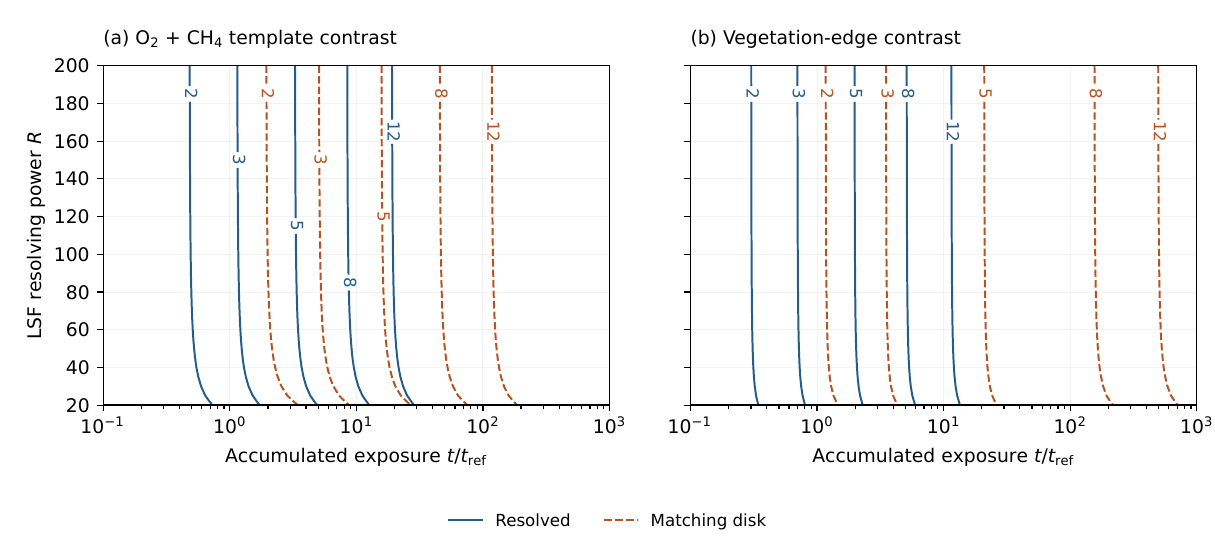}}
\caption{\justifying Expected template-contrast SNR from Eq.~(\ref{eq:detectability}) for the documented ten-cell reflected-light reference: (a) both O$_2$ and CH$_4$ amplitudes increased by 10\%; (b) vegetation-edge amplitude increased by 10\%. Solid blue and dashed orange contours use resolved measurements and their exact disk projection with common Gaussian nuisance priors. Contour labels give $Z$, not posterior odds for life. The 768 measurement channels are fixed while LSF resolving power varies; their noise normalization conserves spectral noise density. Exposure is accumulated per acquired cell in units of an uncalibrated $t_{\rm ref}$, and is not an hour-based mission forecast. The calculation uses the graph response of Eq.~(\ref{eq:reference_response}) and fixed scene morphology.}
\label{fig:detectability}
\end{figure}

Figure~\ref{fig:detectability} compares two fixed reflected-light template
contrasts in the first land-photosynthetic scene of the stated population:
a 10\% increase in both O$_2$ and CH$_4$ amplitudes, or a 10\% increase in
vegetation-edge amplitude from 0.450 to 0.495. The two means are evaluated
directly; all other scene fields remain fixed. For accumulated exposure
$x=t/t_{\rm ref}$ per acquired cell, the expected matched-filter contrast
SNR is
\begin{equation}
 Z_k^2(x,R)=\delta\boldsymbol\mu_k^{\mathsf T}
 \left[x^{-1}C_{\rm stat}(R)+J_{\nu,k}\Sigma_{\nu,k}
 J_{\nu,k}^{\mathsf T}\right]^{-1}\delta\boldsymbol\mu_k .
 \label{eq:detectability}
\end{equation}
Here $\delta\boldsymbol\mu_k$ is the finite hypothesis contrast, and
$J_{\nu,k}\Sigma_{\nu,k}J_{\nu,k}^{\mathsf T}$ marginalizes all unperturbed
coordinates using the Gaussian widths in Table~\ref{tab:fisher_numerical_definition}.
This nuisance covariance is fixed at the fiducial and used in both hypotheses;
no nuisance is fitted again. The disk product applies $\Disk$ to the contrast
and both covariance terms, without extra disk variance. The axis is normalized
exposure, not hours; acquiring all ten cells costs $10xt_{\rm ref}$ before
overheads. Thus $Z_k$ measures conditional template separation, not evidence
for life or a calibrated SGL detection time.

\subsection{Mission requirements and observing sequence}

Target changes require translation, not merely an attitude slew.
Two focal-line positions at the same heliocentric distance have
separation $L=2\bar z\sin(\Delta\theta/2)$, giving $L\simeq285\,\AU$
for $\Delta\theta=30^\circ$ at $550\,\AU$. Within one system,
Eq. (\ref{eq:compression}) gives $L\simeq(\bar z/z_0)s_\perp$:
a projected planet--planet separation $s_\perp=1\,\AU$ maps to
$1.57\times10^4$ km at the fiducial distances. Wide-angle transfers
are not assumed here; transfer time, propulsion, and power require
a separate mission-design calculation.

The simulations translate into the following mission and inference requirements.  The first requirement is a closed count and calibration hierarchy.  In the separate scalar benchmark, $Q_{\rm cor}^{\rm sc}/Q_{\rm exo}^{\rm sc}\simeq7.7\times10^4$; a multiplicative residual $\delta_{\rm flat}$ on the coronal signal produces a false planet-unit signal
\begin{equation}
  \delta_{\rm planet}\simeq\delta_{\rm flat}\frac{Q_{\rm cor}}{Q_{\rm exo}} .
  \label{eq:flat}
\end{equation}
An effective 1 ppm residual is therefore a 0.077 planet-unit signal in the fiducial case; 10 ppm is 0.77 planet units and would overwhelm weak spectral features.  Sub-ppm effective coronal/calibration closure is therefore a primary feasibility gate for SGL spectroscopy, not a downstream calibration refinement.  This is an end-to-end residual requirement after ring extraction, coronal modeling, detector calibration, pointing reconstruction, and template subtraction, not a raw detector-stability specification.

The second requirement is image-plane metrology.  Keeping residual coordinate errors below a fraction $\epsilon_x$ of a sample requires
\begin{equation}
  \sigma_x < \epsilon_x\Delimg
  =0.314\Big(\frac{\epsilon_x}{0.03}\Big)
  \Big(\frac{128}{n}\Big)
  \Big(\frac{30\,\pc}{z_0}\Big)
  \Big(\frac{\bar z}{650\,\AU}\Big){\rm m} .
  \label{eq:metrology}
\end{equation}
This image-plane coordinate tolerance requires in-situ acquisition, repeated registration, measured-coordinate inversion, and spacecraft-state estimation; it is not a prelaunch catalog or attitude requirement alone. At $650\,\AU$ the
one-way light time is about $3.75$ d, requiring autonomous local
control. Equation (\ref{eq:missiontime}) excludes cruise; propulsion,
power, telemetry, cost, and target-yield budgets are not closed here.

The third requirement is temporal sampling.  For an Earth-like rotation period, the equatorial smear during a subexposure is
\begin{equation}
  L_{\rm pix}=\frac{(2\pi R_\oplus/P_{\rm rot})t_{\rm sub}}{2R_\oplus/n},
  \label{eq:smear}
\end{equation}
The adopted $60\,{\rm s}$ subframes at $n=128$ and $30\,{\rm s}$ at $n=256$ each give $L_{\rm pix}=0.279$ for a 24-hour rotation period. Short subexposures do not make a months-long raster instantaneous: surface registration, phase coverage, and evolving clouds require a time-dependent forward model. The accumulated observing dwell is
\begin{equation}
  T_{\rm obs}\simeq f_{\rm oh}\frac{M_pN_\phi n^2t_{\rm samp}}{N_{\rm sc}},
  \label{eq:missiontime}
\end{equation}
where $M_p$ counts registered visits per phase bin at fixed precision, $N_\phi$ counts phase bins, $N_{\rm sc}$ counts interleaved spacecraft, and $f_{\rm oh}$ includes operational overheads. Seasonal monitoring also requires the elapsed seasonal baseline. The retained raster examples assign $t_{\rm samp}=1800\,{\rm s}$, giving $128^2t_{\rm samp}=341.33$ d before overheads. This is the scalar broadband normalization, not a demonstrated $R=70$ channel exposure. Assuming the same assigned dwell for each of 128 simultaneous or sequential channels gives $341$ d or $120$ yr, respectively. These examples isolate multiplexing costs; the absolute channel precision must instead be calculated from Eq.~(\ref{eq:spectral_count_integrals}). Sequential Stokes measurements require their own allocation factor.

\begin{table}[!htbp]
\caption{\justifying Illustrative raster-time accounting with $t_{\rm samp}=1800\,{\rm s}$, $n=128$, and one registered visit at $30\,\pc$ and $650\,\AU$. The assigned scalar broadband dwell is not a calibrated spectral-channel exposure. Rows isolate spectral multiplexing, Stokes allocation, revisits, and spacecraft parallelization. They retain one common assigned dwell to expose arithmetic scaling; actual wavelength-dependent precision requires the radiometric calculation. The thermal branch remains separately normalized.}
\label{tab:products}
\begin{tabular}{p{0.24\textwidth}p{0.40\textwidth}c
p{0.13\textwidth}}
\toprule
Case & Acquisition and architecture & Spacecraft / aperture & Photon dwell before overhead \\
\midrule
Full-resolution reflected image at reference dwell & Single registered scalar image-plane raster using the assigned scalar reference dwell; lower-SNR or coarser acquisition maps can be faster & $N_{\rm sc}=1$, $d=1\,{\rm m}$ & 341 d \\
Assigned Stokes-$I$ raster & Simultaneous 128-channel acquisition at the assigned reference dwell; channel precision not fixed by this table & $N_{\rm sc}=1$, $d=1\,{\rm m}$ & 341 d \\
Aperture trade in reflected cube & Same wavelength branch; scalar $c_W$ benchmark is approximately aperture neutral, while any $A_{\rm eff}^{-1}$ improvement requires a branch-specific optical/reconstruction model & branch-specific & not assigned \\
Sequential-channel strawman & 128 wavelength channels acquired one after another & $N_{\rm sc}=1$, $d=1\,{\rm m}$ & 120 yr \\
Sequential Stokes allocation example & Assumed factor $4/\eta_{\rm pol}^2$ relative to the intensity reference; allocation is counted once & $N_{\rm sc}=1$, $d=1\,{\rm m}$ & $7.6$ yr for $\eta_{\rm pol}=0.7$ \\
Sixteen registered revisits & One phase bin with $M_p=16$ registered cubes & $N_{\rm sc}=1$, $d=1\,{\rm m}$ & 14.9 yr \\
Parallelized revisits & Same $M_p=16$ revisits interleaved across 32 spacecraft in the focal region of the same target & $N_{\rm sc}=32$, $d=1\,{\rm m}$ each & 171 d \\
Short-/thermal-IR targeted spectra & External-occulter or cryogenic long-wavelength branch; must be recomputed from Eq. (\ref{eq:spectral_count_integrals}) rather than assigned the reflected-light dwell & branch-specific & not assigned \\
\bottomrule
\end{tabular}
\end{table}

Combining Eqs.~(\ref{eq:missiontime}) and (\ref{eq:cube_time_scaling}) gives the retained scalar reference law, anchored to the assigned dwell,
\begin{equation}
  T_{\rm refl}\simeq 341\,{\rm d}\; M_pN_\phi f_{\rm seq}\,P_{\rm rec}\,f_{\rm oh}
  \Big(\frac{\snr_{\rm req}}{\snr_{\rm ref}}\Big)^2
  \Big(\frac{R}{70}\Big)
  \Big(\frac{n}{128}\Big)^6
  \Big(\frac{z_0}{30\,\pc}\Big)^4
  \Big(\frac{\bar z}{650\,\AU}\Big)^{-3}
  \Big(\frac{N_{\rm sc}}{1}\Big)^{-1}
  \Big(\frac{B_p}{B_\oplus}\Big)^{-2} .
  \label{eq:observing_scaling}
\end{equation}
The rounded $341$ d anchor is $n^2t_{\rm samp}=341.33$ d for $n=128$ and $t_{\rm samp}=1800$ s; $M_pN_\phi=1$ for one registered cube.  Here $B_p$ is the reflected planet brightness relative to the fiducial Earth analog in the corona/background-dominated limit, $f_{\rm seq}=1$ for simultaneous spectral acquisition and $f_{\rm seq}\simeq N_\lambda$ for sequential channels, and $P_{\rm rec}=(G_{\rm scalar}/G_{\rm actual})^2$ is the reconstruction-covariance exposure penalty at fixed regional precision and acquired sample count.  The $B_p^{-2}$ term is appropriate when the local background and reconstruction covariance dominate; if a branch becomes planet-shot-noise limited the brightness scaling softens toward $B_p^{-1}$.  Equation (\ref{eq:observing_scaling}) makes the main feasibility levers explicit: target distance enters steeply as $z_0^4$, spectral resolution enters approximately linearly, spatial sampling enters as $n^6$ for a full raster, spacecraft parallelization reduces calendar time as $N_{\rm sc}^{-1}$, and the required recovered-SNR enters quadratically.

An Earth analog at $10\,\pc$ is therefore not merely three times easier than the fiducial $30\,\pc$ case.  At fixed $R$, $n$, $\bar z$, reconstruction prescription, and observing branch, the photon dwell decreases by $(10/30)^4=1/81$.  The assigned scalar raster time changes from $341$ d to $4.21$ d before overheads, while the sequential-channel strawman changes from $120$ yr to $1.48$ yr.  The closer target also has a three-times larger image cylinder, $\Dimg=4.01$ km instead of $1.34$ km at $650\,\AU$, and a three-times larger image-plane pitch for the same $n$; this relaxes meter-level metrology but increases the physical scan diameter.  Slew, acquisition, calibration, and station-keeping overheads can therefore become non-negligible for the closest targets even though the photon dwell is much shorter.

\begin{table*}[tbp]
\caption{\justifying Representative design-trade cases from the reflected-light scalar reference law.  These assigned-dwell scalar reference points expose the scaling drivers; their $R_0=70$ anchor is not a radiometrically calibrated channel exposure.  The entries are quoted before operational overheads and before any branch-specific improvements or degradations in throughput, multiplexing, aperture/effective-area law, or reconstruction covariance.  The $10\,\pc$ entries use the same assumptions as the $30\,\pc$ entries except for the $z_0^4$ factor.}
\label{tab:feasibility_cases}
\begin{tabular}{@{}p{0.22\textwidth}p{0.22\textwidth}cc
p{0.34\textwidth}@{}}
\toprule
Case & Scaling relative to fiducial & $30\,\pc$ & $10\,\pc$ & Interpretation \\
\midrule
Assigned scalar raster & $R_0=70$, $n=128$, $N_{\rm sc}=1$, simultaneous acquisition & $341$ d & $4.21$ d & Conditional single-spacecraft reference used to anchor architecture trades. \\
Sequential-channel strawman & Fiducial $\times128$ & $120$ yr & $1.48$ yr & Not a proposed mission mode; included only to show that spectral multiplexing is essential. \\
Reduced-$R$ full raster & $R=35$, $n=128$: $\times0.5$ & $171$ d & $2.10$ d & Lower spectral ambition substantially eases dwell and is a plausible first spectroscopy mode. \\
Coarse reflected map & $R=70$, $n=64$: $\times(1/2)^6$ & $5.33$ d & $1.58$ h & Illustrates that imaging and low-order mapping are much earlier objectives than full high-dimensional spectroscopy. \\
Low-$R$ coarse map & $R=35$, $n=64$: $\times0.5(1/2)^6$ & $2.66$ d & $0.79$ h & Useful as an acquisition and class-mask mode; for nearby targets it may become overhead-limited rather than photon-limited. \\
Sixteen registered revisits & Fiducial $\times16$ & $14.9$ yr & $67.4$ d & Single-spacecraft accumulated dwell for sixteen visits; seasonal timing imposes an additional calendar constraint. \\
Thirty-two-spacecraft revisit campaign & Fiducial $\times16/32$ & $171$ d & $2.10$ d & Illustrates calendar-time relief from parallelization; one possible architecture, not a required baseline. \\
Targeted $8\times8$ regional follow-up after mapping & $64/128^2$ at reference exposure; $\times6.7$ to retain scalar-covariance regional precision & $1.33$--$8.9$ d & $0.39$--$2.6$ h & $64$ samples at $1800\,{\rm s}$ each before the covariance penalty; a nonlocal SGL inverse may need wider sampling. \\
\bottomrule
\end{tabular}
\end{table*}

\begin{table*}[tbp]
\caption{\justifying Feasibility dependency tree for converting the present framework into a mission-performance forecast. Together with Table \ref{tab:feasibility_cases}, this table identifies which long-duration reference cases can be relaxed through architecture trades and which require new validation. The maturity labels refer to the status in this manuscript, not to the general maturity of the underlying fields.}
\label{tab:feasibility_dependency_tree}
\begin{tabular}{@{}p{0.22\textwidth}p{0.24\textwidth}p{0.20\textwidth}p{0.31\textwidth}@{}}
\toprule
Dependency or gate & Why it matters & Status in this manuscript & Closure needed \\
\midrule
SGL monopole geometry and image-cylinder scaling & Sets $\Dimg$, $\Delimg$, and the basic scan geometry & Mature theory / demonstrated analytic use & Include solar multipoles, plasma, finite solar disk, and time-dependent alignment in the flight model. \\
Reflected-light reconstruction & Explicit scalar map response and graph response for information tests; separate scalar $c_W$ radiometric penalty & Controlled reference calculations & Wave-optical SGL+telescope+coronagraph/occulter injection recovery and measured $C_{\rm rec}$. \\
Simultaneous spectral acquisition & Avoids the $128\times$ sequential-channel penalty & Plausible component technology; SGL implementation unvalidated & Annular IFS or dispersive ring extraction with throughput, wavelength registration, and calibration covariance. \\
Sub-ppm effective coronal/calibration closure & A 1 ppm residual is $0.077$ planet units in the scalar count hierarchy & Critical and currently unvalidated as an end-to-end residual & Off-ring sectors, solar monitoring, detector calibration, differential extraction, and template marginalization demonstrated in planet units. \\
External occulter / long-wavelength ring extraction & Enables the $2.4$--$20\,\mu{\rm m}$ architecture-level branch & Speculative for full-band SGL retrieval & Formation flying, diffraction/leakage propagation, cryogenic detectors, thermal-background covariance, and end-to-end throughput. \\
Vector RT, physical clouds, and thermal retrieval & Controls gas/cloud/path-length covariance and thermal-IR interpretation & Required; only surrogate and A--B mismatch used here & DISORT-class or equivalent vector RT, GCM/Earth scenes, cloud heights, BRDF/glint, and Earth-as-exoplanet validation. \\
Surface/mineral/biological library breadth & Controls false positives for pigment-like edges and surface context & Partial; one adversarial mineral edge & Laboratory/field spectra, mixtures, grain size, weathering, evaporites, altered basalts, biological surfaces, and blinded out-of-template tests. \\
Focal-region access and multi-spacecraft operations & Sets cruise, target-line acquisition, calendar time, and revisit cadence & Mission concept / not closed here & Propulsion, navigation, autonomy, communications, station-keeping, and constellation operations beyond $\sim650\,\AU$. \\
\bottomrule
\end{tabular}
\end{table*}

Tables \ref{tab:products} and \ref{tab:feasibility_cases} hold the reference scene and reconstruction prescription fixed to isolate architecture trades. Their dwell estimates require branch-specific calibration before they can set an observing schedule.

Imaging and spectroscopy should also be separated.  Broadband or limited-band imaging is the nearer-term and less demanding SGL product because it can use lower spectral dimension, coarser spatial sampling, and lower per-channel SNR while still delivering surface and cloud morphology.  Table \ref{tab:feasibility_cases} illustrates this distinction: a coarse $64^2$ reflected map is $5.33$ d at 30 pc and $1.58$ h at 10 pc under the scalar reference law, while a low-$R$ coarse map is $2.66$ d and $0.79$ h, respectively.  Full spatially resolved spectroscopy is more expensive because it carries the $R n^6$ full-raster scaling, channel calibration, and reconstruction-covariance penalties; it is also the scientifically transformative mode because it enables regional atmospheric composition, surface context, cloud state, thermal climate, and spatial co-location tests for possible biospheric activity.

The observing program is staged: reflected-light mapping, regional
spectroscopy, and then targeted thermal-IR or polarimetric follow-up.
These modes do not define a single mandatory payload. Hardware choices may evolve before launch; Table \ref{tab:feasibility_dependency_tree} specifies the calibration and validation requirements against which those choices must be evaluated.

\begin{figure}[tbp]
\noindent\makebox[\textwidth][c]{\includegraphics[width=0.84\textwidth]{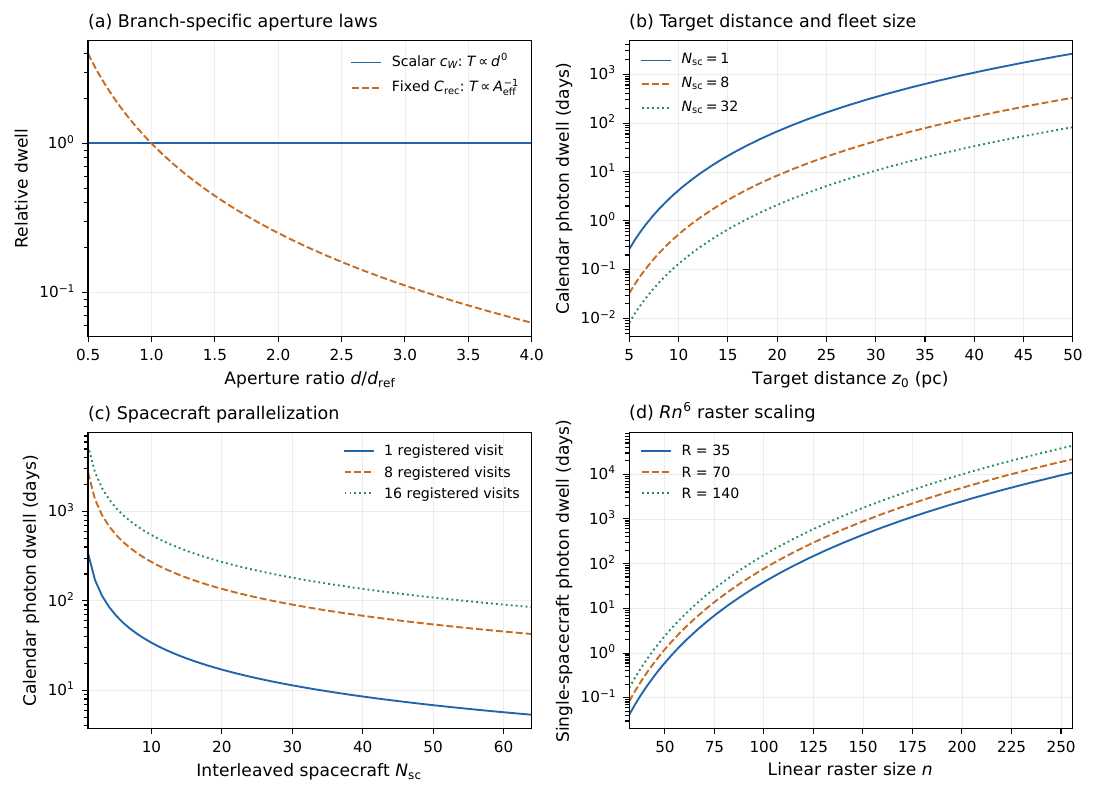}}
  \caption{\justifying Mission-time sensitivity around the fiducial reflected-light reference case.  The assigned scalar normalization in Table~\ref{tab:products} is $341$ d for one spacecraft, $d=1\,{\rm m}$, $z_0=30\,\pc$, $\bar z=650\,\AU$, $n=128$, and reference resolving power $R_0=70$. It illustrates relative design scalings, not a calibrated spectral-cube exposure.  Panel (a) separates two aperture laws rather than mixing them: the scalar aperture-averaged benchmark is approximately aperture neutral because $Q_{\rm exo}^{\rm sc}\propto d^2$, $Q_{\rm cor}^{\rm sc}\propto d^2$, $\snr_C\propto d$, and $g_n\propto d^{-1}$, while a branch with validated fixed reconstruction covariance and larger effective collecting area can approach $T\propto A_{\rm eff}^{-1}$.  Panel (b) shows the $z_0^4$ target-distance dependence for the displayed spacecraft counts; the separate $\bar z^{-3}$ focal-line scaling is given in Eq. (\ref{eq:observing_scaling}).  Panel (c) shows that $N_{\rm sc}$ changes calendar time but not total photon requirement.  Panel (d) shows the $R n^6$ raster scaling at fixed $z_0$, $\bar z$, aperture model, and spacecraft count; this scaling motivates low-$R$ mapping followed by regional spectroscopy.}
  \label{fig:mission_sensitivity}
\end{figure}

Figure \ref{fig:mission_sensitivity} summarizes the same point graphically.  The scaling laws are steep but not binary: target distance enters as $z_0^4$, operating distance along the focal line as $\bar z^{-3}$ in the scalar corona-limited benchmark, spacecraft parallelization as $N_{\rm sc}^{-1}$, resolving power as $R$, and full-raster spatial sampling as $n^6$.  The aperture dependence is branch-specific: the scalar aperture-averaged benchmark is approximately aperture neutral after the $g_n\propto d^{-1}$ reconstruction penalty, whereas a validated optical/effective-area branch with fixed reconstruction covariance can approach $T\propto A_{\rm eff}^{-1}$.  

These time scalings define architecture trades. A 128-channel, 128$\times$128, four-Stokes array occupies $\simeq34$ MB at 32 bits per value, excluding covariance, masks, calibration products, and raw frames. A practical data budget requires structured covariance storage and a telemetry model. Photon dwell, calibration, scan metrology, and reconstruction remain central requirements.

For the short-IR and thermal-IR regimes, the same time accounting must be repeated with wavelength-dependent throughput, detector noise, zodiacal and telescope thermal backgrounds, and long-wavelength ring-extraction covariance.  The $0.40$--$20\um$ performance curves in Figs. \ref{fig:extended_wave} and \ref{fig:spectral_performance_envelope} therefore do not assign the reflected-light 341 d dwell to every wavelength.  They support a staged hierarchy: optical/near-IR imaging and low-$R$ reflected maps first; class-conditioned $0.45$--$2.4\um$ regional spectra next; targeted short-IR and thermal-IR spectra for CH$_4$, CO$_2$, H$_2$O, O$_3$, N$_2$O, temperature, and cloud-top context after the map and covariance model exist.  Imaging is the earlier and less demanding objective.  Spectroscopy is harder, but it is the mode that connects surfaces, atmospheres, clouds, thermal emission, temporal variability, and possible biosignature indicators into a physically testable planetary context.

\begin{figure}[tbp]
\noindent\makebox[\textwidth][c]{\includegraphics[width=0.82\textwidth]{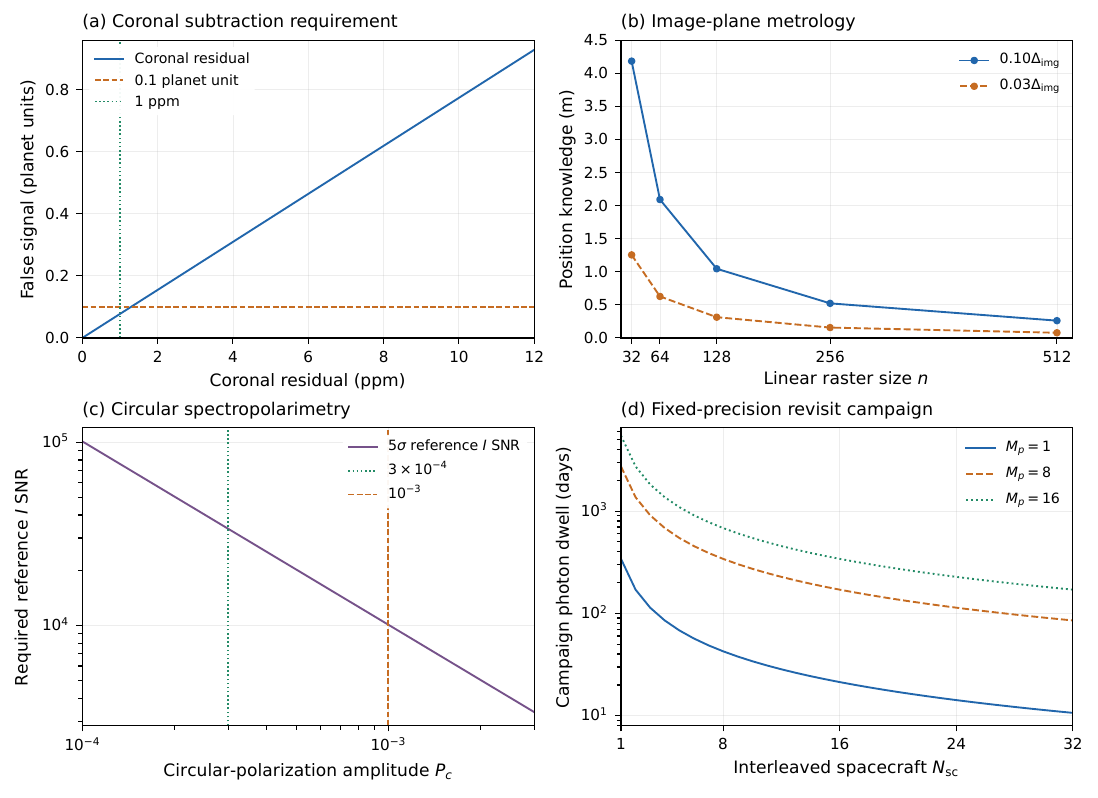}}
  \caption{\justifying Analytical observing requirements for the reflected-light reference branch. Panel (a) converts coronal residuals to planet-signal units using $Q_{\rm cor}/Q_{\rm exo}=6.20\times10^9/(8.01\times10^4)$. Panel (b) gives coordinate-error bounds $0.10\Delimg$ and $0.03\Delimg$ at $30\,\pc$ and $650\,\AU$. Panel (c) uses $m=5$, $\eta_{\rm pol}=0.7$, and $f_V=1/2$ in Eq. (\ref{eq:v_snr}). Panel (d) gives $T=341\,{\rm d}\,M_p/N_{\rm sc}$ for one, eight, and sixteen registered visits at fixed precision per visit, with one phase bin and no overheads. More visits increase campaign dwell; spacecraft parallelization reduces observing time but not a seasonal monitoring baseline. All panels are regenerated from the stated scalar formulas; they do not establish end-to-end calibration or mission feasibility.}
  \label{fig:requirements}
\end{figure}

The final requirement is statistical.  Circular polarimetry is extraordinarily demanding and should be treated as a high-value, systematics-limited extension, not a prerequisite for the basic resolved-spectroscopy claim.  Weak or non-Gaussian channels require a properly marginalized hierarchical retrieval rather than a naive independent-feature score.  Figure \ref{fig:requirements} summarizes the calibration, metrology, polarimetric-SNR, and parallelization requirements discussed above.  These are the engineering conditions under which the statistical framework can become a flight-quality inference pipeline.

\section{Discussion}\label{sec:discussion}

\subsection{What is robust}

Several conclusions do not depend on the precise numerical values in the conditional population information audit or on any secondary classifier diagnostic.  First, disk-integrated spectra are intrinsically mixed measurements.  They are essential for precursor discovery and first characterization, but they do not by themselves determine whether a spectral feature arises from a cloud, surface, gas abundance, path length, or spatially localized biosignature.  Second, resolved spectroscopy supplies class-conditioned spectra and spatial co-location tests.  These are scientifically distinct from imaging-only maps and from disk-integrated spectra.  Third, localized surface biosignatures can benefit strongly from spatial resolution because their disk amplitudes are diluted by surface coverage and clouds.  Fourth, emitted-light spectra supply a different context axis: temperature, cloud-top height, heat redistribution, greenhouse state, and mid-IR molecular bands that cannot be inferred robustly from reflected light alone.  The strongest mature SGL biosignature-context case is therefore likely a combined reflected-light map plus region-conditioned thermal-emission spectra, not either component in isolation.  Fifth, false-positive rejection requires context: O$_2$ must be interpreted with H$_2$O, CO, O$_4$, CO$_2$, stellar UV, and surface/climate state; red-edge-like features must be interpreted with mineral spectra, clouds, geography, and time variability.  Sixth, circular spectropolarimetry is potentially powerful but should be treated as aspirational until systematic calibration is demonstrated.

\subsection{Opportunity cost and relationship to HWO- and LIFE-class observatories}

The SGL mission concept is a resolved-context follow-on observatory rather than a first-line survey.  HWO-class coronagraphic or starshade observations are the natural discovery and first-characterization stage for optical and near-IR disk spectra of nearby terrestrial planets; LIFE-class mid-IR nulling interferometry is the natural disk-integrated route to thermal spectra, temperature context, and molecules such as CO$_2$, H$_2$O, O$_3$, CH$_4$, and N$_2$O in the $5$--$20\,\mu{\rm m}$ range \cite{Astro2020,Quanz2022LIFEI,Hansen2022LIFEIV}.  The SGL adds a different observable: spatially resolved and spatially correlated spectra of a previously identified high-priority planet.  The reference reflected-light cube illustrates the opportunity cost.  An assigned scalar dwell of $1800$ s per sample gives $341$ d for one raster; assigning that dwell to each of 128 sequential channels gives $120$ yr. These arithmetic comparisons expose multiplexing costs and do not set absolute spectral precision.  At fixed recovered-pixel precision the scalar raster law is $T_{\rm cube}^{\rm sc}\propto R n^6 z_0^4\bar z^{-3}/N_{\rm sc}$ with no first-order aperture improvement in the scalar $c_W$ benchmark; any favorable aperture law must come from a specified optical/effective-area branch with its own $C_{\rm rec}$.  The present $C_{\rm rec}$ bracket adds $2.6$--$5.3\times$ dwell for two- to three-times broader Gaussian correlations, or $\simeq6.7\times$ for the illustrative heavy-tailed covariance.  These scalings define the SGL role: HWO/LIFE-like facilities identify and triage promising targets, while the SGL supplies regional spectra, co-location tests, cloud/surface/gas covariance separation, temporal/rotational persistence tests, and spatial context that disk-integrated instruments cannot provide.  Table \ref{tab:hwo_life_sgl} summarizes this opportunity-cost relationship.

\begin{table*}[tbp]
\caption{\justifying Opportunity-cost comparison for SGL resolved-context characterization.  HWO- and LIFE-class observatories provide discovery and disk-spectral triage; the SGL provides the regional spectroscopic follow-on capability needed to test whether candidate biosignatures are spatially coherent with planetary context.}
\label{tab:hwo_life_sgl}
\begin{tabular}{@{}p{0.17\textwidth}p{0.23\textwidth}p{0.23\textwidth}p{0.34\textwidth}@{}}
\toprule
Observatory class & Principal data product & Natural role & What the SGL adds \\
\midrule
HWO-class coronagraph/starshade & Optical/near-IR disk spectra and photometry of nearby terrestrial planets & Discovery, orbit confirmation, first atmospheric and surface-color characterization & Regional optical/near-IR spectra, surface/cloud/gas co-location, and spatial covariance separation. \\
LIFE-class nulling interferometer & Mid-IR disk thermal spectra over the key climate and molecular bands & Thermal characterization, climate context, disk-integrated O$_3$, CO$_2$, H$_2$O, CH$_4$, and N$_2$O constraints & Spatially conditioned thermal-IR follow-up if an occulter/cryogenic SGL branch is validated. \\
SGL resolved-context mission & Scanned Einstein-ring data reconstructed into regional spectral products & Spatially resolved biosignature characterization of one or a few high-priority planets & Tests whether ambiguous disk biosignatures are regionally coherent with clouds, surfaces, water context, temperature, time variability, glint, and false-positive controls. \\
\bottomrule
\end{tabular}
\end{table*}

\subsection{What is model-dependent}

The population gains depend on the analytic band templates, class and field distributions, fixed spatial morphology, graph response, normalized noise, and common priors. The discrepancy-covariance result changes the joint science gain from $1.518$ to $1.486$ nats, but the held-out linearized bias can exceed six local standard deviations. Precision retention is therefore not a test of reliable biosphere inference. The common-data Fisher ordering is analytic; its magnitude and block structure are conditional on the specified experiment. Additional disk covariance produces a separate gain and must not be attributed to spatial resolution.

Multiple scattering couples species-dependent absorption to clouds,
aerosols, and surface reflection
\cite{Stamnes1988,Toon1989,Marley1999,LacisOinas1991}.
The structural Model A--Model B test does not calibrate these errors
or establish unbiased retrievals; covariance inflation cannot remove
a nonzero mean model discrepancy. Independent vector-RT and
Galileo/EPOXI Earth-data tests remain required
\cite{Sagan1993,Robinson2011Earth,Villanueva2018}.

\subsection{False positives and false negatives}

Resolved spectroscopy adds tests of abiotic explanations, but spatial correlation alone is not a biological discriminator.  Abiotic geology produces spatial correlations.  Minerals can correlate with topography, which can correlate with clouds.  Deserts, evaporites, oxidized surfaces, hydrated minerals, altered basalts, hydrothermal deposits, grain-size effects, and weathering products can mimic broad pigment-like slopes or produce spatially coherent spectral structure.  Abiotic oxygen can coexist with misleading surface colors if water loss, CO, and O$_4$ are not constrained.  The adversarial mineral simulation is a first illustrative stress test, not an exhaustive geological library, and the population diagnostic results do not prove robust biological/abiotic discrimination for real surfaces.

False negatives are equally important.  The worked surface example emphasizes a land photosynthetic biosphere; localized pigment, seasonal recurrence, and circular polarization are candidate diagnostics, not a measured ranking of biological evidence.  A marine biosphere, microbial mat, subsurface ecosystem, low-productivity biosphere, pre-oxygenic world, or nonphotosynthetic biosphere may produce weak or absent surface structure.  For a cryptic biosphere, inhabited and abiotic spectra may overlap even when nuisance parameters are well constrained.  The Fisher audit does not quantify the corresponding missed-detection rate.  The proper conclusion is not that the SGL will recognize all life.  It is that the SGL can test a much richer set of hypotheses for life that modifies atmosphere, surface, or time variability at observable levels.

The hypothesis set could also include nightside city lights
\cite{Beatty2022CityLights}, industrial gases such as CF$_4$ and
CCl$_3$F \cite{Lin2014Pollution}, and seasonal agricultural land-use
signatures \cite{LockleyVisioni2021}. City lights require a nightside
source term absent from Eq. (\ref{eq:intensity}); industrial gases
require additional thermal-IR opacity templates; agriculture requires
discrimination from natural vegetation cycles and cloud variability.
None is simulated here, and no SGL detectability forecast is assigned.

\subsection{Role of emitted-light spectroscopy in biosignature context}\label{sec:thermal_context}

The $2.4$--$20\um$ component is a high-priority future observing branch.  Its scientific leverage is different from the reflected-light branch.  Reflected spectra provide albedo, Rayleigh scattering, O$_2$/O$_4$, short-near-IR H$_2$O/\allowbreak CH$_4$/\allowbreak CO$_2$/\allowbreak CO, surface color, pigment/mineral slopes, glint candidates, and cloud morphology.  Thermal spectra provide the planet's own energy budget: brightness temperature, day-night and regional heat redistribution, cloud-top temperature, greenhouse state, H$_2$O at $6.3\um$, O$_3$ at $9.6\um$, CO$_2$ at $15\um$, CH$_4$ at $7.7\um$, and N$_2$O near $7.8$ and $17\um$.  These observables are not redundant.  For example, O$_3$ in reflected light and O$_3$ at $9.6\um$ test different combinations of abundance, temperature, cloud state, and vertical weighting; CO$_2$ at $2.0$ and $15\um$ test different parts of the column and climate context; and a vegetation-like reflected edge is more credible if it occurs in regions with plausible thermal environment and recurrent cloud/liquid-water context.

The reason the present population audit remains reflected-light only is validation status, not scientific priority.  A defensible thermal-IR retrieval requires a state vector and covariance model that the present surrogate does not yet contain: vertical temperature-pressure profiles, thermal emission and scattering, cloud-top height and emissivity, surface thermal inertia, day-night heat transport, line-by-line or correlated-$k$ opacity over $5$--$20\um$, cryogenic detector noise, telescope thermal emission, occulter leakage, and chromatic wave-optical SGL reconstruction.  

Registered reflected and thermal spectra constrain complementary nuisance directions. Reflected light couples surface and molecular features to clouds, aerosols, path length, and albedo; thermal emission couples temperature and cloud-top state to vertical structure and emissivity. Joint regional measurements test whether water/cloud context, surface class, climate, and disequilibrium support a physically coherent biosphere interpretation.

Figure~\ref{fig:thermal_branch_trade} combines the contrast crossover, continuum SNR, cumulative acquisition, and explicit instrument/background sensitivity. In the declared reference, favorable thermal channels require less acquisition than the $3$--$5\um$ transition; the common coronal floor makes some transition targets unreachable. The joint reflected-plus-emitted state vector addresses surface, water, redox, greenhouse, and temperature degeneracies, but demonstrating those constraints requires thermal-emission retrieval and a validated long-wavelength optical/reconstruction response.

Three computations would promote parts of the framework toward external mission-performance or biosignature-retrieval claims: a DISORT-class or comparable vector multiple-scattering radiative-transfer replacement for the surrogate $C_{\rm RT}$, a wave-optical SGL+telescope+occulter injection-recovery calculation that measures $C_{\rm rec}$, and a blinded out-of-template Earth-data injection-recovery test.  The blinded Earth-data test is the lowest-friction next step because it exercises the existing retrieval and information-audit pipeline against data not generated by the same modeling family, while still leaving the full wave-optical and thermal-instrument validation for later.

\subsection{SGL-specific limitations}

Figure \ref{fig:reconstruction_covariance} quantifies this sensitivity. Gaussian correlation lengths two to three times the reference reduce regional gain to $0.62$--$0.43$ of the scalar value, costing $2.6$--$5.3$ in exposure. The illustrative power-law tail, not derived from the SGL inverse, reduces $G_{\rm reg}$ from $7.77$ to $3.00$: a $61\%$ gain loss and $6.7$-fold exposure penalty. Flight-relevant reconstruction must determine the appropriate covariance.

Table~\ref{tab:load_bearing_approximations} identifies the assumptions of each numerical reference. The graph response controls the population Jacobian and covariance; the scalar convolution controls the map/index test; the analytical $c_W$ penalty and prescribed regional covariance control the continuum exposure estimates.

\begin{table*}[tbp]
\caption{\justifying Load-bearing approximations in the present framework and the validation needed before conditional information gains can be interpreted as external forecasts.}
\label{tab:load_bearing_approximations}
\begin{tabular}{@{}p{0.16\textwidth}p{0.28\textwidth}p{0.22\textwidth}p{0.31\textwidth}@{}}
\toprule
Approximation & Current treatment & Where it enters & Validation required \\
\midrule
Radiative-transfer model form & Structural A--B mismatch supplies a provisional $C_{m,AB}^{\rm RT}$; no DISORT-class RT yet & $J_m$, gas/surface/cloud coupling, $C_m^{\rm RT}$ & Vector multiple-scattering RT and Earth-as-exoplanet tests. \\
SGL optical/reconstruction model & Core-regularized scalar convolution for maps; graph $H,W$ in Eq.~(\ref{eq:reference_response}) for Fisher tests; separate scalar $c_W$ radiometry & Mean response and covariance in each stated experiment & Wave-optical SGL+telescope+occulter propagation and injection recovery. \\
Cloud and atmospheric structure & Fixed analytic cloud and path model & Gas-band depths, apparent column, nuisance covariance & Physical cloud fields, cloud heights, GCM/Earth scene ensembles. \\
Surface/mineral diversity & Simplified end members and one adversarial mineral edge & Surface and mineral blocks, $C_m^{\rm geo}$ & Laboratory and field spectra, mixtures, weathering, and grain-size variation. \\
Population generator & Classes and perturbations from one specified surrogate family & Marginal Fisher gains and held-out linearized shifts & Blinded out-of-template population tests with independent forward models. \\
\bottomrule
\end{tabular}
\end{table*}

Independent validation should fix the retrieval basis, nuisance model, and decision thresholds before evaluating held-out scenes from a separate radiative-transfer model. The scenes should span realistic cloud heights, mixed surfaces, weathering, grain size, pigments, and geological mimics, including cases outside the retrieval template family. Report parameter bias, interval coverage, false-positive and missed-detection rates, and sensitivity to scene selection. Earth-as-an-exoplanet data provide an empirical check. These tests would extend the present conditional methods result to claims of biological/abiotic separability.

Instrument validation requires wave-optical SGL+telescope+occulter injection recovery with solar multipoles and plasma, chromatic ring extraction, host-star/exozodiacal leakage, pointing errors, and time-dependent scans \cite{TuryshevToth2023Extended,TothTuryshev2023Rotating,Turyshev2026SGLExoimage}. The test must retain the reconstructed mean response, propagate coronal, detector, and thermal covariance, and compare recovered biases and regional precision with the predicted acquisition time. Thermal and polarimetric claims additionally require emission and modulation/demodulation models. The SGL gain does not remove coronal shot noise; Eq. (\ref{eq:flat}) gives the effective residual requirement. These demonstrations delimit future physical and mission claims rather than the Gaussian information identities established here.

The $0.45$--$2.40\um$ simulated cube is one branch of the $0.40$--$20\um$ observability problem. Internal coronagraphy requires demonstrated ring/limb separation, throughput, and leakage covariance; the $2\um$ CO$_2$ and $2.35\um$ CO bands already challenge a meter-class aperture. External occultation can extend smaller telescopes into the mid-IR, with different throughput, thermal, detector, and coronal covariance \cite{TuryshevToth2022Spectral}. The additional CH$_4$, CO$_2$, O$_3$, H$_2$O, and N$_2$O bands in Table~\ref{tab:thermal_case} motivate that extension. Converting the photon-budget envelope into thermal retrievals requires cryogenic/background control, thermal-emission RT, and wave-optical reconstruction; reflected-light population results do not establish those capabilities.

The strongest observing strategy is hierarchical.  A precursor observatory should first identify a nearby terrestrial planet with promising habitability and disk-integrated biosignatures.  The SGL mission should then acquire broadband maps, low-resolution Stokes-$I$ spectral cubes, cloud and rotation monitoring, class-conditioned spectra, and only then targeted deep spectra or polarimetry of the most diagnostic regions.  The spectral-performance envelope in Fig. \ref{fig:spectral_performance_envelope} makes the reason quantitative: first-generation full-planet cubes are naturally low-to-moderate resolution products, whereas $R>100$ spectroscopy, circular polarimetry, or thermal-IR measurements are better treated as targeted regional modes after the map and covariance model exist.  The first product should be a calibrated cube with a measured response and covariance sufficient for model comparison.

\section{Conclusions}\label{sec:conclusions}

This paper frames SGL astrobiology as covariance-aware resolved spectroscopic inference, not as a stand-alone survey or a synthetic classification exercise.  HWO- and LIFE-class observatories are the natural discovery and first-characterization stage: they can identify promising nearby terrestrial planets and obtain disk-integrated optical, near-IR, and mid-IR spectra.  The SGL role is complementary: its proposed resolved-context capability would test whether atmospheric disequilibrium, liquid-water context, cloud state, surface spectra, temperature diagnostics, temporal recurrence, and possible polarization signatures are spatially coherent on the planet itself.

The central contribution is a covariance-aware framework for assessing possible biospheres through class-conditioned spectra, regional cloud/gas/surface covariance, surface-mineral context, and co-location tests.  Equation (\ref{eq:contrast_loss}) identifies precisely which contrast between specified biological and abiotic hypotheses is discarded by disk integration under a common Gaussian covariance.  This complements the conditional Fisher precision audit without identifying parameter-volume reduction with evidence for life.  Figure~\ref{fig:information_gain} gives $1.518$ nats for the joint gas/vegetation/mineral block and $1.486$ nats after adding trained surrogate-discrepancy covariance. The larger all-parameter gain is dominated by response and cloud coordinates. Independent radiative transfer and bias/coverage tests are needed before interpreting these precision gains as reliable biosphere constraints.

The regional-spectroscopy feasibility argument is likewise conditional.  In the scalar benchmark a $1800$ s dwell gives a high convolved-ring statistic but only order-unity recovered broadband-pixel SNR, so the science product is a covariance-limited regional spectrum rather than a collection of independent high-SNR pixels.  If reconstruction correlations are two to three times broader than the scalar benchmark, an $8\times8$ regional coadd loses $38$--$57\%$ of its scalar SNR gain and requires $2.6$--$5.3$ times the dwell for the same regional precision.  An illustrative heavy-tailed covariance bracket loses $61\%$ of the scalar SNR gain and requires $\simeq6.7$ times the dwell.  This quantifies $C_{\rm rec}$ as a load-bearing mission requirement: regional spectroscopy remains the correct inference route, but its efficiency is set by the measured reconstruction covariance.

The spectral-performance calculation provides the observing hierarchy.  The reported numerical branch is a $0.45$--$2.40\,\mu{\rm m}$ Stokes-$I$ reflected-light surrogate/reconstruction calculation.  The $0.40$--$20\,\mu{\rm m}$ SGL observability envelope shows why a mature mission should preserve a path to optical imaging, emitted-light spectroscopy, full-band spectroscopy, and targeted polarimetry, especially through an external-occulter or equivalent long-wavelength ring-extraction architecture.  The $5$--$20\,\mu{\rm m}$ thermal regime may ultimately be one of the strongest SGL biosignature-context channels because it supplies temperature, cloud-top, greenhouse, O$_3$, CO$_2$, H$_2$O, CH$_4$, and N$_2$O information that reflected light cannot supply.  However, Stokes $Q/U/V$ retrievals and $2.4$--$20\,\mu{\rm m}$ thermal-IR biosignature retrievals are not demonstrated population forecasts in this paper; they are architecture-level observing modes that require polarimetric systematics closure, external-occultation or equivalent ring extraction, cryogenic/background covariance, wave-optical long-wavelength reconstruction, and a validated thermal-emission retrieval model.

The SGL is technically demanding: it requires access to the solar gravitational focal region, controlled solar-corona residuals, accurate image-plane navigation, simultaneous or highly multiplexed spectral acquisition, validated reconstruction covariance, and branch-specific optical or occulter closure.  

The SGL should not be described as an instrument that directly proves life.  Its value is that it could provide the spatial, spectral, temporal, and environmental context needed to assess whether an exoplanet is consistent with habitability, a complex biosphere, or possible biological activity.  A mature SGL data set would test whether atmospheric disequilibrium, liquid-water indicators, cloud state, surface heterogeneity, vegetation-like edges, glint, temperature structure, greenhouse state, and possible polarization signatures are co-located and physically coherent, or whether a conservative abiotic explanation is preferred.  This paper develops the covariance-aware framework, full-band radiometric bookkeeping, conditional mission-time sensitivity laws, and validation criteria needed to define that observing program.  Next steps are simultaneous spectral acquisition, sub-ppm effective coronal calibration, focal-region access, wave-optical SGL injection recovery, measured $C_{\rm rec}$, independent radiative-transfer and geological false-positive tests, physical cloud and thermal-IR retrieval models, and an observing sequence matched to the branch-specific $R n^6 z_0^4\bar z^{-3}$ scaling and the validated aperture/effective-area law.

\section*{Data and Code Availability}
All numerical data products are synthetic. The numerical source data, parameter definitions, and executable calculations for every figure and table are supplied as accompanying material. They include the scene cubes and masks, spectral-index pairs, Fisher and covariance arrays, continuum channel rates, and validation results. No observational exoplanet data are used.

\section*{Acknowledgments} 
The work described here was carried out at the Jet Propulsion Laboratory, California Institute of Technology, Pasadena, California, under a contract with the National Aeronautics and Space Administration.
\textcopyright 2026. California Institute of Technology. Government sponsorship acknowledged.

\appendix
\section{Numerical traceability}\label{app:traceability}

Tables~\ref{tab:covariance_branches}--\ref{tab:radiometric_references} distinguish the covariance framework from the executed population, scalar-scene, spectral-contrast, and continuum calculations. Table~\ref{tab:fisher_numerical_definition} specifies the population coordinates and priors; Table~\ref{tab:radiometric_references} specifies both absolute count configurations. Normalized inference noise has no assigned physical exposure.

For an explicitly ordered fitted vector, central differences at an interior fiducial point give
\begin{equation}
 [J_m(h)]_{:i}=\frac{f_m(\pars_0+h_i e_i)-f_m(\pars_0-h_i e_i)}{2h_i}.
 \label{eq:fisher_difference_specification}
\end{equation}
All 14 fitted coordinates are dimensionless and have untruncated Gaussian priors centered at zero. Log and logit coordinates keep the physical amplitudes admissible. Perturbations retain the same scene, spatial support, and covariance, recomputing the parameter-dependent mean and response. Halving the stated steps changes the tested Jacobians by at most $6.2\times10^{-7}$ relative and gains by $4.0\times10^{-6}$ nat. A complex-step derivative check differs by at most $1.9\times10^{-7}$ relative. Covariance derivatives are not part of this fixed-covariance Fisher experiment.

With $C_m=L_mL_m^T$ and a common positive-definite Gaussian prior covariance $\Sigma_P=L_PL_P^T$, a dimensionless stable evaluation is
\begin{equation}
 A_m=L_m^{-1}J_m L_P,\qquad
 \Delta I_m=\tfrac12\big[\ln|I+A_m^TA_m|-\ln|I+A_d^TA_d|\big].
 \label{eq:whitened_information_gain}
\end{equation}
The common prior determinant cancels. Cholesky solves and $\ln|I+A^TA|=\sum_k\ln[1+\sigma_k(A)^2]$ avoid determinant products. Marginal precision satisfies $F_{A|N}=[(F^{-1})_{AA}]^{-1}$; using $F_{AA}$ would condition on the other parameters. Independent posterior-covariance evaluations reproduce all 768 scene/mode/covariance gains and 3,840 marginal-block gains within $10^{-11}$ nat. All data covariances are positive definite; the largest condition number is $325$. Dense and structured covariance solves agree to numerical precision. Full-data Fisher information before and after the invertible $W$ agrees within $6.7\times10^{-16}$ relative.

A mean of local population gains is
\begin{equation}
 \overline{\Delta I_m}=\sum_{q=1}^{N_{\rm pop}}w_q\,
 \frac12\ln\frac{|F_{m,q}|}{|F_{d,q}|},\qquad
 w_q\geq0,\quad\sum_qw_q=1.
 \label{eq:population_gain_aggregation}
\end{equation}
It differs from a gain formed from averaged Fisher matrices. Here $N_{\rm pop}=96$, $w_q=1/96$. The standard error is $[\sum_{c=1}^6s_c^2/(6^2\,16)]^{1/2}$, where $s_c^2$ is the within-class sample variance of the gains. Prefixes of 4, 8, and 16 scenes per class give $6.33\pm0.14$, $6.45\pm0.09$, and $6.53\pm0.06$ nats; these assess sampling precision within this declared population. They do not establish convergence over real planet diversity.

The class order is land-photosynthetic, cryptic inhabited, wet abiotic, abiotic oxygen-rich, desert/mineral, and cloudy ocean. In the gas order of Table~\ref{tab:fisher_numerical_definition}, the class amplitude rows and surface/cloud references are
\begin{equation}
 X_{\rm cls}=\begin{pmatrix}
 1&1&1&0.8&1&0.08&1\\
 0.6&0.6&1&0.5&1&0.08&0.4\\
 0.05&0.03&1&0.05&1.5&0.7&0.03\\
 2&1.5&0.12&0.02&1.2&1.5&3\\
 0.2&0.1&0.15&0.05&1.3&0.5&0.2\\
 0.8&0.8&1.2&0.2&1&0.1&0.7
 \end{pmatrix},\qquad
 (o,v,m,c,\tau_c)_{\rm cls}=\begin{pmatrix}
 .45&.70&.10&.45&10\\
 .55&.02&.15&.55&12\\
 .60&0&.25&.50&10\\
 .15&0&.45&.25&7\\
 .05&0&.85&.20&6\\
 .90&0&.05&.80&18
 \end{pmatrix}.
 \label{eq:class_reference_inputs}
\end{equation}
These are scenario coefficients, not occurrence rates or measured abundances. Generate each class sequentially, with row-major cell ordering. For each scene, draw a $10\times5$ array of independent standard normals and left-multiply by the Cholesky factor of $K_x+10^{-12}I$, where $(K_x)_{pq}=\exp[-|x_p-x_q|^2/(2\,0.6^2)]$ and $x_p$ denotes the two-dimensional projected position. Let its columns be $z_0,\ldots,z_4$ and $s(u)=(1+e^{-u})^{-1}$. The fixed scene fields are
\begin{align}
 o_p&=s[\operatorname{logit}(o_{\rm cls})+0.8z_{0p}],\nonumber\\
 v_p&=v_{\rm cls}(1-o_p)s(z_{1p}),\quad
 m_p=m_{\rm cls}(1-o_p)s(-z_{1p}),\nonumber\\
 f_{c,p}&=s[\operatorname{logit}(c_{\rm cls})+0.5z_{2p}+0.25z_{0p}],\nonumber\\
 \tau_{c,p}&=\tau_{c,\rm cls}e^{0.15z_{4p}},\quad h_{c,p}=0.6z_{2p}.
 \label{eq:reference_scene_fields}
\end{align}
Then draw seven independent normals $u_j$ and set $X_{0,j}=X_{{\rm cls},j}e^{0.12u_j}$. Only water has a spatial multiplier, $\exp[0.20(o_p-\bar o)+0.08z_{3p}]$. Evaluation and training seeds are given in Sec.~\ref{sec:simulations}. The background surface is $A_{0,p}=0.03o_p+(1-o_p)[0.15+0.03\ln(\lambda/0.55\um)]$; $A_v=0.04+b_v s[(\lambda/\um-0.705)/0.018]$ and $A_m=0.15+b_m s[(\lambda/\um-0.72)/0.07]-0.06e^{-(\lambda/\um-0.9)^2/(2\,0.08^2)}$. Model B uses $(1-v_p-m_p)A_{0,p}+v_pA_v+m_pA_m$. The fitted transformations are in Table~\ref{tab:fisher_numerical_definition}.

The long-range noise control adds $0.5\exp[-|x_p-x_q|^2/(2\,0.9^2)]$ to $WW^T$ before disk projection. The uniform-source null repeats each disk Jacobian in all ten cells with unchanged covariance; it is a statistical sufficiency test, not a phase-$60^\circ$ planetary brightness map. Prior tests multiply every width by $0.5$ or $2$ at fixed likelihood. The model-discrepancy covariance is the centered 24-scene estimator in Eq.~(\ref{eq:ab_rt_covariance}), with no empirical shrinkage or added eigenvalue floor; the positive-definite baseline covariance stabilizes its rank-23 update.

For a declared read schedule, write a channel's count variance as $V_a(t)=v_at+r_a^2+s_a^2t^2$: $v_a$ contains Poisson count variance and any time-proportional read variance, $r_a^2$ is fixed read variance, and $s_a$ is residual count-rate uncertainty. For target recovered-pixel SNR $\nu$, let $D_a=(g_nQ_a^{\rm exo})^2-\nu^2s_a^2$. Then
\begin{equation}
 t_a=\frac{\nu^2v_a+\sqrt{\nu^4v_a^2+4D_a\nu^2r_a^2}}{2D_a},
 \qquad D_a>0.
 \label{eq:noise_aware_dwell}
\end{equation}
For positive Poisson variance, no finite exposure reaches the target when $D_a\leq0$. The radiometric calculation solves Eq.~(\ref{eq:noise_aware_dwell}) within each admissible integer-read interval with $r_a^2=N_{\rm fr}N_{\rm pix}\sigma_r^2$. Regional solutions divide the statistical terms by $G^2$ while leaving $s_a$ unchanged. Independent substitution checks all 4,673 finite roots at their actual read counts; 619 pixel cases are correctly identified as floor limited. With $R_p=6371\,{\rm km}$, the geometry gives $D_{\rm img}=1338.46\,{\rm m}$ and $g_{128}=0.181993$ for $d=0.4\,{\rm m}$. The prescribed $8\times8$ covariance gains are $7.77046$, $4.83024$, $3.36384$, and $3.00024$; they describe statistical coaddition, not the common floor.

The normalized focal intensity is $p_A(\psi)=\pi d^2[2J_1(u)/u]^2/(4\lambda^2)$, $u=\pi d\psi/\lambda$, and $\mathcal E(\theta,\lambda)=\int_{|\boldsymbol\psi|<\theta_{\rm ap}}p_A(|\boldsymbol\psi-\boldsymbol\theta|)d^2\boldsymbol\psi$. For the host, $\beta=a_p\sin60^\circ/z_0$, $u_\star=\beta/\theta_E$, $\theta_\pm=(\sqrt{\beta^2+4\theta_E^2}\pm\beta)/2$, and $\mu_\pm=(u_\star^2+2)/(2u_\star\sqrt{u_\star^2+4})\pm1/2$. Its unlensed photon irradiance is $\pi(R_\odot/z_0)^2B_\lambda(T_\star)\lambda/(hc)$, multiplied by $\mu_+\mathcal E(\theta_+,\lambda)+\mu_-\mathcal E(\theta_-,\lambda)$ and the same area/throughput/QE. Host finite-disk averaging is checked against this point-source reference.

The diffuse $I_\nu$ is specified after occultation, converted to $I_\lambda=I_\nu c/\lambda^2$ per metre, and integrated over $\pi\theta_{\rm ap}^2$; the effective optics contribution uses $0.10B_\lambda(40\,{\rm K})$ in the same entrance-radiance convention. Detector sampling is fixed within each arm at $\lambda_{\min}/(2d)$, with two spectral pixels per channel: $N_{\rm pix}=\lceil2\pi\theta_{\rm ap}^2/[\lambda_{\min}/(2d)]^2\rceil$, evaluated at the geometric channel center. This determines dark current and read variance. Partial-width channels remain in passband sums and appear as hollow circles; full-width curves are segmented by arm. Diagnostic rows use channels centered exactly at the stated wavelength via Eq.~(\ref{eq:channel_bounds}).

Rate integration uses 16-point Gauss quadrature per channel, an optical grid of 641 logarithmic wavelengths, 384 coronal radial nodes out to $256\theta_\odot$, 128 Airy chord nodes, and 24 radial nodes for the elliptic large-source pupil average. The finite-$J_0^2$ disk/pupil check uses 48 radial and 512 azimuthal nodes. Refinement and independent two-dimensional Airy/source checks agree at the stated tolerances; the Airy acceptance check differs by less than $1.9\times10^{-9}$ absolute. All 945 channel rows across the four resolving powers tile their bands and sum each physical background once. These numerical checks validate the idealized reference, not an external-occultation or thermal-retrieval instrument.

\paragraph{Scalar-cube and index reference.}
Figures~\ref{fig:spectral_cube} and \ref{fig:retrieved_maps} use a finite scalar
benchmark on $128^2$ source pixels, $x_i=(i+1/2-64)/64$, embedded in a $256^2$
zero source field at phase $60^\circ$. The first class in Eq.~(\ref{eq:class_reference_inputs}) fixes the surface,
cloud, and gas coefficients, without global gas scatter; the analytic spectrum
uses Table~\ref{tab:fisher_numerical_definition}. Five independent
PCG64 normal fields (seed $824609$), smoothed by a six-pixel Gaussian with
reflecting boundaries and four-standard-deviation truncation, are centered
and standardized over the illuminated disk and inserted into
Eq.~(\ref{eq:reference_scene_fields}) and its water multiplier. The $128$ channels span $0.45$--$2.40\,\mu\mathrm m$ with
$R=70$ and $16$ spectral quadrature subbins.
The normalized kernel and fixed inverse are
\[
 K_{uv}=\frac{(u^2+v^2+0.5^2)^{-1/2}\,\mathbf1_{u^2+v^2\le64^2}}
 {\sum_{uv}(u^2+v^2+0.5^2)^{-1/2}\,\mathbf1_{u^2+v^2\le64^2}},\quad
 \widetilde W=\frac{\widetilde H^*}{|\widetilde H|^2+10^{-4}},\quad
 \mathsf R=\mathsf W\mathsf H .
\]
The finite core, cutoff, and periodic inverse define this scalar illustration;
the zero margin prevents forward-source wrap. Independent Gaussian noise has
$\sigma_0=10^{-4}$ in normalized radiance per channel and full-grid pixel
(PCG64 seed $824610$), without an absolute exposure assignment. Thus
$\sigma_{\rm rec}^2=\sigma_0^2\langle|\widetilde W|^2\rangle
=9.09219\times10^{-6}$ and the covariance-lag array is
$\sigma_0^2\operatorname{IFFT}_2(|\widetilde W|^2)$.
The map residual is $(\widehat S-\mathsf RS)/\sigma_{\rm rec}$;
the separate $(\mathsf R-\mathsf I)S$ bias has RMS $0.003163$ over the
selected source cube. Response-centered residual RMS is $0.99896$.

Index windows below are in $\mu\mathrm m$. For window $[\lambda_L,\lambda_U]$,
channel weights are $w_a=|[\lambda_a^-,\lambda_a^+]\cap[\lambda_L,\lambda_U]|/(\lambda_U-\lambda_L)$.
Depth is $1-B/C$, with $C$ linearly
interpolating the two continuum-window means to the band midpoint; the red
edge is $(N-R)/(N+R)$. All $7,872$ pixels satisfying
$\mu_{\rm obs},\mu_\star\ge0.2$ are retained, without noisy selection,
continuum clipping, or index clipping. No recovered denominator is nonpositive.
Source and reconstructed indices use identical weights and masks.
\begin{center}
\begin{tabular}{llll}
\hline
Index & Band & Left continuum & Right continuum\\
\hline
O$_2$ & $0.750$--$0.770$ & $0.725$--$0.745$ & $0.780$--$0.800$\\
H$_2$O & $0.920$--$0.970$ & $0.865$--$0.895$ & $1.000$--$1.030$\\
CH$_4$ & $2.250$--$2.330$ & $2.140$--$2.190$ & $2.370$--$2.400$\\
Red edge & $R:0.650$--$0.680$ & $N:0.790$--$0.830$ & ---\\
\hline
\end{tabular}
\end{center}
Pearson correlation, unweighted RMSE against source indices, and ordinary
least-squares trends are evaluated over this common mask; spatially correlated
pixels are not treated as independent uncertainty samples. The four
$(r,\mathrm{RMSE})$ pairs are $(0.846,0.0129)$, $(0.971,0.0098)$,
$(0.991,0.0064)$, and $(0.973,0.0135)$ for O$_2$, H$_2$O, red edge, and
CH$_4$, respectively. Independent noise draws reproduce the analytic
variance within $0.18\%$; doubling spectral quadrature changes the tested
spectra by $3.6\times10^{-6}$ relative.

\paragraph{Spectral-contrast reference.}
For Fig.~\ref{fig:detectability}, the scene is the first land-photosynthetic
draw from PCG64 seed 824601, selected before evaluation. The gas alternative
sets $\theta_1=\theta_4=\ln1.10$; the edge alternative sets
$\theta_8=\operatorname{logit}(0.495/0.60)-\operatorname{logit}(0.450/0.60)
=0.451985$. All remaining coordinates use their Table~\ref{tab:fisher_numerical_definition}
prior widths and derivative steps. At each $R=20,25,\ldots,200$, the analytic
spectrum is convolved with a Gaussian LSF of log-wavelength FWHM $1/R$ and
averaged over $N=768$ equal-log bins spanning 0.45--2.40~$\mu$m. The statistical
covariance is $C_{\rm stat}=\sigma_N^2(WW^{\mathsf T})\otimes I_N$, with
$\sigma_N=0.004\sqrt{N/128}$; this scaling conserves the white-noise spectral
density under bin refinement. Four integration subbins per channel are used.
Across $R=20,70,200$ and $x=0.1,1,10,1000$, doubling the channel count changes
$Z$ by at most $2.1\times10^{-4}$ relative, and doubling the quadrature
density changes it by $1.3\times10^{-5}$. At $R=70$ and $x=1$, resolved/disk
$Z$ is 2.758/1.521 for the gas contrast and 3.549/1.869 for the edge contrast.

\begin{table*}[tbp]
\caption{\justifying Covariance-aware data products and validation requirements for an SGL biosignature-inference observation.  The branch list adapts high-contrast-spectroscopy covariance logic to the SGL cube while retaining SGL-specific reconstruction and ring-extraction terms.}
\label{tab:covariance_branches}
\begin{tabular}{@{}p{0.15\textwidth}p{0.22\textwidth}p{0.26\textwidth}p{0.33\textwidth}@{}}
\toprule
Branch & Native origin & Cube-space manifestation & Required validation \\
\midrule
Photon / detector & Planet, corona, zodiacal light, dark current, readout & Diagonal in independent counts; correlated after reconstruction & Count-rate tables and propagated $\mathsf W C_\epsilon\mathsf W^T$. \\
SGL reconstruction & Inverse operator and regularization & Spatially correlated $C_{\rm rec}$ & Injection recovery through an SGL PSF and scan library. \\
Coronal residual & Structured K/F corona and subtraction & Spatio-spectral-temporal covariance & Solar monitoring, off-ring sectors, and calibrated subtraction residuals. \\
Ring extraction & Coronagraph or occulter leakage & Extraction matrix and leakage covariance & End-to-end diffraction, occulter, and detector propagation. \\
Calibration & Flat field, gain, throughput, visit modes & Low-rank gray, color, and visit modes & Laboratory calibration and on-orbit closure. \\
Wavelength / LSF & Dispersion, focus, LSF and focus & Derivative weighted $C_\lambda+C_{\rm LSF}$ & Wavelength and LSF injection tests. \\
Astrophysical & Clouds, exozodi, host star, geology & Nuisance subspace $C_{\rm astro}$ & Earth/GCM scenes, host-star spectra, and spectral libraries. \\
RT model form & Missing multiple scattering, thermal emission, BRDF & $C_{\rm RT}={\rm Cov}(\delta\data_{\rm RT})$ & Vector RT and Earth-as-exoplanet data. \\
Polarimetry & Modulation, leakage, instrumental polarization & Stokes cross-covariance $C_{\rm pol}$ & Dual-beam and modulator null tests. \\
\bottomrule
\end{tabular}
\end{table*}

\begin{table*}[tbp]
\caption{\justifying Figure provenance and active numerical assumptions. Each quantitative figure is generated from its stated reference; scalar maps, population inference, spectral contrasts, and absolute continuum counts have distinct noise conventions.}
\label{tab:figure_covariance_map}
\begin{tabular}{@{}p{0.05\textwidth}p{0.27\textwidth}p{0.34\textwidth}p{0.26\textwidth}@{}}
\toprule
Fig. & Product & Active model/covariance & Statistic\\
\midrule
\ref{fig:framework} & Geometry, spectra, and sampling & Analytic geometry; normalized continuum rate shapes; explicit voxel dimensions & Geometric requirements and array size.\\
\ref{fig:spectral_cube} & Scalar radiance reconstruction & Specified core-regularized kernel, Wiener response, propagated Gaussian noise & Response-centered residuals and regularization bias.\\
\ref{fig:spectral_model} & Analytic bands and absolute rates & New Fisher templates; separate continuum/background budget & Optical depth, channel rates and SNR.\\
\ref{fig:extended_wave} & Extraction and backgrounds & Airy acceptance, coronal/host/detector noise, common floor & Acceptance, geometry and continuum SNR.\\
\ref{fig:reconstruction_covariance} & Regional coaddition & Explicit Gaussian/power-law covariance brackets & Equal-weight coadd gain.\\
\ref{fig:spectral_performance_envelope} & Absolute spectral envelope & New continuum model, scalar $g_n$, read schedule, coronal floor & Channel SNR and per-sample exposure.\\
\ref{fig:thermal_branch_trade} & $R=50$ thermal comparison & Same physical reference; instrument/background perturbations & Contrasts, exact regional totals and sensitivity.\\
\ref{fig:retrieved_maps} & Recovered spectral indices & Same scalar cube, fixed windows and geometric support & Pearson correlation, RMSE, and least-squares trends.\\
\ref{fig:regional_spectra} & Six-class spectra and requirements & New population means and continuum counts & Spectra, SNR and polarization requirements.\\
\ref{fig:information_gain} & Population information audit & Eqs.~(\ref{eq:reference_response})--(\ref{eq:rhoc}), common priors, trained discrepancy & Total, joint science, and marginal block gains.\\
\ref{fig:detectability} & Fixed-contrast resolution sweep & Exact disk projection; Gaussian noise and common nuisance priors & Matched-filter noncentrality versus $t/t_{\rm ref}$ and $R$.\\
\ref{fig:mission_sensitivity} & Scalar mission-time laws & Separate broadband normalization and raster scalings & Conditional relative dwell.\\
\ref{fig:requirements} & Calibration, metrology, polarimetry & Scalar count ratios and engineering tolerances & Analytical requirements.\\
\bottomrule
\end{tabular}
\end{table*}

\begin{table*}[tbp]
\caption{\justifying Complete 14-coordinate Fisher reference. All fiducials and prior means are zero; independent untruncated Gaussian widths $s_i$ and centered steps $h_i$ are dimensionless. $s(u)$ is the logistic function. Below, each idealized optical-depth template is $\widehat\tau_j(\lambda)=\sum_b a_b\exp[-(\lambda-\lambda_b)^2/(2\sigma_b^2)]$; triplets list $(\lambda_b,\sigma_b,a_b)$ with wavelengths in $\mu$m. These analytic bands are not measured opacities.}
\label{tab:fisher_numerical_definition}
\small
\begin{tabular}{@{}p{0.20\textwidth}p{0.59\textwidth}cc@{}}
\toprule
Ordered coordinates & Transformation / fiducial physical quantity & $s_i$ & $h_i$\\
\midrule
$1$--$7$: log gas amplitudes & $X_j=X_{0,j}e^{\theta_j}$; order O$_2$, O$_3$, H$_2$O, CH$_4$, CO$_2$, CO, O$_4$ & $0.6$ & $0.005$\\
$8$: vegetation edge & $b_v=0.60s[\operatorname{logit}(0.75)+\theta_v]$, $b_{v,0}=0.45$ & $0.7$ & $0.005$\\
$9$: mineral edge & $b_m=0.30s[\operatorname{logit}(2/3)+\theta_m]$, $b_{m,0}=0.20$ & $0.7$ & $0.005$\\
$10$: cloud fraction & $f_c=s[\operatorname{logit}(f_{c,0})+\theta_c]$ & $0.5$ & $0.005$\\
$11$: cloud path & $p_{\rm cld}=0.42e^{\theta_p}$ & $0.25$ & $0.005$\\
$12$: gray gain & Multiplicative $e^{\theta_g}$ & $0.10$ & $0.001$\\
$13$: spectral color & Multiplicative $e^{\theta_\lambda\ln[\lambda/(\sqrt{0.45\times2.4}\um)]}$ & $0.10$ & $0.001$\\
$14$: response scale & $H=e^{-0.8e^{\theta_b}L}$, fixed $W$ in Eq.~(\ref{eq:reference_response}) & $0.3$ & $0.005$\\
\bottomrule
\end{tabular}
\medskip
\begin{tabular}{@{}lp{0.81\textwidth}@{}}
\toprule
Template & Gaussian triplets\\
\midrule
O$_2$ & $(0.69,0.004,0.025)$; $(0.76,0.004,0.10)$\\
O$_3$ & $(0.60,0.09,0.05)$\\
H$_2$O & $(0.94,0.025,0.12)$; $(1.13,0.03,0.13)$; $(1.38,0.06,0.60)$; $(1.90,0.09,0.70)$\\
CH$_4$ & $(0.89,0.015,0.01)$; $(1.15,0.02,0.03)$; $(1.65,0.035,0.20)$; $(2.30,0.05,0.25)$\\
CO$_2$ & $(1.60,0.02,0.04)$; $(2.01,0.04,0.16)$\\
CO & $(2.35,0.025,0.18)$\\
O$_4$ & $(0.477,0.008,0.018)$; $(0.577,0.01,0.025)$; $(1.06,0.015,0.04)$; $(1.27,0.02,0.07)$\\
\bottomrule
\end{tabular}
\end{table*}

\begin{table*}[tbp]
\caption{\justifying Absolute continuum references for Figures~\ref{fig:spectral_performance_envelope} and \ref{fig:thermal_branch_trade}. Both use $R_p=6371\,{\rm km}$, $R_\odot=695700\,{\rm km}$, $GM_\odot=1.3271244\times10^{20}\,{\rm m^3\,s^{-2}}$, $z_0=30\,\pc$, $\bar z=650\,\AU$, $d=0.4\,{\rm m}$, and the source/optical model of Eqs.~(\ref{eq:planet_fullband_spectrum})--(\ref{eq:corona_reference}). Rates are predicted detected electrons per second, not measured calibration values. Differences between configurations arise from channelization and detector sampling, not a fitted amplitude.}
\label{tab:radiometric_references}
\small
\begin{tabular}{@{}p{0.43\textwidth}p{0.23\textwidth}p{0.23\textwidth}@{}}
\toprule
Quantity & Figure~\ref{fig:spectral_performance_envelope} & Figure~\ref{fig:thermal_branch_trade}\\
\midrule
Main resolving power; full-band channel count & $70$; $276$ & $50$; $197$\\
Modeled and reference-sum passband & $0.40$--$20\um$ & $0.40$--$20\um$\\
$K_{\rm occ}$; collecting area & $1$; $0.125664\,{\rm m^2}$ & Same\\
Source/pupil gain; ideal solar transmission & $2.26485\times10^6$; $0$ & Same\\
Throughput $\tau$ by arm: $0.4$--$2.4$, $2.4$--$5$, $5$--$20\um$ & $0.40,0.30,0.25$ & Same\\
Quantum efficiency by arm & $0.80,0.75,0.65$ & Same\\
Dark current [e$^-\,{\rm s^{-1}\,pixel^{-1}}$] by arm & $0.001,0.01,0.05$ & Same\\
Read noise [e$^-\,{\rm pixel^{-1}}$] by arm & $2,5,10$ & Same\\
Accumulated reference; subframe duration & $300\,{\rm s}$; $60\,{\rm s}$ ($30\,{\rm s}$ at $n=256$) & Same\\
Planet full-band rate & $20344.9$ & $20344.9$\\
Coronal full-band rate & $8.67035\times10^8$ & $8.67035\times10^8$\\
Unmasked host full-band rate & $3.03487\times10^8$ & $3.03487\times10^8$\\
Diffuse / optics full-band rates & $324.460$ / $1981.45$ & Same\\
Dark full-band rate & $2006.03$ & $1432.09$\\
Total background full-band rate & $1.1705268\times10^9$ & $1.1705262\times10^9$\\
Planet / background counts in $300\,{\rm s}$ & $6.10348\times10^6$ / $3.51158\times10^{11}$ & Same at displayed precision\\
Planet / optical-background photons in $300\,{\rm s}$, after throughput and before QE & $9.16219\times10^6$ / $4.54854\times10^{11}$ & Same at displayed precision\\
$10\um$ centered channel bounds & $10e^{\pm1/140}\um$ & $10e^{\pm1/100}\um$\\
$10\um$ planet / corona / host rates & $168.856$ / $1.15297\times10^6$ / $52448.0$ & $236.402$ / $1.61417\times10^6$ / $73429.3$\\
$10\um$ total background / detector pixels & $1.20544\times10^6$ / $236$ & $1.68762\times10^6$ / $236$\\
$10\um$ ring / $n=128$ pixel SNR, $300$ s & $2.663$ / $0.485$ & $3.151$ / $0.574$\\
$10\um$ pixel exposure for SNR $3$ [h] & $3.193$ & $2.280$\\
$10\um$ regional total, 64 samples, $G=7.77$ [h] & $3.385$ & $2.417$\\
Common coronal rate floor; statistical gains & $10^{-7}Q_a^{\rm cor}$; $3.00024,7.77046,8$ & Same\\
\bottomrule
\end{tabular}
\end{table*}

The full-band reference sums are independent of channel width up to quadrature error for optical components; dark and read terms depend on the declared detector sampling. At $10\um$, the wider $R=50$ channel raises the ring SNR from $2.663$ to $3.151$ at $300\,{\rm s}$ and reduces the pixel exposure from $3.193$ to $2.280$ h. This provides an internal cross-configuration check. Neither reference is normalized to the separate $1\,{\rm m}$ scalar benchmark in Eqs.~(\ref{eq:qexo})--(\ref{eq:qcor}).

\FloatBarrier

\bibliographystyle{apsrev4-2}

%

\end{document}